\documentclass[phd,tocprelim]{cornell}
\renewcommand{\caption}[1]{\singlespacing\hangcaption{#1}\normalspacing}

\let\ifpdf\relax
\usepackage[colorlinks=true, linkcolor=blue, citecolor=blue,
    urlcolor=blue, menucolor=blue]{hyperref}

\usepackage[letterpaper]{geometry}  % margin, paper size etc.
\usepackage{bm, bbm}   % bm for bold math, bbm for \mathbbm{1}
\usepackage{amsmath}   % for \overset, \underset, \text, \lVert, \rVert
\usepackage{amssymb}   % amssymb or amsfonts for \mathbb
\usepackage{graphicx}  % \includegraphics, \resizebox. graphicx is better than graphics.
\usepackage{cases}     % \begin{cases} for stack of equations after the bracket
\usepackage{multirow}  % \multirow and \multicolumn
\usepackage{longtable} % \begin{longtable} for long tables.
\usepackage{caption}   % \caption*{}, \captionof (while using \resize)
\usepackage{float}     % prevent table from floating by \begin{table}[H]
\usepackage{adjustbox} % \begin{adjustbox} to resize the table
\usepackage[title]{appendix} % more control of appendix
\usepackage{hhline}    % \hhline double lines in tables
\usepackage{array}     % vertical center in the table \begin{tabular}{ | m{1in} |... }
\usepackage{url}       % include url by \url{}
\usepackage{lmodern}   % (Optional) make \fontsize{15}{18}\selectfont available.

\newcommand{\vone}{\mathbbm{1}}      % command to write vector 1 in vector form
\newcommand{\norm}[1]{\left\| #1 \right\|}   % command to write norm
\newcommand{\w}{\widetilde}          % command to write wide tilde

\title{The Adaptive Multi-Factor Model and the Financial Market}
\author{Liao Zhu}
\conferraldate {May}{2020}
\degreefield {Ph.D.}
\copyrightholder{Liao Zhu}
\copyrightyear{2020}

\begin{document}

\maketitle
\makecopyright

\begin{abstract}
Modern evolvements of the technologies have been leading to a profound influence on the financial market. The introduction of constituents like Exchange-Traded Funds, and the wide-use of advanced technologies such as algorithmic trading, results in a boom of the data which provides more opportunities to reveal deeper insights. However, traditional statistical methods always suffer from the high-dimensional, high-correlation, and time-varying instinct of the financial data. In this dissertation, we focus on developing techniques to stress these difficulties. With the proposed methodologies, we can have more interpretable models, clearer explanations, and better predictions.

We start from proposing a new algorithm for the high-dimensional financial data -- the Groupwise Interpretable Basis Selection (GIBS) algorithm, to estimate a new Adaptive Multi-Factor (AMF) asset pricing model, implied by the recently developed Generalized Arbitrage Pricing Theory, which relaxes the convention that the number of risk-factors is small. We first obtain an adaptive collection of basis assets and then simultaneously test which basis assets correspond to which securities. Since the collection of basis assets is large and highly correlated, high-dimension methods are used. The AMF model along with the GIBS algorithm is shown to have significantly better fitting and prediction power than the Fama-French 5-factor model.

Next, we do the time-invariance tests for the $\beta$'s for both the AMF model and the FF5 in various time periods. We show that for nearly all time periods with length less than 6 years, the $\beta$ coefficients are time-invariant for the AMF model, but not the FF5 model. The $\beta$ coefficients are time-varying for both AMF and FF5 models for longer time periods. Therefore, using the dynamic AMF model with a decent rolling window (such as 5 years) is more powerful and stable than the FF5 model.

We also successfully provide a new explanation of the well-known low-volatility anomaly which pervades in the finance literature for a long time. We use the Adaptive Multi-Factor (AMF) model estimated by the Groupwise Interpretable Basis Selection (GIBS) algorithm to find those basis
assets significantly related to low and high volatility portfolios. These two portfolios load on very different factors, which indicates that volatility is not an independent risk, but that it is related
to existing risk factors. The out-performance of the low-volatility portfolio is due to the (equilibrium) performance of these loaded risk factors. For completeness, we compare the AMF model with the traditional Fama-French 5-factor (FF5) model, documenting the superior
performance of the AMF model.
\end{abstract}

\begin{biosketch}
Liao Zhu is a Ph.D. in the Department of Statistics and Data Science, Cornell University. Before that he was in the School of Gifted Young, the University of Science and Technology of China, where he finished his B.S. degree majoring statistics in 2015.
\end{biosketch}

\begin{dedication}
This document is dedicated to all Cornell graduate students.
\end{dedication}

\begin{acknowledgements}
I would like to thank my advisor - professor Martin Wells and my co-advisor - professor Robert Jarrow for their encouragements, insights and supports during my Ph.D. career. I am extremely grateful to their mentorship, which has a long-last influence to my life. I would also like to thank my committee members, professor David Mimno, David Matteson and David Ruppert for their generous support and advice.

I would like to thank Dr. Manny Dong for helping with the data access. Thanks to my collaborators, professor Sumanta Basu, Dr. Rinald Murataj, etc. Thanks to my parents, my friends, and all the supports from Cornell University.
\end{acknowledgements}

\contentspage
\tablelistpage
\figurelistpage

\normalspacing \setcounter{page}{1} \pagenumbering{arabic}
\pagestyle{cornell} \addtolength{\parskip}{0.5\baselineskip}

\chapter{Introduction}
Modern evolvements of the technologies have been leading to a profound influence on the financial market. The introduction of constituents like Exchange-Traded Funds, and the wide-use of advanced technologies such as algorithmic trading, results in a boom of the data which provides more opportunities to reveal deeper insights. However, traditional statistical methods always suffer from the high-dimensional, high-correlation, and time-varying instinct of the financial data. In this dissertation, we focus on developing techniques to stress these difficulties. With the proposed methodologies, we can have more interpretable models, clearer explanations, and better predictions.

We start from proposing a new algorithm for the high-dimensional \\financial data, the \textbf{Groupwise Interpretable Basis Selection (GIBS) algorithm} to estimate a new \textbf{Adaptive Multi-Factor (AMF) asset pricing model}, implied by the generalized arbitrage pricing theory (APT) recently developed by Jarrow and Protter (2016) \cite{jarrow2016positive} and Jarrow (2016) \cite{jarrow2016bubbles}. \footnote{The Appendix \ref{ap: summary_apt} provides a brief summary of the generalized APT.} This generalized APT is derived in a continuous-time, continuous trading economy imposing only the assumptions of frictionless markets, competitive markets, and the existence of a martingale measure.\footnote{By results contained in Jarrow and Larsson (2012) \cite{jarrow2012meaning}, this is equivalent to the economy satisfying no free lunch with vanishing risk (NFLVR) and no dominance (ND). See Jarrow and Larrson (2012) \cite{jarrow2012meaning} for the relevant definitions.} As such, this generalized APT includes both Ross's (1976) \cite{ross1976arbitrage} static APT and Merton's (1973) \cite{merton1973intertemporal} inter-temporal CAPM as special cases.

Compared to PCA based methods that construct risk-factors
from linear combinations of \textit{various} stocks, which are consequently
often difficult to interpret, our GIBS algorithm consists of a set of \textit{interpretable
and tradeable} basis assets. The new model explains more variation
in \textit{realized} returns. It is important to note here that in
a model of realized returns, some of the basis assets will reflect
idiosyncratic risks that do not earn a non-zero risk premium. Those
basis assets that have non-zero excess expected returns are the relevant
risk-factors identified in the traditional estimation methodologies.
While a few recent papers adopt high-dimensional estimation methods
for modeling the cross-section of expected returns and an associated
parsimonious representation of the stochastic discount function \cite{bryzgalova2015spurious,feng2020taming},
our empirical test is specifically designed to align with the generalized
APT model's implications using basis assets and realized returns.

To test the generalized APT, we first obtain the collection of all
possible basis assets. Then, we provide a simultaneous test, security
by security, of which basis assets are significant for each security.
However, there are several challenges that must be overcome to execute
this estimation. First, in the security return regression using the
basis assets as independent variables, due to the assumption that
the regression coefficients ($\beta$'s) are constant, it's necessary
to run the estimation over a small time window because the $\beta$'s
are likely to change over longer time windows. This implies that the
number of sample points may be less than the number of independent
variables ($p>n$). This is the so-called high-dimension regime, where
the ordinary least squares (OLS) solution no longer holds. Second,
the collection of basis assets selected for investigation will be
highly correlated. And, it is well known that large correlation among
independent variables causes difficulties (redundant basis assets
selected, low fitting accuracy, etc., see \cite{hebiri2012how,vandegeer2013lasso})
in applying the Least Absolute Shrinkage and Selection Operator (LASSO).
To address these difficulties, we propose a novel and hybrid algorithm
-- the GIBS algorithm, for identifying basis assets that are different
from the traditional variance-decomposition approach. The GIBS algorithm
takes advantage of several high-dimensional methodologies, including
prototype clustering, LASSO, and the ``1se rule'' for prediction.

We investigate Exchange-Traded Funds (ETFs) since they inherit and
aggregate the basis assets from their constituents. In recent years
there are more than one thousand ETFs, so it is reasonable to believe
that one can obtain the basis assets from the collection of ETFs in
the CRSP database, plus the Fama-French 5 factors. Consider the market
return, one of the Fama-French 5 factors. It can be duplicated by
ETFs such as the SPDR S\&P 500 ETF, the Vanguard S\&P 500 ETF, etc.
Therefore, it is also reasonable to believe that the remaining ETFs
can represent other basis assets as well. The AMF model along with the GIBS algorithm is shown to have significantly better fitting and prediction power than the Fama-French 5-factor model.

Next, we do the time-invariance tests for the $\beta$'s for both the AMF model and the FF5 in various time periods. The intercept (arbitrage) tests show that there are no significant non-zero intercepts in either AMF or FF5 model, which validates the 2 models. We show that the constant-beta assumption holds in the AMF model in all time periods with length less than 6 years and is quite robust regardless of the start year.

However, even for short time periods, FF5 sometimes gives very unstable estimation, especially in the financial crisis. This indicates that AMF is more insightful and can capture the risk-factors to explain the market shift during the financial crisis.

For time periods with length longer than 6 years, both AMF and FF5 fail to provide time-invariance $\beta$'s. However, the $\beta$'s estimate by the AMF is more time-invariant than the FF5 for nearly all time periods. This shows the superior performance of the AMF model.

Considering the two results above, using the dynamic AMF model with a decent rolling window (such as 5 years) is more powerful and stable than the FF5 model.

We also successfully provides a new explanation of the well-known low-volatility anomaly which pervades in the finance literature for a long time. The low-risk anomaly contradicts accepted APT or CAPM theories that higher risk portfolios earn higher returns. The low-risk anomaly is
not a recent empirical finding but an observation documented by a
a large body of literature dating back to the 1970s. Despite its longevity,
the academic community differs over the causes of the anomaly. The
two main explanations are: 1) it is due to leverage constraints that
retail, pension and mutual fund investors face which limits their
ability to generate higher returns by owning lower risk stocks, and
2) it is due to behavioral biases ranging from the lottery demand
for high beta stocks, beating index benchmarks with a limit to arbitrage,
and the sell-side analysts over-bias on high volatility stocks' earnings.

In this paper, we study the low-volatility anomaly from a new perspective
based on the Adaptive Multi-Factor (AMF) model proposed in the paper
by Zhu et al. (2018) \cite{zhu2020high} using the recently developed
Generalized Arbitrage Pricing Theory (see Jarrow and Protter (2016)
\cite{jarrow2016positive}). In Zhu et al. (2018) \cite{zhu2020high},
basis assets (formed from the collection of Exchange Traded Funds
(ETF)) are used to capture risk factors in \textit{realized} returns
across securities. Since the collection of basis assets is large and
highly correlated, high-dimension methods (including the LASSO and
prototype clustering) are used. This paper employs the same methodology
to investigate the low-volatility anomaly. We find that high-volatility
and low-volatility portfolios load on different basis assets, which
indicates that volatility is not an independent risk. The out-performance
of the low-volatility portfolio is due to the (equilibrium) performance
of these loaded risk factors. For completeness, we compare the AMF
model with the traditional Fama-French 5-factor (FF5) model, documenting
the superior performance of the AMF model.

\clearpage
\chapter{High Dimensional Estimation, Basis Assets, pand Adaptive Multi-Factor Models}

\section{Introduction}

The purpose of this paper is to proposes a new algorithm for the high-dimensional \\financial data, the \textbf{Groupwise Interpretable Basis Selection (GIBS) algorithm} to estimate a new \textbf{Adaptive Multi-Factor (AMF) asset pricing model}, implied by the generalized arbitrage pricing theory (APT) recently developed by Jarrow and Protter (2016) \cite{jarrow2016positive} and Jarrow (2016) \cite{jarrow2016bubbles}. \footnote{The Appendix \ref{ap: summary_apt} provides a brief summary of the generalized APT.} This generalized APT is derived in a continuous time, continuous trading economy imposing only the assumptions of frictionless markets, competitive markets, and the existence of a martingale measure.\footnote{By results contained in Jarrow and Larsson (2012) \cite{jarrow2012meaning}, this is equivalent to the economy satisfying no free lunch with vanishing risk (NFLVR) and no dominance (ND). See Jarrow and Larrson (2012) \cite{jarrow2012meaning} for the relevant definitions.} As such, this generalized APT includes both Ross's (1976) \cite{ross1976arbitrage} static APT and Merton's (1973) \cite{merton1973intertemporal} inter-temporal CAPM as special cases.

The generalized APT has four advantages over the traditional APT and
the inter-temporal CAPM. First, it derives the \emph{same} form of
the empirical estimation equation (see Equation (\ref{eq: amf_theory_ols_subset})
below) using a weaker set of assumptions, which are more likely to
be satisfied in practice.\footnote{The stronger assumptions in Ross's APT are: (i) a realized return
process consisting of a finite set of common factors and an idiosyncratic
risk term across a countably infinite collection of assets, and (ii)
no infinite asset portfolio arbitrage opportunities; in Merton's ICAPM
they are assumptions on (i) preferences, (ii) endowments, (ii) beliefs
and information, and (iv) those necessary to guarantee the existence
of a competitive equilibrium. None of these stronger assumptions are
needed in Jarrow and Protter (2016) \cite{jarrow2016positive}.} Second, the no-arbitrage relation is derived with respect to realized
returns, and not with respect to expected returns. This implies, of
course, that the error structure in the estimated multi-factor model
is more likely to lead to a larger $R^{2}$ and to satisfy the standard
assumptions required for regression models. Third, the set of basis
assets and the implied risk-factors are tradeable under the generalized
APT, implying their potential observability. Fourth, since the space
of random variables generated by the uncertainty in the economy is
infinite dimensional, the implied basis asset representation of any
security's return is parsimonious and sparse. Indeed, although the
set of basis assets is quite large (possibly infinite dimensional),
only a finite number of basis assets are needed to explain any assets'
realized return and different basis assets apply to different assets.
This last insight is certainly consistent with intuition since an
Asian company is probably subject to different risks than is a U.S.
company. Finally, adding a non-zero alpha to the no-arbitrage relation
in realized return space enables the identification of arbitrage opportunities.
This last property is also satisfied by the traditional APT and the
inter-temporal CAPM.

The generalized APT is important for practice because it provides
an exact identification of the relevant set of basis assets characterizing
a security's \emph{realized} (emphasis added) returns. This enables
a more accurate risk-return decomposition facilitating its use in
trading (identifying mispriced assets) and for risk management. Taking
expectations of this realized return relation with respect to the
martingale measure determines which basis assets are \emph{risk-factors},
i.e. which basis assets have non-zero expected excess returns (risk
premiums) and represent systematic risk. Since the traditional models
are nested within the generalized APT, an empirical test of the generalized
APT provides an alternative method for testing the traditional models
as well. One of the most famous empirical representations of a multi-factor
model is given by the Fama-French (2015) \cite{fama2015five}
five-factor model (FF5), see also \cite{fama1993common, fama1992cross}.
Recently, Harvey, Liu, and Zhu (2016) \cite{harvey2016and}
reviewed the literature on the estimation of factor models, the collection
of risk-factors employed, and argued for the need to use an alternative
statistical methodology to sequentially test for new risk-factors.
Our paper provides one such alternative methodology using the collection
of basis assets to determine which of these earn risk premium.

Since the generalized APT is a model for \textit{realized} returns
that allows different basis assets to affect different stocks differently,
an empirical test of this model starts with slightly different goals
than tests of conventional asset pricing models (discussed above)
whose implications are only with respect to \textit{expected} returns
and risk-factors. First, instead of searching for a few common risk-factors
that affect the entire cross-section of expected returns, as in the
conventional approach, we aim to find an exhaustive set of basis assets,
while maintaining parsimony for each individual stock (and hopefully
for the cross-section of stocks as well), using the GIBS algorithm we propose here. This alternative approach
has the benefit of increasing the explained variation in our time
series regressions. Second, as a direct implication of the estimated
realized return relation, the cross-section of expected returns is
uniquely determined. This implies, of course, that the collection
of risk-factors will be those basis assets with non-zero expected
excess returns (i.e. they earn non-zero risk premium).\footnote{An investigation of the cross-section of expected returns implied
by the generalized APT model estimated in this paper is a fruitful
area for future research. }

In addition, compared to PCA based methods that construct risk-factors
from linear combination of \textit{various} stocks, which are consequently
often difficult to interpret, our GIBS algorithm consists of a set of \textit{interpretable
and tradeable} basis assets. The new model explains more variation
in \textit{realized} returns. It is important to note here that in
a model of realized returns, some of the basis assets will reflect
idiosyncratic risks that do not earn non-zero risk premium. Those
basis assets that have non-zero excess expected returns are the relevant
risk-factors identified in the traditional estimation methodologies.
While a few recent papers adopt high-dimensional estimation methods
for modeling the cross-section of expected returns and an associated
parsimonious representation of the stochastic discount function \cite{bryzgalova2015spurious,feng2020taming},
our empirical test is specifically designed to align with the generalized
APT model's implications using basis assets and realized returns.

To test the generalized APT, we first obtain the collection of all
possible basis assets. Then, we provide a simultaneous test, security
by security, of which basis assets are significant for each security.
However, there are several challenges that must be overcome to execute
this estimation. First, in the security return regression using the
basis assets as independent variables, due to the assumption that
the regression coefficients ($\beta$'s) are constant, it's necessary
to run the estimation over a small time window because the $\beta$'s
are likely to change over longer time windows. This implies that the
number of sample points may be less than the number of independent
variables ($p>n$). This is the so-called high-dimension regime, where
the ordinary least squares (OLS) solution no longer holds. Second,
the collection of basis assets selected for investigation will be
highly correlated. And, it is well known that large correlation among
independent variables causes difficulties (redundant basis assets
selected, low fitting accuracy etc., see \cite{hebiri2012how,vandegeer2013lasso})
in applying the Least Absolute Shrinkage and Selection Operator (LASSO).
To address these difficulties, we propose a novel and hybrid algorithm
-- the GIBS algorithm, for identifying basis assets which are different
from the traditional variance-decomposition approach. The GIBS algorithm
takes advantage of several high-dimensional methodologies, including
prototype clustering, LASSO, and the ``1se rule'' for prediction.

We investigate Exchange-Traded Funds (ETFs) since they inherit and
aggregate the basis assets from their constituents. In recent years
there are more than one thousand ETFs, so it is reasonable to believe
that one can obtain the basis assets from the collection of ETFs in
the CRSP database, plus the Fama-French 5 factors. Consider the market
return, one of the Fama-French 5 factors. It can be duplicated by
ETFs such as the SPDR S\&P 500 ETF, the Vanguard S\&P 500 ETF, etc.
Therefore, it is also reasonable to believe that the remaining ETFs
can represent other basis assets as well.

We group the ETFs into different asset classes and use prototype clustering
to find good representatives within each class that have low pairwise
correlations. This reduced set of ETFs forms our potential basis assets.
After finding this set of basis assets, we still have more basis assets
than observations ($p>n$), but the basis assets are no longer highly
correlated. This makes LASSO an appropriate approach to determine
which set of basis assets are important for a security's return. To
be consistent with the literature, we fit an OLS regression on each
security's return with respect to its basis assets (that are selected
by LASSO) to perform an intercept ($\alpha$) and a goodness of fit
test. The importance of these tests are discussed next.

As noted above, the intercept test can be interpreted as a test of
the generalized APT under the assumptions of frictionless, competitive,
and arbitrage-free markets (more formally, the existence of an equivalent
martingale measure). The generalized APT abstracts from market microstructure
frictions, such as bid-ask spreads and execution speeds (costs), and
strategic trading considerations, such as high-frequency trading.
To be consistent with this abstraction, we study returns over a weekly
time interval, where the market microstructure frictions and strategic
trading considerations are arguably less relevant. Because the generalized
APT ignores market microstructure considerations, we label it a ``large-time
scale'' model.

If we fail to reject a zero alpha, we accept this abstraction, thereby
providing support for the assertion that the frictionless, competitive,
and arbitrage-free market construct is a good representation for ``large
time scale'' security returns. If the model is accepted, a goodness
of fit test quantifies the explanatory power of the model relative
to the actual time series variations in security returns. A ``good''
model is one where the model error (the difference between the model's
predictions and actual returns) behaves like white noise with a ``small''
variance. The adjusted $R^{2}$ provides a good metric of comparison
in this regard. Conversely, if we reject a zero alpha, then this is
evidence consistent with either: (i) that microstructure considerations
are necessary to understand ``large time scale'' models, or (ii)
that there exist arbitrage opportunities in the market. This second
possibility is consistent with the generalized APT being a valid description
of reality, but where markets are inefficient. To distinguish between
these two alternatives, we note that a non-zero intercept enables
the identification of these ``alleged'' arbitrage opportunities,
constructed by forming trading strategies to exploit the existence
of these ``positive alphas''. The implementation of these trading
strategies enables a test between these two alternatives.

Here is a brief summary of our results.
\begin{itemize}
\item The AMF model gives fewer significant intercepts (alphas) as compared to the Fama-French 5-factor model (percentage of companies with non-zero intercepts from 6.22\% to 3.86\% ). For both models, considering the False Discovery Rate, we cannot reject the hypothesis that the intercept is zero for all securities in the sample. This implies that historical security returns are consistent with the behavior implied by ``large-time scale'' models.
\item In an Goodness-of-Fit test comparing the Fama-French 5-factor and the AMF model, the AMF model has a substantially larger In-Sample Adjusted $R^{2}$ and the difference of goodness-of-fit of two models are significant. Furthermore, the AMF model increased the Out-of-Sample $R^2$ for the prediction by $ 24.07\% $. This supports the superior performance of the generalized APT in characterizing security returns.
\item As a robustness test, for those securities whose intercepts were non-zero (although insignificant), we tested the AMF model to see if positive alpha trading strategies generate arbitrage opportunities. They do not, thereby confirming the validity of the generalized APT.
\item The estimated GIBS algorithm selects 182 basis assets for the AMF model. All of these basis assets are significant for some stock, implying that a large number of basis assets are needed to explain security returns. On average each stock is related to only 2.98 basis assets, with most stocks having between $1\sim15$ significant basis assets. Cross-validation results in the Section \ref{sec: amf_comparison} are consistent with our sparsity assumption. Furthermore, different securities are related to different basis assets, which can be seen in Table \ref{tab: amf_perc_sgn} and the Heat Map in Figure \ref{fig: amf_heatmap}. Again, these observations support the validity of the generalized APT.
\item To identify which of the basis assets are risk-factors, we compute the average excess returns on the relevant basis assets over the sample period. These show that 77.47\% of the basis assets are risk-factors, earning significant risk premium.
\item Comparison of GIBS with the alternative methods discussed in Section \ref{sec: amf_comparison} shows the superior performance of GIBS. The comparison between GIBS and GIBS + FF5 shows that some of the FF5 factors are overfitting noise in the data.
\end{itemize}
\medskip{}

More recently, insightful papers by Kozak, Nagel, and Santosh (2017,
2018) \cite{kozak2018interpreting, kozak2017shrinking} proposed alternative
methods for analyzing risk-factors models. As Kozak et al. (2018)
\cite{kozak2018interpreting} note, if the ``risk-factors'' are
considered as a variance decomposition for a large amount of stocks,
one can always find that the number of important principal components
is small. However, this may not imply that there are only a small
number of relevant risk-factors because the Principal Component Analysis
(PCA) can either mix the underlying risk-factors together or separate
them into several principal components. It may be that there are a
large number of risk-factors, but the ensemble appears in only a few
principal components. The sparse PCA method used in \cite{kozak2018interpreting}
removes many of the weaknesses of traditional PCA, and even gives an interpretation of the risk-factors as stochastic discount risk-factors. However all these methods still suffer from the problems (low interpretability, low prediction accuracy, etc.) inherited from the variance decomposition framework. An alternative approach, the one we use here, is to abandon variance decomposition methods and to use high-dimensional methods instead,
such as prototype clustering to select basis assets as the ``prototypes'' or ``center'' of the groups they are representing. The proposed GIBS method gives much clearer interpretation, and much better prediction accuracy.

A detailed comparison of alternative methods that address the difficulties
of high-dimension and strong correlation id given in Section \ref{sec: amf_comparison}
below. Other methods include Elastic Net or Ridge Regression to deal
with the correlation. Elastic Net and Ridge Regression handle multicollinearity
by adding penalties to make the relevant matrix invertible. However,
neither of these methods considers the underlying cluster structure,
which makes it hard to interpret the selected basis assets. This is
the reason for the necessity of prototype clustering in the first
step, since it gives an interpretation of the selected basis assets
as the ``prototypes'' or the ``centers'' of the groups they are
representing. After the prototype clustering, we use a modified version
of LASSO (use in the GIBS algorithm) instead of either Elastic Nets
or Ridge Regression. The reason is that, compared to GIBS, alternative
methods achieve a much lower prediction goodness-of-fit (see Table
\ref{tab: amf_compare_methods}), but select more basis assets (see Figure
\ref{fig: amf_compare_n_factors}), which overfits and makes the model
less interpretable.

The tuning parameter that controls sparsity in LASSO, $\lambda$,
is traditionally selected by cross-validation and with this $\lambda$,
the model selects an average of 15.66 basis assets for each company.
However, as shown in the comparison Section \ref{sec: amf_comparison},
this overfits the noise in the data when compared with the GIBS algorithm
with respect to Out-of-Sample $R^{2}$. The reasons for the poor performance
of this cross-validation is discussed in Section \ref{sec: amf_comparison}
below. Consequently, to control against overfitting, we use the ``1se
rule'' along with the threshold that the number of basis assets can
not exceed 20.

The Adaptive Multi-Factor (AMF) model estimated by the GIBS algorithm
in this paper is shown to be consistent with the data and superior
to the Fama-French 5-factor model. An outline of this paper is as
follows. Section \ref{sec: amf_high_dim_method} describes the high-dimensional
statistical methods used in this paper and Section \ref{sec: amf_model}
presents the Adaptive Multi-Factor (AMF) model to be estimated. Section
\ref{sec: gibs_algo} gives the proposed GIBS algorithm to estimate
the model and Section \ref{sec: amf_results} presents the empirical
results. Section \ref{sec: amf_comparison} discusses the reason we
chose our method and provide a detailed comparison over alternative
methods. Section \ref{sec: amf_risk_premium} discusses the risk premium
of basis assets and Section \ref{sec: amf_illustration} presents some
illustrative examples. Section \ref{sec: amf_conclusion} concludes.
All codes are written in R and are available upon request.

\section{High-Dimensional Statistical Methodology}
\label{sec: amf_high_dim_method}

Since high-dimensional statistics is relatively new to the finance
literature, this section reviews the relevant statistical methodology.

\subsection{Preliminaries and Notations}
\label{sec: notations}

Let $\norm{\bm{v}}_{q}$ denote the standard $l_{q}$ norm of a vector
$\bm{v}$ of dimension $p\times1$, i.e.
\[
\norm{\bm{v}}_{q}=\begin{cases}
(\sum_{i}|\bm{v}_{i}|^{q})^{1/q} & \text{if}\;\;0<q<\infty\\
\#\{i:\bm{v}_{i}\neq0\}, & \text{if}\;\;q=0\\
\max_{i}|\bm{v}_{i}| & \text{if}\;\;q=\infty.
\end{cases}
\]
Suppose $\bm{\beta}$ is also a vector with dimension $p\times1$,
a set $S\subseteq\{1,2,...,p\}$, then $\beta_{S}$ is a $p\times1$
vector with i-th element
\[
(\bm{\beta}_{S})_{i}=\begin{cases}
\beta_{i}, & \text{if}\;\;i\in S\\
0, & otherwise.
\end{cases}
\]

Here the index set $S$ is called the support of $\bm{\beta}$, in
other words, $supp(\bm{\beta})=\{i:\bm{\beta}_{i}\neq0\}$. Similarly,
if $\bm{X}_{n\times m} $ is a matrix instead of a vector, for any index set $ S \subseteq\{1,2,..., m\} $, use $\bm{X}_{S}$ to denote the the columns of $\bm{X}$ indexed by $S$. Denote $\vone_{n}$ as a $n\times1$ vector with all elements being 1, $\bm{J_{n}}=\vone_{n}\vone_{n}'$ and $\bm{\bar{J}_{n}}=\frac{1}{n}\bm{J_{n}}$. $\bm{I_{n}}$ denotes the identity matrix with diagonal 1 and 0 elsewhere. The subscript
$n$ is always omitted when the dimension $n$ is clear from the context. The notation $\#S$ means the number of elements in the set $S$.

\subsection{Minimax Prototype Clustering and Lasso Regressions}
\label{sec: proto_lasso}

This section describes the prototype clustering to be used to deal
with the problem of high correlation among the independent variables
in our LASSO regressions. To remove unnecessary independent variables,
using clustering methods, we classify them into similar groups and
then choose representatives from each group with small pairwise correlations.
First, we define a distance metric to measure the similarity between
points (in our case, the returns of the independent variables). Here,
the distance metric is related to the correlation of the two points,
i.e.
\begin{equation}
d(r_{1},r_{2})=1-|corr(r_{1},r_{2})|\label{eq: corr_dist}
\end{equation}
where $r_{i}=(r_{i,t},r_{i,t+1},...,r_{i,T})'$ is the time series
vector for independent variable $i=1,2$ and $corr(r_{1},r_{2})$
is their correlation. Second, the distance between two clusters needs
to be defined. Once a cluster distance is defined, hierarchical clustering
methods (see \cite{kaufman2009finding}) can be used to organize the
data into trees.

In these trees, each leaf corresponds to one of the original data
points. \\ Agglomerative hierarchical clustering algorithms build trees
in a bottom-up \\ approach, initializing each cluster as a single point,
then merging the two closest clusters at each successive stage. This
merging is repeated until only one cluster remains. Traditionally,
the distance between two clusters is defined as either a complete
distance, single distance, average distance, or centroid distance.
However, all of these approaches suffer from interpretation difficulties
and inversions (which means parent nodes can sometimes have a lower
distance than their children), see Bien, Tibshirani (2011)\cite{bien2011hierarchical}.
To avoid these difficulties, Bien, Tibshirani (2011)\cite{bien2011hierarchical}
introduced hierarchical clustering with prototypes via a minimax linkage
measure, defined as follows. For any point $x$ and cluster $C$,
let
\begin{equation}
d_{max}(x,C)=\max_{x'\in C}d(x,x')
\end{equation}
be the distance to the farthest point in $C$ from $x$. Define the
\emph{minimax radius} of the cluster $C$ as
\begin{equation}
r(C)=\min_{x\in C}d_{max}(x,C)
\end{equation}
that is, this measures the distance from the farthest point $x\in C$
which is as close as possible to all the other elements in C. We call
the minimizing point the \emph{prototype} for $C$. Intuitively, it
is the point at the center of this cluster. The \emph{minimax linkage}
between two clusters $G$ and $H$ is then defined as
\begin{equation}
d(G,H)=r(G\cup H).
\end{equation}
Using this approach, we can easily find a good representative for
each cluster, which is the prototype defined above. It is important
to note that minimax linkage trees do not have inversions. Also, in
our application as described below, to guarantee interpretable and
tractability, using a single representative independent variable is
better than using other approaches (for example, principal components
analysis (PCA)) which employ linear combinations of the independent
variables.

The LASSO method was introduced by Tibshirani (1996) \cite{tibshirani1996regression}
for model selection when the number of independent variables ($p$)
is larger than the number of sample observations ($n$). The method
is based on the idea that instead of minimizing the squared loss to
derive the Ordinary Least Squares (OLS) solution for a regression, we should add to the loss
a penalty on the absolute value of the coefficients to minimize the
absolute value of the non-zero coefficients selected. To illustrate
the procedure, suppose that we have a linear model
\begin{equation}
\bm{y}=\bm{X\beta}+\bm{\epsilon}\qquad\text{where}\qquad\bm{\epsilon}\sim N(0,\sigma_{\epsilon}^{2}\bm{I}),\label{eq: lasso}
\end{equation}
$\bm{X}$ is an $n\times p$ matrix, $\bm{y}$ and $\bm{\epsilon}$
are $n\times1$ vectors, and $\bm{\beta}$ is a $p\times1$ vector.

The LASSO estimator of $\bm{\beta}$ is given by
\begin{equation}
\bm{\hat{\beta}}_{\lambda}=\underset{\bm{\beta}\in\mathbb{R}^{p}}{\arg\min}\left\{ \frac{1}{2n}\norm{\bm{y}-\bm{X\beta}}_{2}^{2}+\lambda\norm{\bm{\beta}}_{1}\right\} \label{eq: lasso_beta}
\end{equation}
where $\lambda>0$ is the tuning parameter, which determines the magnitude
of the penalty on the absolute value of non-zero $\beta$'s. In this
paper, we use the R package \textit{glmnet} \cite{friedman2010regularization}
to fit LASSO.

In the subsequent estimation, we will only use a modified version
of LASSO as a model selection method to find the collection of important
independent variables. After the relevant basis assets are selected,
we use a standard Ordinary Least Squares (OLS) regression on these variables to test for the
goodness of fit and significance of the coefficients. More discussion
of this approach can be found in Zhao, Shojaie, Witten (2017) \cite{zhao2017defense}.

In this paper, we fit the prototype clustering followed by a LASSO
on the prototype basis assets selected. The theoretical justification
for this approach can be found in \cite{reid2018general} and \cite{zhao2017defense}.

\section{The Adaptive Multi-Factor Model}

\label{sec: amf_model}

This section presents the Adaptive Multi-Factor (AMF) asset pricing model that is
estimated using the high-dimensional statistical methods just discussed.
Given is a frictionless, competitive, and arbitrage free market. In
this setting, a dynamic generalization of Ross's (1976) \cite{ross1976arbitrage}
APT and Merton's (1973) \cite{merton1973intertemporal} ICAPM derived by Jarrow
and Protter (2016) \cite{jarrow2016positive} implies that the following
relation holds for any security's \emph{realized }return:
\begin{equation}
R_{i}(t)-r_{0}(t)=\sum_{j=1}^{p}\beta_{i,j}\big[r_{j}(t)-r_{0}(t)\big]
=\bm{\beta}_{i}' [\bm{r}(t)-r_{0}(t)\vone]
\label{eq: amf_linear model}
\end{equation}
where at time $t$, $R_{i}(t)$ denotes the return of the \textit{i-th} security
for $1\leq i\leq N$ (where $N$ is the number of securities), $r_{j}(t)$
denotes the return used as the \textit{j-th} basis asset for $1\leq j\leq p$,
$r_{0}(t)$ is the risk free rate, $\bm{r}(t)=(r_{1}(t),r_{2}(t),...,r_{p}(t))'$
denotes the vector of security returns, $\vone$ is a column vector
with every element equal to one, and $\bm{\beta}_{i}=(\beta_{i,1},\beta_{i,2},...,\beta_{i,p})'$.

This generalized APT requires that the basis assets are represented
by traded assets. In Jarrow and Protter (2016) \cite{jarrow2016positive}
the collection of basis assets form an algebraic basis that spans
the set of security payoffs at the model's horizon, time $T$. No
arbitrage, i.e. the existence of a martingale measure, implies that
this same basis set applies to the returns over intermediate time
periods $t\in[0,T]$, which yields the basis asset risk-return relation
given in Equation (\ref{eq: amf_linear model}). It is important to
emphasize that this no-arbitrage relation is for realized returns,
not expected returns. Realized returns are the objects to which asset
pricing estimation is applied. Secondly, the no-arbitrage relation
requires the additional assumptions of frictionless and competitive
markets. Consequently, this asset pricing model abstracts from market
micro-structure considerations. For this reason, this model structure
is constructed to understand security returns over larger time intervals
(days or weeks) and not intra-day time intervals where market micro-structure
considerations apply.

Consistent with this formulation, we use traded ETFs for the basis
assets. In addition, to apply the LASSO method, for each security
$i$ we assume that only a small number of the $\beta_{i,j}$ coefficients
are non-zero ($\beta_{i}$ has the sparsity property). Lastly, to
facilitate estimation, we also assume that the $\beta_{i,j}$ coefficients
are constant over time, i.e. $\beta_{i,j}(t)=\beta_{i,j}$. This assumption
is an added restriction, not implied by the theory. It is only a reasonable
approximation if the time period used in our estimation is not too
long (we will return to this issue subsequently).

To empirically test our model, both an intercept $\alpha_{i}$ and
a noise term $\epsilon_{i}(t)$ are added to Equation (\ref{eq: amf_linear model}),
that is,
\begin{equation}
R_{i}(t)-r_{0}(t)=\alpha_{i}+\sum_{j=1}^{p}\beta_{i,j}(t)\big[r_{j}(t)-r_{0}(t)\big]
+\epsilon_{i}(t)=\alpha+\bm{\beta}_{i}'[\bm{r}(t)-r_{0}(t)\vone]+\epsilon_{i}(t)
\label{eq: amf_testable}
\end{equation}
where $\epsilon_{i}(t)\overset{iid}{\sim}N(0,\sigma_{i}^{2})$ and
$1\leq i\leq N$.

The error term is included to account for noise in the data and ``random''
model error, i.e. model error that is unbiased and inexplicable according
to the independent variables included within the theory. If our theory
is useful in explaining security returns, this error should be small
and the adjusted $R^{2}$ large. The $\alpha$ intercept is called
\emph{Jensen's alpha}. Using the recent theoretical insights of Jarrow
and Protter (2016) \cite{jarrow2016positive}, the intercept test
can be interpreted as a test of the generalized APT under the assumptions
of frictionless, competitive, and arbitrage-free markets (more formally,
the existence of an equivalent martingale measure). As noted above,
this approach abstracts from market microstructure frictions, such
as bid-ask spreads and execution speeds (costs), and strategic trading
considerations, such as high-frequency trading. To be consistent with
this abstraction, we study returns over a weekly time interval, where
the market microstructure frictions and strategic trading considerations
are arguably less relevant. If we fail to reject a zero alpha, we
accept this abstraction, thereby providing support for the assertion
that the frictionless, competitive, and arbitrage-free market construct
is a good representation of ``large time scale'' security returns.
If the model is accepted, a goodness of fit test quantifies the explanatory
power of the model relative to the actual time series variations in
security returns. The adjusted $R^{2}$ provides a good test in this
regard. The GRS test in \cite{gibbons1989test} is usually an excellent
procedure for testing intercepts but it is not appropriate in the
LASSO regression setting.

Conversely, if we reject a zero alpha, then this is evidence consistent
with either: (i) that microstructure considerations are necessary
to understand ``large time scale'' as well as ``short time scale''
returns or (ii) that there exist arbitrage opportunities in the market.
This second possibility is consistent with the generalized APT being
a valid description of reality, but where markets are inefficient.
To distinguish between these two alternatives, we note that a non-zero
intercept enables the identification of these ``alleged'' arbitrage
opportunities, constructed by forming trading strategies to exploit
the existence of these ``positive alphas.''

Using weekly returns over a short time period necessitates the use
of high-dimensional statistics. To understand why, consider the following.
For a given time period $(t,T)$, letting $n=T-t+1$, we can rewrite
Equation (\ref{eq: amf_testable}) using time series vectors as
\begin{equation}
\bm{R}_{i}-\bm{r}_{0}=\alpha_{i}\vone_{n}+(\bm{r}
-\bm{r}_{0}\vone'_{p})\bm{\beta}_{i}+\bm{\epsilon}_{i}
\label{eq: amf_vec}
\end{equation}
where $ 1 \leq i \leq N $, $\bm{\epsilon}_{i}\sim N(0,\sigma_{i}^{2}\bm{I}_{n})$ and {\small
\begin{equation}
\bm{R}_{i}=\begin{pmatrix}R_{i}(t)\\
R_{i}(t+1)\\
\vdots\\
R_{i}(T)
\end{pmatrix}_{n\times1},\,
\bm{r}_{0}=\begin{pmatrix}r_{0}(t)\\
r_{0}(t+1)\\
\vdots\\
r_{0}(T)
\end{pmatrix}_{n\times1},\,
\bm{\epsilon}_{i}=\begin{pmatrix}\epsilon_{i}(t)\\
\epsilon_{i}(t+1)\\
\vdots\\
\epsilon_{i}(T)
\end{pmatrix}_{n\times1}
\end{equation}

\begin{equation}
\bm{\beta}_{i}=\begin{pmatrix}\beta_{i,1}\\
\beta_{i,2}\\
\vdots\\
\beta_{i,p}
\end{pmatrix}_{p\times1}, \,
\bm{r}_{i}=\begin{pmatrix}r_{i}(t)\\
r_{i}(t+1)\\
\vdots\\
r_{i}(T)
\end{pmatrix}_{n\times1},\,
\bm{r}(t)=\begin{pmatrix}r_{1}(t)\\
r_{2}(t)\\
\vdots\\
r_{p}(t)
\end{pmatrix}_{p\times1}\,
\end{equation}

\begin{equation}
\bm{r}=(\bm{r}_{1},\bm{r}_{2},...,\bm{r}_{p})_{n\times p}=\begin{pmatrix}\bm{r}(t)'\\
\bm{r}(t+1)'\\
\vdots\\
\bm{r}(T)'
\end{pmatrix}_{n\times p}, \,
\bm{R} = (\bm{R}_1, \bm{R}_2, ..., \bm{R}_N)
\label{eq: amf_vec_end}
\end{equation}
}Recall that we assume that the coefficients $\beta_{ij}$ are constants.
This assumption is only reasonable when the time period $(t,T)$ is
small, say three years, so the number of observations $n\approx150$
given we employ weekly data. Therefore, our sample size $n$ in this
regression is substantially less than the number of basis assets $p$.

We fit the GIBS algorithm to select the
basis assets set $S$ ($S$ is derived near the end). Then, the model
becomes
\begin{equation}
\bm{R}_{i}-\bm{r}_{0}=\alpha_{i}\vone_{n}+(\bm{r}_{S}
-(\bm{r}_{0})_{S}\vone'_{p})(\bm{\beta}_{i})_{S}+\bm{\epsilon}_{i}\quad.
\label{eq: amf_theory_ols_subset}
\end{equation}
Here, the intercept and the significance of each basis asset can be
tested, making the identifications $\bm{y}=\bm{R}_{i}-\bm{r}_{0}$
and $\bm{X}=\bm{r}-\bm{r_{0}}\vone_{p}'$ in Equation (\ref{eq: lasso}). Goodness of fit tests, comparisons of the in-sample adjusted $R^{2}$, and prediction out-of-sample $R^2$ \cite{campbell2008predicting} can be employed.

An example of Equation (\ref{eq: amf_theory_ols_subset}) is the Fama-French
(2015) \cite{fama2015five} five-factor model where all
of the basis assets are risk-factors, earning non-zero expected excess
returns. Here, the five traded risk-factors are: (i) the market portfolio
less the spot rate of interest ($R_{m}-R_{f}$), (ii) a portfolio
representing the performance of small (market capital) versus big
(market capital) companies ($SMB$), (iii) a portfolio representing
the performance of high book-to-market ratio versus small book-to-market
ratio companies ($HML$), (iv) a portfolio representing the performance
of robust (high) profit companies versus that of weak (low) profits
($RMW$), and (v) a portfolio representing the performance of firms
investing conservatively and those investing aggressively ($CMA$),
i.e.
\begin{equation}
\begin{array}{ll}
R_{i}(t)-r_{0}(t) =
& \alpha_{i}+\beta_{mi}(R_{m}(t)-r_{0}(t))+\beta_{si} SMB(t)+\beta_{hi} HML(t)\\
& + \beta_{ri} RMW(t)+\beta_{ci} CMA(t)+\epsilon_{i}(t).
\end{array}\label{eq: ff5}
\end{equation}
The key difference between the Fama-French five-factor and Equation
(\ref{eq: amf_theory_ols_subset}) is that Equation (\ref{eq: amf_theory_ols_subset}) allows distinct
securities to be related to different basis assets, many of which
may not be risk-factors, chosen from a larger set of basis assets
than just these five. In fact, we allow the number of basis assets
$p$ to be quite large (e.g. over one thousand), which enables the
number of non-zero coefficients $\bm{\beta}_{i}$ to be different
for different securities. As noted above, we also assume the coefficient
vector $\bm{\beta}_{i}$ to be sparse. The traditional literature,
which includes the Fama-French five-factor model, limits the regression
to a small number of risk-factors. In contrast, using the LASSO method,
we are able to fit our model using time series data when $p>n$, as
long as the $\beta_{i}$ coefficients are sparse and the basis assets
are not highly correlated. As noted previously, we handle this second
issue via clustering methods.

\section{The Estimation Procedure (GIBS algorithm)}

\label{sec: gibs_algo}

This section discusses the estimation procedure for the basis asset implied Adaptive Multi-Factor (AMF) model. To overcome the high-dimension and high-correlation difficulties, we propose a \textbf{Groupwise Interpretable Basis Selection (GIBS) algorithm} to empirically estimate the AMF model. The details are given in this section, and the sketch of the GIBS algorithm is shown in Table \ref{tab: gibs_algo} at the end of this section.

The data consists of security returns and all the ETFs available in
the CRSP database over the three year time period from January 2014
to December 2016. The same approach can be used in other time periods
as well. However, in earlier time periods, there were less ETFs. In
addition, in the collection of basis assets we include the five Fama-French
factors. A security is included in our sample only if it has prices
available for more than 80\% of all the trading weeks. For easy comparison,
companies are classified according to the first 2 digits of their
SIC code (a detailed description of SIC code classes can be found
in Appendix \ref{sec: amf_comp_classes}).

Suppose that we are given $p_{1}$ tradable basis assets $\bm{r}_{1},\bm{r}_{2},...,\bm{r}_{p_{1}}$.
In our investigation, these are returns on traded ETFs, and for comparison
to the literature, the Fama-French 5 factors. Using recent year data,
the number of ETFs is large, slightly over 1000 ($p_{1}\approx1000$).
Since these basis assets are highly-correlated, it is problematic
to fit $\bm{R}_{i}-\bm{r}_{0}$ directly on these basis assets using
a LASSO regression. Hence, we use the Prototype Clustering method
discussed in Section \ref{sec: proto_lasso} to reduce the number
of basis assets by selecting low-correlated representatives. Then,
we fit a modified version of the LASSO regression to these low-correlated representatives.
This improves the fitting accuracy and also selects a sparser and
more interpretable model.

For notation simplicity, denote
\begin{equation}
  \bm{Y}_{i}=\bm{R}_{i}-\bm{r}_{0}\,, \quad \bm{X}_{i}=\bm{r}_{i}-\bm{r}_{0}\,, \quad
  \bm{Y}=\bm{R}-\bm{r}_{0}\,, \quad \bm{X}=\bm{r}-\bm{r}_{0}
\end{equation}
where the definition of $\bm{R}_i, \, \bm{R}, \, \bm{r}_i, \, \bm{r}$ are in equation (\ref{eq: amf_vec} - \ref{eq: amf_vec_end}). Let $\bm{r}_{1}$ denote the market return. It is easy to check that most of the ETF basis assets $\bm{X}_{i}$ are correlated with $\bm{X}_{1}$ (the market return minus the risk free rate). We note that this pattern is not true for the other four Fama-French factors. Therefore, we
first orthogonalize every other basis asset to $\bm{X}_{1}$ before
doing the clustering and the LASSO regression. By orthogonalizing with
respect to the market return, we apvoid choosing redundant basis assets
similar to it and meanwhile, increase the accuracy of fitting. Note
that for OLS, projection does not affect the estimation since it only
affects the coefficients, not the estimated $\bm{\hat{y}}$. However,
in the LASSO, projection does affect the set of selected basis assets
because it changes the magnitude of shrinking.
Thus, we compute
\begin{equation}
\bm{\widetilde{X}}_{i}=(\bm{I}-P_{\bm{X}_{1}})\bm{X}_{i}
=(\bm{I}-\bm{X}_{1}(\bm{X}'_{1}\bm{X}_{1})^{-1}\bm{X}_{1}')\bm{X}_{i}
\qquad\text{where}\quad2\leq i\leq p_{1}
\label{eq: proj1}
\end{equation}
where $P_{\bm{X}_{1}}$ denotes the projection operator. Denote the
vector
\begin{equation}
\bm{\widetilde{X}}=(\bm{X}_{1},\bm{\widetilde{X}}_{2},\bm{\widetilde{X}}_{3},...,
\bm{\widetilde{X}}_{p_{1}}).
\label{eq: proj2}
\end{equation}
Note that this is equivalent to the residuals after regressing other
basis assets on the market return minus the risk free rate.

The transformed ETF basis assets $\bm{\widetilde{X}}$ contain highly
correlated members. We first divide these basis assets into categories
$A_{1},A_{2},...,A_{k}$ based on their financial interpretation. Note that $A\equiv\cup_{i=1}^{k}A_{i}=\{1,2,...,p_{1}\}$. The list of categories with more descriptions can be found in Appendix \ref{ap: etf_classes}. The categories are (1) bond/fixed income, (2) commodity, (3) currency, (4) diversified portfolio, (5) equity, (6) alternative ETFs, (7) inverse, (8) leveraged, (9) real estate, and (10) volatility.

Next, from each category we need to choose a set of representatives.
These representatives should span the categories they are from, but
also have low correlation with each other. This can be done by using
the prototype-clustering method with distance defined by equation
(\ref{eq: corr_dist}), which yield the ``prototypes'' (representatives) within each cluster (intuitively, the prototype is at the center of each cluster) with low-correlations.

Within each category, we use the prototype clustering methods previously
discussed to find the set of representatives. The number of representatives
in each category can be decided according to a correlation threshold.
(Alternatively, we can also use the PCA dimension or other parameter
tuning methods to decide the number of prototypes. Note that even
if we use the PCA dimension to suggest the number of prototypes to
keep, the GIBS algorithm does not use any linear combinations of factors
as PCA does). This gives the sets $B_{1},B_{2},...,B_{k}$ with $B_{i}\subset A_{i}$
for $1\leq i\leq k$. Denote $B\equiv\cup_{i=1}^{k}B_{i}$. Although
this reduction procedure guarantees low-correlation between the elements
in each $B_{i}$, it does not guarantee low-correlation across the
elements in the union $B$. So, an additional step is needed, which
is prototype clustering on $B$ to find a low-correlated representatives
set $U$. Note that $U\subseteq B$. Denote $p_{2}\equiv\#U$. The list
of all ETFs in the set $U$ is given in Appendix \ref{ap: etf_representatives} Table \ref{tab: amf_etf_representatives}. This is still a large set with $p_{2}=182$.

Recall from the notation Section \ref{sec: notations} that $\bm{\w{X}}_{U}$ means the columns of the matrix $\bm{\w{X}}$ indexed by the set $U$. Since basis assets in $\bm{\widetilde{X}}_{U}$ are not highly correlated, a LASSO regression can be applied.  By equation (\ref{eq: lasso_beta}), we have that
\begin{equation}
\bm{\widetilde{\beta}}_{i}=\underset{\bm{\beta}_{i}\in\mathbb{R}^{p},(\bm{\beta}_{i})_{j}
=0(\forall j\in U^{c})}{\arg\min}\left\{ \frac{1}{2n}\norm{\bm{Y}_{i}-\bm{\widetilde{X}\beta}_{i}}_{2}^{2}
+\lambda\norm{\bm{\beta}_{i}}_{1}\right\}
\end{equation}
where $U^{c}$ denotes the complement of $U$. However, here we use
a different $\lambda$ compared to the traditional LASSO. Normally
the $\lambda$ of LASSO is selected by the cross-validation. However
this will overfit the data as discussed in Section \ref{sec: amf_comparison}.
So here we use a modified version of the $\lambda$ select rule and
set
\begin{equation}
  \lambda=\max\{\lambda_{1se},min\{\lambda:\#supp(\bm{\widetilde{\beta}}_{i})\leq20\}\}
\end{equation}
where $\lambda_{1se}$ is the $\lambda$ selected by the ``1se rule''. The ``1se rule'' gives the most regularized model such that error is within one standard error of the minimum error achieved by the cross-validation (see \cite{friedman2010regularization, simon2011Regularization, tibshirani2012strong}). Further discussion of the choice of $\lambda$ can be found in Section \ref{sec: amf_comparison}.

Therefore we can derive the the set of basis assets selected as
\begin{equation}
S_i \equiv supp(\bm{\widetilde{\beta}}_{i})
\label{eq: factor_select}
\end{equation}

Next, we fit an Ordinary Least Squares (OLS) regression on the selected basis assets. Since this
is an OLS regression, we use the original basis assets $\bm{X}_{S_{i}}$
rather than the orthogonalized basis assets with respect to the market
return $\bm{\widetilde{X}}_{S_i}$. In this way, we construct the set of basis assets $S_{i}$.

Note that here we can also add the Fama-French 5 factors into $S_{i}$ if not selected, which will be also discussed in Section \ref{sec: amf_comparison} as the GIBS + FF5 model. This is included to compare our results with the literature. However, the comparison results between the GIBS and the GIBS + FF5 model in Section \ref{sec: amf_comparison} show that adding back Fama-French 5 factors into $S_{i}$ results in overfitting and should be avoided. Hence, the GIBS algorithm emplyed herein doesn't include the Fama-French 5 factors if they are not selected in the procedure above.
%%_{i}\ensuremath{}\ensuremath{}\ensuremath{}\ensuremath{}\ensuremath{}

%Note that after the above procedure there might be some risk-factors
%in the set $\widetilde{S}'$ still correlated with the market return
%($\bm{X}_{1}$) because of that we projected out the market return
%before removing the high-correlated columns and that we pick about
%20 risk-factors for each company with LASSO (some weak risk-factors
%may still be selected if there are not enough strong risk-factors.)
%Therefore, we need to remove these weak risk-factors, in other word,
%removing risk-factors which are still highly correlated to $\bm{X}_{1}$
%from the set $\widetilde{S}_{i}'$.

The following OLS regression is used to estimate $\bm{\hat\beta}_{i}$,
the OLS estimator of $\bm{\beta}_{i}$ in
\begin{equation}
\bm{Y}_{i}=\alpha_{i}\vone_{n}
+\bm{X}_{S_{i}}(\bm{\beta}_{i})_{S_{i}}
+\bm{\epsilon}_{i}.
\label{eq: amf_ols_subset}
\end{equation}
Note that $supp(\bm{\hat\beta}_{i})\subseteq {S}_{i}$. The
adjusted $R^{2}$ is obtained from this estimation. Since we are in the
OLS regime, significance tests can be performed on $\bm{\hat\beta}_{i}$.
This yields the significant set of coefficients
\begin{equation}
S_{i}^{*}\equiv\{j:P_{H_{0}}(|\bm{\beta}_{i,j}|\geq|\bm{\hat\beta}_{i,j}|)<0.05\}
\quad\text{where}\quad H_{0}:\text{True value}\,\,\bm{\beta}_{i,j}=0.
\label{eq: factor_sgn}
\end{equation}

Note that the significant basis asset set is a subset of the selected basis asset set. In another word,
\begin{equation}
S_i^* \subseteq supp(\bm{\hat\beta}_i) \subseteq S_i \subseteq \{1, 2, ..., p\} .
\end{equation}

To sum up, the sketch of the GIBS algorithm is shown in Table \ref{tab: gibs_algo}. Recall from the notation Section \ref{sec: notations} that for an index set $ S \subseteq \{1, 2, ..., p\} $, $\bm{\w{X}}_{S}$ means the columns of the matrix $\bm{\w{X}}$ indexed by the set $S$.
\begin{table}[H]
\center
\begin{tabular}{|| l ||}
\hhline{|=|}
\\[-1.6ex]
\textbf{The Groupwise Interpretable Basis Selection (GIBS) algorithm }  \\[1ex]
\hhline{|=|}
\\[-1.6ex]
Inputs: Stocks to fit $\bm{Y}$ and basis assets $\bm{X}$.
\\[0.6ex]
\hline
\\[-1.6ex]
1. Derive $\bm{\widetilde{X}}$ using $\bm{X}$ and the equation (\ref{eq: proj1}, \ref{eq: proj2}).
\\[0.6ex]
2. Divide the transformed basis assets $\bm{\widetilde{X}}$ into k groups $A_{1},A_{2},\cdots A_{k}$ by a
\\[0.6ex]
\hspace{0.4cm} financial interpretation.
\\[0.6ex]
3. Within each group, use prototype clustering to find prototypes $ B_i \subset A_i $.
\\[0.6ex]
4. Let $ B = \cup_{i = 1}^k B_i $, use prototype clustering in B to find prototypes $ U \subset B $.
\\[0.6ex]
5. For each stock $\bm{Y}_{i}$, use a modified version of LASSO to reduce $\bm{\widetilde{X}}_{U}$
\\[0.6ex]
\hspace{0.4cm} to the selected basis assets $\bm{\widetilde{X}}_{S_{i}}$
\\[0.6ex]
6. For each stock $ \bm{Y}_i $, fit linear regression on $ \bm{X}_{S_i} $.
\\[0.6ex]
\hline
\\[-1.6ex]
Outputs: Selected factors $ S_i $, significant factors $ S_i^*$, and coefficients in step 6.
\\[0.6ex]
\hhline{|=|}
\end{tabular}
\caption{The sketch of Groupwise Interpretable Basis Selection (GIBS) algorithm}
\label{tab: gibs_algo}
\end{table}

It is also important to understand which basis assets affect which
securities. Given the set of securities is quite large, it is more
reasonable to study which classes of basis assets affect which classes
of securities. The classes of basis assets are given in Appendix \ref{ap: etf_classes},
and the classes of securities classified by the first 2 digits of
their SIC code are in Appendix \ref{sec: amf_comp_classes}.
For each security class, we count the number of significant basis
asset classes as follows.

Recall that $N$ is the number of securities. Denote $l$ to be the
number of security classes. Denote the security classes by $C_{1},C_{2},C_{3},...,C_{l}$
where $\bigcup_{d=1}^{l}C_{d}=\{1,2,...,N\}$. Recall that the number
of basis assets is $p$. Let the number of basis asset classes be
$m$. Let the basis asset classes be denoted $F_{1},F_{2},F_{3},...,F_{m}$
where $\bigcup_{b=1}^{m}F_{b}=\{1,2,...,p\}$ and $p$ is the number
of basis assets which were significant for at least one of the security
i. Also recall that $ S_i^* \subseteq \{1, 2, ..., p\} $ in equation (\ref{eq: factor_sgn}). Denote the significant count matrix to be $\bm{A}=\{a_{b,d}\}_{m\times l}$
where
\begin{equation}
a_{b,d}=\sum_{i\in C_{d}} \#\{ S_i^* \cap F_{b} \}.
\end{equation}
That is, each element $a_{b,d}$ of matrix $\bm{A}$ is the number
of significant basis assets in basis asset class $b$, selected by
securities in class $d$. Finally, denote the proportion matrix to
be $\bm{G}=\{g_{b,d}\}_{m\times l}$ where
\begin{equation}
g_{b,d}=\frac{a_{b,d}}{\sum_{1\leq j\leq m}a_{j,d}}.\label{eq: perc_table}
\end{equation}
In other words, each element $g_{b,d}$ of matrix $\bm{G}$ is the
proportion of significant basis assets in basis asset class $b$ selected
by security class $d$ among all basis assets selected by security
class $d$. Note that the elements in each column of $\bm{G}$ sum
to one.

\section{Estimation Results}

\label{sec: amf_results}

Our results show that the GIBS algorithm selects a total of 182
basis assets from different sectors for at least one company. And
all of these 182 basis assets are significant in the second stage
OLS regressions after the GIBS selection for at least one company.
This validates our assumption that the total number of basis assets
is large; much larger than 10 basis assets, which is typically the
maximum number of basis asset with non-zero risk premiums (risk-factors)
seen in the literature (see Harvey, Liu, Zhu (2016) \cite{harvey2016and}).

\begin{figure}[H]
\center \vspace{0in}
\includegraphics[width=0.8\textwidth]{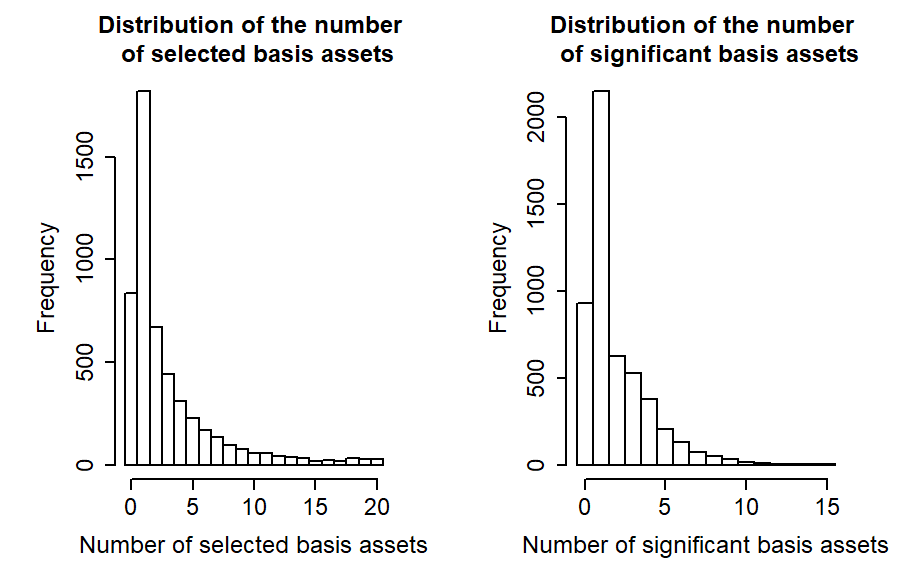}
\caption{Distribution of the number of basis assets. The left figure shows the histogram of the number of basis assets selected by GIBS. On the right we report the histogram of number of basis assets significant in the second-step OLS regression at 5\% level.}
\label{fig: amf_n_factors} \vspace{0in}
\end{figure}

In addition, the results validate our sparsity assumption, that each
company is significantly related to only a small number of basis assets
(at a 5\% level of significance). Indeed, for each company an average
of 2.98 basis assets are selected by GIBS and an average of 1.92 basis assets show significance in the second stage OLS regression. (Even using the traditional cross-validation method with overfitting discussed in Section \ref{sec: amf_comparison}, only an average of 15.66 basis assets are selected.) In other words, the average number of elements in $S_i$ (see Equation \ref{eq: factor_select}) is 2.98 and the average number of elements in $S_{i}$ (see Equation \ref{eq: factor_sgn}) is 1.92. Figure \ref{fig: amf_n_factors}
shows the distribution of the number of basis assets selected by GIBS and the number of basis assets that are significant in the second stage OLS regression. As depicted, most securities have between $1\sim15$ significant basis assets. Thus high dimensional methods are appropriate and necessary here.

\begin{figure}[H]
\centering
\includegraphics[width=\textwidth]{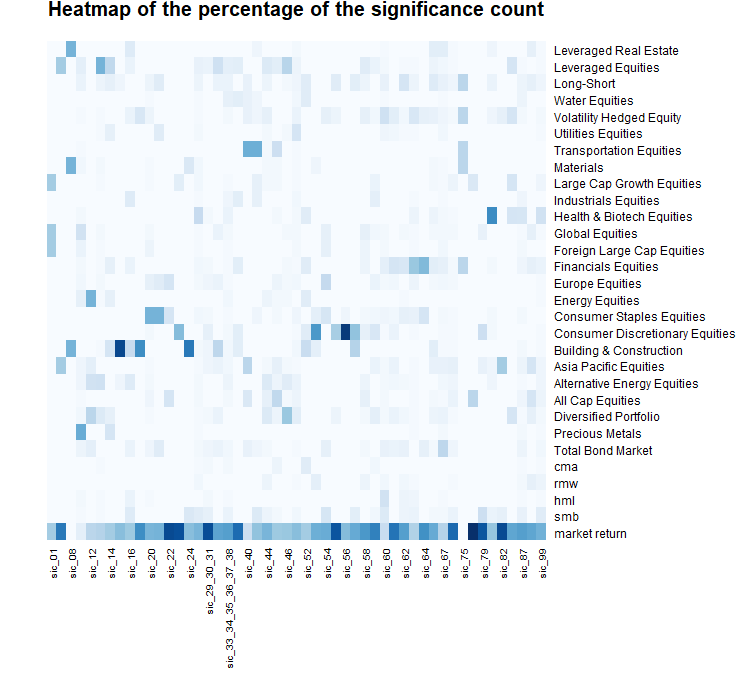}
\caption{Heat map of percentage of significance count. Each grid are related
to the percentage of basis assets selected in the corresponding sector
(shown as the row name) by the company group (classified by the SIC
code shown as the column name). The figure shows that different company
groups may choose some basis assets in common, but also tends to select
different sectors of basis assets.}
\label{fig: amf_heatmap}
\end{figure}

Table \ref{tab: amf_perc_sgn} provides the matrix $\bm{G}$ in percentage. Each
grid is $100\cdot g_{b,d}$ where $g_{b,d}$ is defined in equation
(\ref{eq: perc_table}). Figure \ref{fig: amf_heatmap} is a heat map from which we can
visualize patterns in Table \ref{tab: amf_perc_sgn}. The darker the grid, the
larger the percentage of significant basis assets. As indicated, different
security classes depend on different classes of basis assets, although
some basis assets seem to be shared in common. Not all of the Fama-French
5 risk-factors are significant in presence of the additional basis
assets in our model. Only the market portfolio shows a strong significance
for nearly all securities. The emerging market equities and the money
market ETF basis assets seem to affect many securities as well. As
shown, all of the basis assets are needed to explain security returns
and different securities are related to a small number of different
basis assets.

\begin{table}[H]
%\resizebox{6in}{!}{ %\vspace*{-1cm} \hspace*{-1cm}
%\resizebox{!}{0.49\textwidth}{
%\begin{minipage}{0.95\textwidth}
\centering
\begin{adjustbox}{width=\textwidth, totalheight=\textheight,keepaspectratio}
\begin{tabular}{|| *{18}{c|} |}
\hhline{|*{18}{=}|}
\multirow{2}{*}{ETF Class}  & \multicolumn{17}{c||}{SIC First 2 digits}\tabularnewline
\multirow{2}{*}{} & 01 & 07 & 08 & 10 & 12 & 13 & 14 & 15 & 16 & 17 & 20 & 21 & 22 & 23 & 24 & 25-28 & 29-31 \\
\hhline{|*{18}{=|}|}
market return & 25 & 50 & 0 & 7 & 20 & 21 & 25 & 30 & 26 & 44 & 32 & 33 & 62 & 62 & 30 & 33 & 62 \\
   \hline
smb & 0 & 0 & 0 & 1 & 0 & 1 & 0 & 0 & 9 & 0 & 0 & 0 & 0 & 0 & 10 & 8 & 6 \\
   \hline
hml & 0 & 0 & 0 & 1 & 0 & 1 & 0 & 0 & 4 & 0 & 1 & 0 & 0 & 0 & 0 & 1 & 2 \\
   \hline
rmw & 0 & 0 & 0 & 0 & 0 & 0 & 0 & 0 & 0 & 0 & 0 & 0 & 0 & 0 & 0 & 3 & 0 \\
   \hline
cma & 0 & 0 & 0 & 0 & 0 & 1 & 0 & 0 & 0 & 0 & 0 & 0 & 0 & 0 & 0 & 1 & 2 \\
   \hline
Total Bond Market & 0 & 0 & 0 & 4 & 7 & 0 & 0 & 0 & 4 & 0 & 3 & 8 & 0 & 0 & 0 & 2 & 3 \\
   \hline
Precious Metals & 0 & 0 & 0 & 36 & 0 & 0 & 12 & 0 & 0 & 0 & 0 & 0 & 0 & 0 & 0 & 1 & 0 \\
   \hline
Diversified Portfolio & 0 & 0 & 0 & 2 & 20 & 9 & 6 & 0 & 0 & 0 & 1 & 0 & 0 & 0 & 0 & 1 & 2 \\
   \hline
All Cap Equities & 0 & 0 & 0 & 1 & 0 & 0 & 0 & 0 & 0 & 0 & 3 & 0 & 12 & 0 & 0 & 2 & 0 \\
   \hline
Alternative Energy Equities & 0 & 0 & 0 & 3 & 13 & 14 & 0 & 3 & 9 & 0 & 2 & 0 & 0 & 0 & 0 & 2 & 0 \\
   \hline
Asia Pacific Equities & 0 & 25 & 0 & 3 & 7 & 0 & 0 & 0 & 0 & 0 & 1 & 0 & 0 & 0 & 0 & 1 & 3 \\
   \hline
Building \& Construction & 0 & 0 & 33 & 0 & 0 & 1 & 12 & 63 & 17 & 44 & 0 & 0 & 0 & 0 & 50 & 2 & 3 \\
   \hline
Consumer Discrtnry. Equities & 0 & 0 & 0 & 0 & 0 & 0 & 0 & 0 & 0 & 0 & 2 & 0 & 0 & 31 & 0 & 0 & 6 \\
   \hline
Consumer Staples Equities & 0 & 0 & 0 & 0 & 0 & 0 & 0 & 0 & 0 & 0 & 33 & 33 & 12 & 0 & 0 & 3 & 2 \\
   \hline
Energy Equities & 0 & 0 & 0 & 6 & 33 & 1 & 6 & 0 & 0 & 0 & 0 & 0 & 0 & 0 & 0 & 0 & 0 \\
   \hline
Europe Equities & 0 & 0 & 0 & 2 & 0 & 2 & 0 & 0 & 0 & 0 & 6 & 8 & 12 & 0 & 0 & 2 & 3 \\
   \hline
Financials Equities & 0 & 0 & 0 & 1 & 0 & 1 & 6 & 0 & 4 & 0 & 0 & 0 & 0 & 0 & 0 & 2 & 2 \\
   \hline
Foreign Large Cap Equities & 25 & 0 & 0 & 4 & 0 & 1 & 0 & 0 & 0 & 0 & 4 & 0 & 0 & 0 & 0 & 1 & 0 \\
   \hline
Global Equities & 25 & 0 & 0 & 14 & 0 & 2 & 0 & 0 & 0 & 0 & 2 & 0 & 0 & 0 & 0 & 1 & 0 \\
   \hline
Health \& Biotech Equities & 0 & 0 & 0 & 1 & 0 & 0 & 0 & 0 & 0 & 0 & 0 & 0 & 0 & 0 & 0 & 17 & 2 \\
   \hline
Industrials Equities & 0 & 0 & 0 & 0 & 0 & 1 & 0 & 0 & 9 & 0 & 0 & 0 & 0 & 0 & 0 & 1 & 0 \\
   \hline
Large Cap Growth Equities & 25 & 0 & 0 & 0 & 0 & 1 & 0 & 0 & 0 & 0 & 1 & 0 & 0 & 8 & 0 & 2 & 0 \\
   \hline
Materials & 0 & 0 & 33 & 2 & 0 & 2 & 0 & 0 & 0 & 0 & 1 & 0 & 0 & 0 & 10 & 2 & 0 \\
   \hline
Transportation Equities & 0 & 0 & 0 & 1 & 0 & 0 & 0 & 0 & 0 & 0 & 0 & 0 & 0 & 0 & 0 & 1 & 0 \\
   \hline
Utilities Equities & 0 & 0 & 0 & 0 & 0 & 1 & 6 & 0 & 0 & 0 & 0 & 8 & 0 & 0 & 0 & 2 & 0 \\
   \hline
Volatility Hedged Equity & 0 & 0 & 0 & 1 & 0 & 1 & 0 & 0 & 4 & 11 & 4 & 0 & 0 & 0 & 0 & 1 & 0 \\
   \hline
Water Equities & 0 & 0 & 0 & 0 & 0 & 0 & 0 & 0 & 0 & 0 & 1 & 0 & 0 & 0 & 0 & 1 & 0 \\
   \hline
Long-Short & 0 & 0 & 0 & 4 & 0 & 4 & 6 & 3 & 0 & 0 & 4 & 8 & 0 & 0 & 0 & 4 & 0 \\
   \hline
Leveraged Equities & 0 & 25 & 0 & 7 & 0 & 33 & 19 & 0 & 4 & 0 & 0 & 0 & 0 & 0 & 0 & 5 & 5 \\
   \hline
Leveraged Real Estate & 0 & 0 & 33 & 0 & 0 & 0 & 0 & 0 & 9 & 0 & 0 & 0 & 0 & 0 & 0 & 0 & 0 \\
\hhline{|*{18}{=}|}
\end{tabular}
%\end{minipage}
%}

\end{adjustbox}
\caption{Table for the percentage of significance count. The table provides the matrix $\bm{G}$ in percentage (each grid is $100\cdot g_{b,d}$ where $g_{b,d}$ is defined in equation (\ref{eq: perc_table})). Each grid is the percentage of the basis asset selected in the corresponding sector (shown as the row name) by the company group (classified by the SIC
code shown as the column name). Note that the elements in each column add up to 100, which means 100\% (maybe be slightly different from 100 due to the rounding issue). The percent signs are omitted to save space.}
\label{tab: amf_perc_sgn}
\end{table}

%\hspace*{-1cm}
\begin{table}[H]
%\resizebox{!}{0.5\textwidth}{
%\begin{minipage}{0.95\textwidth}
\centering
\begin{adjustbox}{width=\textwidth, totalheight=\textheight,keepaspectratio}
\begin{tabular}{|| *{18}{c|} |}
\hhline{|*{18}{=}|}
\multirow{2}{*}{ETF Class}  & \multicolumn{17}{c||}{SIC First 2 digits}\tabularnewline
\multirow{2}{*}{} & 32 & 33-38 & 39 & 40 & 42 & 44 & 45 & 46 & 47-51 & 52 & 53 & 54 & 55 & 56 & 57 & 58 & 59 \\
\hhline{|*{18}{=|}|}
market return & 38 & 40 & 54 & 15 & 28 & 33 & 26 & 27 & 30 & 25 & 34 & 35 & 60 & 30 & 36 & 42 & 48 \\
   \hline
smb & 0 & 4 & 0 & 0 & 9 & 4 & 2 & 0 & 2 & 0 & 0 & 12 & 0 & 0 & 7 & 0 & 2 \\
   \hline
hml & 0 & 1 & 0 & 0 & 0 & 0 & 0 & 0 & 0 & 0 & 0 & 0 & 0 & 0 & 0 & 0 & 0 \\
   \hline
rmw & 0 & 0 & 0 & 0 & 0 & 4 & 0 & 0 & 1 & 0 & 7 & 0 & 0 & 0 & 0 & 4 & 0 \\
   \hline
cma & 0 & 2 & 0 & 0 & 0 & 0 & 4 & 0 & 0 & 8 & 0 & 0 & 0 & 0 & 0 & 0 & 2 \\
   \hline
Total Bond Market & 5 & 2 & 0 & 5 & 3 & 2 & 0 & 0 & 1 & 0 & 0 & 6 & 0 & 0 & 0 & 2 & 0 \\
   \hline
Precious Metals & 0 & 0 & 0 & 0 & 0 & 0 & 0 & 0 & 1 & 0 & 0 & 0 & 0 & 0 & 0 & 0 & 0 \\
   \hline
Diversified Portfolio & 5 & 0 & 0 & 0 & 0 & 10 & 4 & 27 & 7 & 0 & 0 & 0 & 4 & 0 & 0 & 2 & 7 \\
   \hline
All Cap Equities & 0 & 2 & 0 & 10 & 0 & 6 & 19 & 2 & 4 & 0 & 0 & 0 & 0 & 0 & 0 & 6 & 0 \\
   \hline
Alternative Energy Equities & 0 & 4 & 0 & 0 & 0 & 10 & 4 & 8 & 5 & 0 & 0 & 0 & 0 & 0 & 0 & 4 & 0 \\
   \hline
Asia Pacific Equities & 5 & 2 & 0 & 20 & 0 & 2 & 6 & 2 & 1 & 0 & 0 & 0 & 0 & 0 & 0 & 0 & 4 \\
   \hline
Building \& Construction & 19 & 3 & 8 & 0 & 0 & 0 & 0 & 0 & 2 & 17 & 7 & 0 & 0 & 0 & 21 & 0 & 0 \\
   \hline
Consumer Discrtnry. Equities & 0 & 2 & 8 & 0 & 0 & 0 & 0 & 0 & 1 & 8 & 41 & 0 & 24 & 67 & 29 & 6 & 11 \\
   \hline
Consumer Staples Equities & 0 & 1 & 0 & 0 & 3 & 0 & 2 & 0 & 3 & 0 & 7 & 12 & 0 & 3 & 0 & 2 & 2 \\
   \hline
Energy Equities & 0 & 2 & 0 & 0 & 0 & 4 & 2 & 2 & 0 & 8 & 0 & 0 & 0 & 0 & 0 & 0 & 0 \\
   \hline
Europe Equities & 5 & 1 & 0 & 0 & 0 & 0 & 4 & 2 & 3 & 0 & 0 & 18 & 0 & 0 & 0 & 0 & 4 \\
   \hline
Financials Equities & 0 & 1 & 8 & 0 & 0 & 0 & 0 & 2 & 1 & 8 & 0 & 0 & 0 & 0 & 0 & 4 & 0 \\
   \hline
Foreign Large Cap Equities & 0 & 1 & 0 & 0 & 0 & 0 & 0 & 0 & 2 & 0 & 0 & 6 & 0 & 0 & 0 & 4 & 0 \\
   \hline
Global Equities & 5 & 1 & 0 & 0 & 0 & 0 & 0 & 2 & 2 & 0 & 0 & 6 & 0 & 0 & 0 & 6 & 2 \\
   \hline
Health \& Biotech Equities & 0 & 5 & 0 & 0 & 0 & 0 & 2 & 0 & 2 & 8 & 0 & 0 & 0 & 0 & 0 & 0 & 0 \\
   \hline
Industrials Equities & 0 & 3 & 8 & 0 & 6 & 0 & 0 & 0 & 2 & 0 & 0 & 0 & 0 & 0 & 0 & 0 & 7 \\
   \hline
Large Cap Growth Equities & 0 & 1 & 0 & 0 & 6 & 2 & 2 & 0 & 1 & 0 & 0 & 0 & 0 & 0 & 0 & 0 & 4 \\
   \hline
Materials & 0 & 0 & 0 & 0 & 0 & 2 & 2 & 0 & 1 & 0 & 3 & 0 & 0 & 0 & 0 & 0 & 0 \\
   \hline
Transportation Equities & 0 & 0 & 0 & 35 & 34 & 2 & 15 & 0 & 1 & 0 & 0 & 0 & 0 & 0 & 0 & 0 & 0 \\
   \hline
Utilities Equities & 0 & 0 & 0 & 0 & 0 & 2 & 0 & 2 & 12 & 0 & 0 & 0 & 0 & 0 & 0 & 0 & 0 \\
   \hline
Volatility Hedged Equity & 0 & 2 & 0 & 5 & 3 & 6 & 0 & 0 & 4 & 0 & 0 & 6 & 4 & 0 & 0 & 6 & 2 \\
   \hline
Water Equities & 0 & 6 & 8 & 5 & 3 & 0 & 0 & 0 & 2 & 8 & 0 & 0 & 0 & 0 & 0 & 0 & 0 \\
   \hline
Long-Short & 5 & 5 & 0 & 5 & 0 & 4 & 0 & 2 & 3 & 8 & 0 & 0 & 8 & 0 & 7 & 4 & 0 \\
   \hline
Leveraged Equities & 14 & 7 & 8 & 0 & 0 & 10 & 8 & 21 & 4 & 0 & 0 & 0 & 0 & 0 & 0 & 8 & 4 \\
   \hline
Leveraged Real Estate & 0 & 0 & 0 & 0 & 3 & 0 & 0 & 0 & 2 & 0 & 0 & 0 & 0 & 0 & 0 & 0 & 0 \\
   \hline
\hhline{|*{18}{=}|}
\end{tabular}
%\end{minipage}
%}
\end{adjustbox}
\caption*{Table \ref{tab: amf_perc_sgn} continued.}
\end{table}

%\hspace*{-1cm}
\begin{table}[H]
\centering
\begin{adjustbox}{width=\textwidth, totalheight=\textheight,keepaspectratio}
\begin{tabular}{|| *{18}{c|} |}
\hhline{|*{18}{=}|}
\multirow{2}{*}{ETF Class}  & \multicolumn{17}{c||}{SIC First 2 digits}\tabularnewline
\multirow{2}{*}{}  & 60 & 61 & 62 & 63 & 64 & 65 & 67 & 70-73 & 75 & 78 & 79 & 80 & 82 & 83 & 87 & 89 & 99 \\
\hhline{|*{18}{=|}|}
market return & 16 & 51 & 40 & 22 & 44 & 36 & 21 & 55 & 0 & 70 & 60 & 27 & 62 & 38 & 40 & 38 & 34 \\
   \hline
smb & 10 & 0 & 4 & 5 & 0 & 2 & 2 & 4 & 0 & 0 & 15 & 5 & 6 & 0 & 5 & 0 & 8 \\
   \hline
hml & 14 & 0 & 4 & 3 & 0 & 0 & 1 & 0 & 0 & 0 & 0 & 0 & 0 & 0 & 1 & 0 & 1 \\
   \hline
rmw & 3 & 0 & 3 & 1 & 0 & 0 & 1 & 1 & 0 & 0 & 0 & 0 & 0 & 0 & 2 & 6 & 4 \\
   \hline
cma & 0 & 0 & 0 & 0 & 0 & 0 & 0 & 0 & 0 & 0 & 0 & 0 & 0 & 0 & 1 & 0 & 1 \\
   \hline
Total Bond Market & 5 & 5 & 5 & 0 & 6 & 2 & 19 & 3 & 0 & 0 & 0 & 0 & 0 & 0 & 4 & 0 & 3 \\
   \hline
Precious Metals & 0 & 0 & 1 & 1 & 0 & 0 & 2 & 0 & 0 & 0 & 0 & 0 & 0 & 0 & 2 & 0 & 1 \\
   \hline
Diversified Portfolio & 2 & 5 & 3 & 1 & 0 & 4 & 4 & 1 & 0 & 0 & 0 & 0 & 0 & 12 & 1 & 6 & 2 \\
   \hline
All Cap Equities & 1 & 0 & 0 & 1 & 0 & 2 & 1 & 4 & 0 & 20 & 0 & 0 & 0 & 0 & 3 & 12 & 4 \\
   \hline
Alternative Energy Equities & 2 & 2 & 2 & 1 & 0 & 0 & 2 & 3 & 0 & 0 & 0 & 3 & 0 & 0 & 1 & 0 & 3 \\
   \hline
Asia Pacific Equities & 2 & 5 & 1 & 1 & 0 & 5 & 6 & 6 & 0 & 0 & 5 & 3 & 25 & 0 & 4 & 12 & 3 \\
   \hline
Building \& Construction & 0 & 0 & 0 & 0 & 0 & 7 & 1 & 1 & 0 & 0 & 0 & 0 & 0 & 0 & 2 & 0 & 1 \\
   \hline
Consumer Discrtnry. Equities & 1 & 5 & 1 & 1 & 0 & 0 & 1 & 1 & 0 & 0 & 15 & 2 & 0 & 0 & 0 & 0 & 1 \\
   \hline
Consumer Staples Equities & 3 & 2 & 6 & 5 & 12 & 0 & 1 & 1 & 0 & 0 & 0 & 2 & 0 & 0 & 2 & 0 & 1 \\
   \hline
Energy Equities & 0 & 0 & 1 & 0 & 0 & 0 & 0 & 0 & 0 & 0 & 0 & 0 & 0 & 0 & 0 & 0 & 1 \\
   \hline
Europe Equities & 2 & 2 & 1 & 4 & 0 & 2 & 2 & 2 & 0 & 0 & 0 & 0 & 0 & 0 & 0 & 0 & 1 \\
   \hline
Financials Equities & 7 & 12 & 11 & 27 & 31 & 7 & 6 & 2 & 20 & 0 & 0 & 2 & 0 & 0 & 3 & 6 & 4 \\
   \hline
Foreign Large Cap Equities & 2 & 0 & 0 & 1 & 0 & 0 & 2 & 1 & 0 & 0 & 0 & 0 & 0 & 0 & 2 & 0 & 1 \\
   \hline
Global Equities & 2 & 0 & 2 & 1 & 0 & 2 & 2 & 1 & 0 & 0 & 5 & 0 & 0 & 0 & 1 & 6 & 1 \\
   \hline
Health \& Biotech Equities & 1 & 0 & 0 & 4 & 0 & 4 & 1 & 1 & 0 & 0 & 0 & 45 & 0 & 12 & 11 & 0 & 13 \\
   \hline
Industrials Equities & 0 & 0 & 0 & 1 & 0 & 0 & 1 & 0 & 0 & 0 & 0 & 0 & 0 & 0 & 3 & 0 & 0 \\
   \hline
Large Cap Growth Equities & 0 & 0 & 0 & 0 & 0 & 2 & 1 & 4 & 0 & 10 & 0 & 0 & 0 & 12 & 0 & 0 & 3 \\
   \hline
Materials & 0 & 0 & 1 & 0 & 0 & 2 & 1 & 0 & 20 & 0 & 0 & 0 & 0 & 0 & 0 & 0 & 1 \\
   \hline
Transportation Equities & 0 & 0 & 0 & 0 & 0 & 0 & 0 & 1 & 20 & 0 & 0 & 0 & 0 & 0 & 0 & 0 & 1 \\
   \hline
Utilities Equities & 3 & 2 & 1 & 1 & 0 & 0 & 3 & 0 & 0 & 0 & 0 & 0 & 0 & 0 & 0 & 0 & 1 \\
   \hline
Volatility Hedged Equity & 15 & 7 & 3 & 11 & 6 & 5 & 3 & 2 & 20 & 0 & 0 & 3 & 6 & 12 & 2 & 0 & 2 \\
   \hline
Water Equities & 1 & 0 & 0 & 0 & 0 & 0 & 1 & 0 & 0 & 0 & 0 & 0 & 0 & 0 & 4 & 0 & 1 \\
   \hline
Long-Short & 5 & 0 & 12 & 4 & 0 & 9 & 5 & 3 & 20 & 0 & 0 & 5 & 0 & 0 & 4 & 6 & 3 \\
   \hline
Leveraged Equities & 2 & 0 & 1 & 1 & 0 & 4 & 2 & 1 & 0 & 0 & 0 & 0 & 0 & 12 & 1 & 0 & 1 \\
   \hline
Leveraged Real Estate & 1 & 0 & 1 & 0 & 0 & 7 & 7 & 1 & 0 & 0 & 0 & 3 & 0 & 0 & 0 & 6 & 1 \\
\hhline{|*{18}{=|}|}
\end{tabular}
\end{adjustbox}
\caption*{Table \ref{tab: amf_perc_sgn} continued.}
\end{table}

%\clearpage
\subsection{Intercept Test}

\begin{figure}[H]
\center \vspace{0cm}
\includegraphics[width=0.9\textwidth]{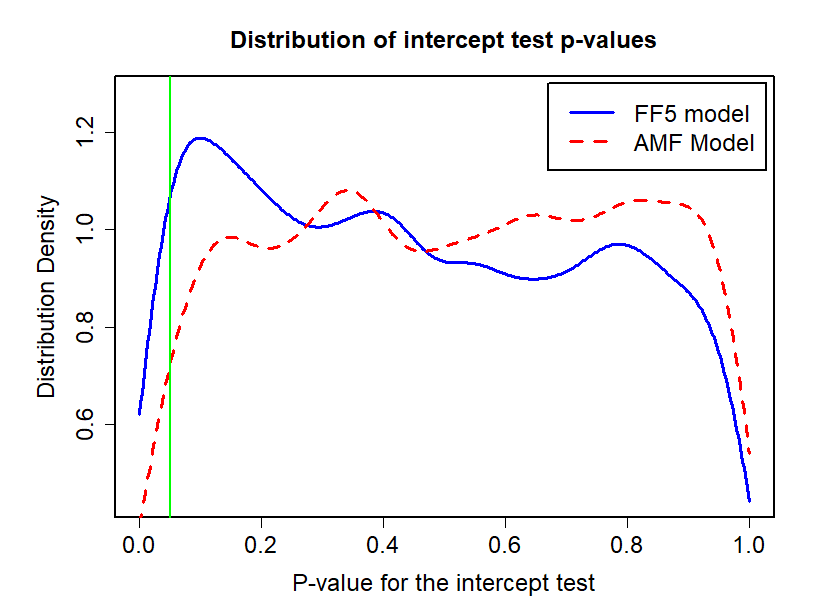}
\captionof{figure}{Comparison of intercept test p-values for the Fama-French 5-factor (FF5) model and the Adaptive Multi-factor (AMF) model.}
\label{fig: amf_int}
\end{figure}

This section provides the tests for a zero intercept. Using the Fama-French
5-factor model as a comparison, Figure \ref{fig: amf_int} compares the intercept
test p-values between our basis asset implied Adaptive Multi-factor (AMF) model and Fama-French 5-factor (FF5) model. As indicated, 6.22\% (above 5\%) of the securities have significant intercepts in the FF5 model, while 3.86\% (below 5\%) of the securities in AMF have significant $\alpha$'s. This may suggests that the AMF model is more insightful than the FF5 model, since AMF reveals more relevant factors and makes the intercept closer to 0.

Since we replicate this test for about 5000 stocks in the CRSP database,
it is important to control for a False Discovery Rate (FDR) because even
if there is a zero intercept, a replication of 5000 tests will have
about 5\% showing false significance. We adjust for the false discovery rate using the Benjamini-Hochberg (BH) procedure \cite{benjamini1995controlling}
and the Benjamini-Hochberg-Yekutieli (BHY) procedures \cite{benjamini2001control}.
The BH method does not account for the correlation between tests,
while the BHY method does. In our case, each test is done on an individual
stock, which may have correlations. So the BHY method is more appropriate
here.

Chordia et al. (2017) \cite{chordia2017p} suggests that the false
discovery proportion (FDP) approach in \cite{romano2007control} should
be applied rather than a false discovery rate procedure. The BH approach
only controls the expected value of FDP while FDP controls the family-wise
error rate directly. %We use the R package \textit{StepwiseTest} \cite{hsu2016stepwisetest}
%to perform the FDP test, which gives all zeros for the intercept test,
%meaning that no true discovery (significant non-zero intercept) was
%found. In other word, in our analysis, FDP gives the same results
%as BHY.
There is another test worth mentioning, which is the GRS test. The
GRS test in \cite{gibbons1989test} is usually an excellent procedure
for testing intercepts. However, these two tests are not appropriate
in the high-dimensional regression setting as in our case. To be specific, these
tests are implicitly based on the assumption that all companies are
only related to the same small set of basis assets. Here the ``small''
means there are many fewer basis assets than observations. Our setting
is more general since we may have more basis assets than observations,
although each company is related to only a small number of basis assets,
different companies may be related to different sets of basis assets.
The GRS is unable to handle setting.

As noted earlier, Table \ref{tab: amf_int} shows that 3.86\% of stocks
in the multi-factor model have p-values for the intercept t-test of
less than 0.05. While in the FF5 model, this percentage is 6.22\%.
After using the BHY method to control for the false discovery rate,
we can see that the q-values (the minimum false discovery rate needed
to accept that this rejection is a true discovery, see \cite{storey2002direct})
for both models are almost 1, indicating that there are no significant
non-zero intercepts. All the significance shown in the intercept tests
is likely to be false discovery. This is the evidence that both
models are consistent with the behavior of ``large-time scale''
security returns.

\begin{table}[H]
\center %
\begin{tabular}{|| *{5}{r|} |}
\hhline{|*{5}{=}}
\multirow{2}{*}{Value Range}  & \multicolumn{4}{c||}{Percentage of Stocks (\%)}\tabularnewline
\multirow{2}{*}{}  & FF5 p-val & AMF p-val & FF5 FDR q-val & AMF FDR q-val \\
\hhline{|*{5}{=|}}
0 - 0.05 & 6.22 & 3.86 & 0.00 & 0.00 \\
\hline
0.05 - 0.9 & 84.88 & 85.33 & 0.02 & 0.00 \\
\hline
0.9 - 1 & 8.90 & 10.81 & 99.98 & 100.00 \\
\hhline{|*{5}{=|}}
\end{tabular}
\caption{Intercept Test with control of false discovery rate. The first column is the value range of p-values or q-values listed in the other columns. The other 4 columns are related to p-values and False Discovery Rate (FDR) q-values for the FF5 model and the AMF model. For each column, we listed the percentage of companies with values within each value range. It is clear that nearly all rejections of zero-alpha are false discoveries. }
\label{tab: amf_int}
\end{table}

%\clearpage
\subsection{In-Sample and Out-of-Sample Goodness-of-Fit}

\begin{figure}[H]
\center \vspace{0cm}
 \includegraphics[width=0.8\textwidth]{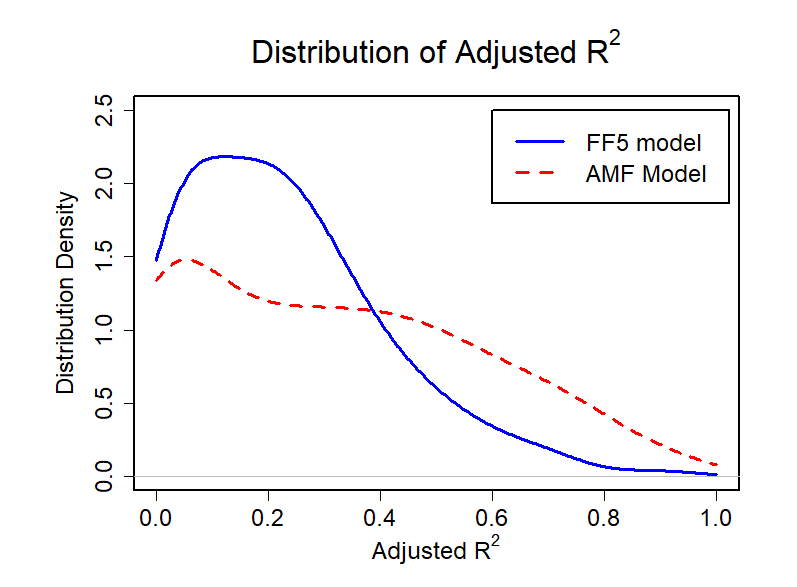}
 \captionof{figure}{Comparison
of adjusted $R^{2}$ for the Fama-French 5-factor (FF5) model and
the Adaptive Multi-factor (AMF) model}
\label{fig: amf_adj_r2}
\end{figure}

This section tests to see which model fits the data best. Figure
\ref{fig: amf_adj_r2} compares the distribution of the adjusted $R^{2}$'s (see \cite{theil1961economic}) between the AMF and the FF5 model. As indicated, the AMF model has more explanatory power. The mean adjusted $R^{2}$ for the AMF model is 0.319 while that for the FF5 model is 0.229. The AMF model increases the adjusted $R^2$ by 39.2\% compared to the FF5.

We next perform an F-test, for each security, to show that there is
a significant difference between the goodness-of-fit of the AMF %(equation \ref{eq: amf_ols_subset})
and the FF5 model. %(equation \ref{eq: ff5}).
Since we need a nested comparison for an F-test, we compare the results
between FF5 and GIBS + FF5 (which is including FF5 factors back to
GIBS for fitting if any of the FF5 factors are not selected). In our
case, the FF5 is the restricted model, having $p-r_{1}$ degrees of
freedom and a sum of squared residuals $SS_{R}$, where $r_{1}=5$.
The AMF is the full model, having $p-r_{1}-r_{2}$ degrees of freedom
(where $r_{2}$ is the number of basis assets selected in addition
to FF5) and a sum of squared residuals $SS_{F}$. Under the null-hypothesis
that FF5 is the true model, we have
\begin{equation}
F_{obs}=\frac{(SS_{R}-SS_{F})/r_{2}}{SS_{F}/(p-r_{1}-r_{2})}\overset{H_{0}}{\sim}F_{r_{2},p-r_{1}-r_{2}}.
\end{equation}

There are 5132 stocks in total. For 1931 (37.63\%) of them, the GIBS
algorithm only selects some of the FF5 factors, so for these stocks,
GIBS + FF5 does not give extra information. However, for 3201 (62.37\%)
of them, the GIBS algorithm does select ETFs outside of the FF5 factors.
For these stocks, we do the F test to check whether the difference
between the two models are significant, in other words, whether AMF
gives a significantly better fit. As shown in Table \ref{tab: amf_f_test},
for 97.72\% of the stocks, the AMF model fits better than the FF5
model.

Again, it is important to test the False Discovery Rate (FDR). Table
\ref{tab: amf_f_test} contains the p-values and the false discovery rate
q-values using both the BH method and BHY methods. As indicated, for
most of the stocks, the AMF is significantly better than the FF5 model,
even after considering the false discovery rate. For 97.72\% of stocks,
the AMF model is better than the FF5 at the significance level of
0.05. After considering the false discovery rate using the strict
BHY method which includes the correlation between tests, there is
still 90.16\% of stocks significant with q-values less than 0.05.
Even if we adjust our false discover rate q-value significance level
to 0.01, there is still 83.29\% of the stocks showing a significant
difference. As such, this is strong evidence supporting the multi-factor
model's superior performance in characterizing security returns.

\begin{table}[H]
\center %
\begin{tabular}{||c|r|r|r||}
\hhline{|=|===|}
\multirow{2}{*}{Value Range}  & \multicolumn{3}{c||}{Percentage of Stocks}\tabularnewline
\multirow{2}{*}{}  & p-value  & BH q-value  & BHY q-value\tabularnewline
\hline
$ 0 \sim 0.01 $ & 93.44\% & 93.06\% & 83.29\% \\
\hline
$ 0 \sim 0.05 $ (Significant) & 97.72\% & 97.53\% & 90.16\% \\
\hline
$ 0.05 \sim 1 $ (Non-Significant) & 2.28\% & 2.47\% & 9.84\% \\
\hhline{|=|=|=|=|}
\end{tabular}\caption{F test with control of false discovery rate. We do the F test and
report its p-value, q-values for each company. The first column is
the value range of p-values or q-values listed in the other columns.
In the other three columns we report percentage of companies with
p-value, BH method q-value, and BHY method q-value in each value range.
The table shows that for most companies the increment of goodness
of fit is very significant.}
\label{tab: amf_f_test}
\end{table}

Apart from the In-Sample goodness of fit results, we also compare
the Out-of-Sample goodness of fit of the FF5 and AMF model in the
prediction time period. We use the two models to predict the return
of the following week and report the Out-of-Sample $R^{2}$ for the
prediction (see Table \ref{tab: amf_compare_methods}). The Out-of-Sample
$R^{2}$ (see \cite{campbell2008predicting}) is used to measure the
predictive accuracy of a model. The Out-of-Sample $R^{2}$ for the
FF5 is $0.030$, while that for the AMF is $0.038$. That is, the
AMF model increased the Out-of-Sample $R^{2}$ by $24.07\%$ compared
to the FF5 model. The AMF model shows its superior performance by
giving a more accurate prediction using an even lower number of factors,
which is also strong evidence against overfitting. The AMF model provides
additional insight when compared to the FF5.

%\clearpage
\subsection{Robustness Test}

As a robustness test, for those securities whose intercepts were non-zero,
we tested the basis asset implied multi-factor model to see if positive
alpha trading strategies generate arbitrage opportunities. To construct
the positive alpha trading strategies, we use the data from the year
2017 as an out-of-sample period. Recall that the previous analysis
was over the time period 2014 to 2016. As explained above, we fit
the AMF model using the data up to the last week of 2016. We then
ranked the securities by their alphas from positive to negative. We
take the top 50\% of those with significant (p-val less than 0.05)
positive alphas and form a long-only equal-weighted portfolio with
\$1 in initial capital. Similarly, take the bottom 50\% of those with
significant negative alphas and form a short-only equal-weighted portfolio
with -\$1 initial capital. Then, each week over 2017, we update the
two portfolios by re-fitting the AMF model and repeating the same
construction. Combining the long-only and short-only portfolio forms
a portfolio with 0 initial investment. If the alphas represent arbitrage
opportunities, then the combined long and short portfolio's change
in value will always be non-negative and strictly positive for some
time periods.

\begin{figure}[H]
\center \includegraphics[width=0.65\linewidth]{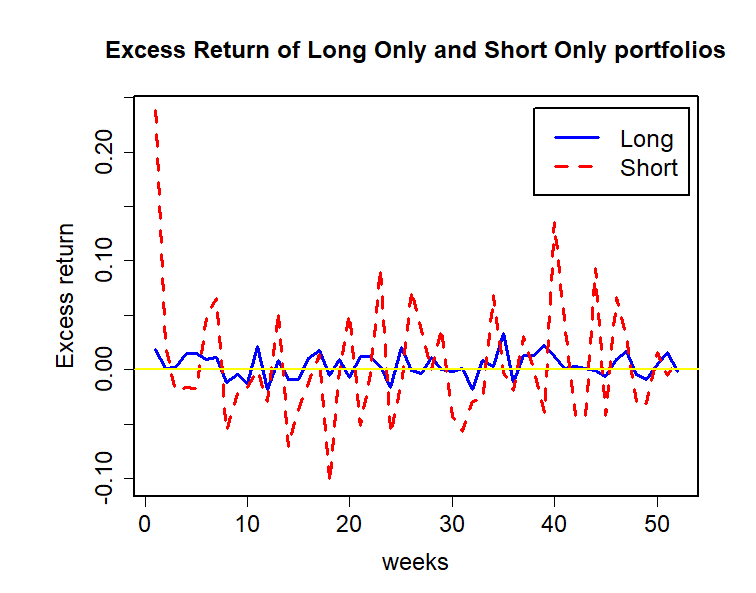} \captionof{figure}{Returns of Long-only and Short-only Portfolio}
\label{fig: amf_returns}
\end{figure}

\begin{figure}[H]
\center \includegraphics[width=0.65\linewidth]{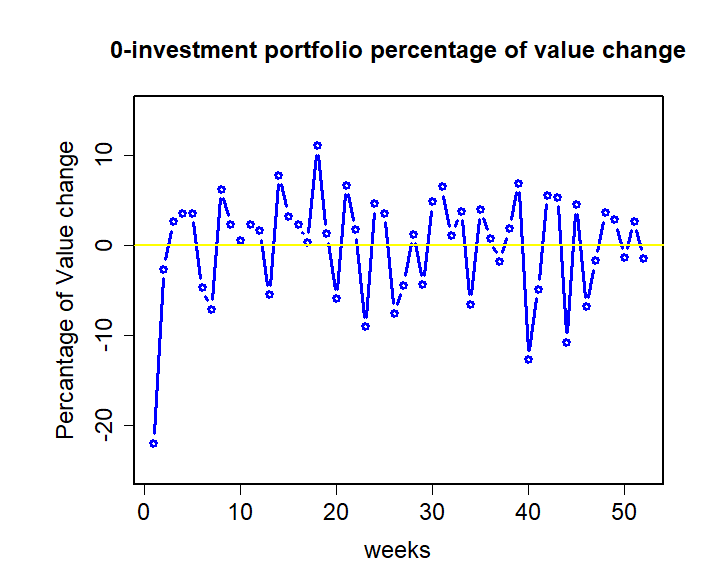}
\captionof{figure}{Percentage of Value Change of 0-Investment portfolio}
\label{fig: amf_value_change}
\end{figure}

The results of the arbitrage tests are shown in Figures \ref{fig: amf_returns}
and \ref{fig: amf_value_change}. As indicated, the change in value of the
0-investment portfolio randomly fluctuates on both sides of 0. This
rejects the possibility that the positive alpha trading strategy is
an arbitrage opportunity. Thus, this robustness test confirms our
previous intercept test results, after controlling for a false discovery
rate. Although not reported, we also studied different quantiles from
10\% to 40\% and they give similar results.

%\clearpage{}

%\begin{figure}
%\vspace{0cm}
% %
%\begin{minipage}[c]{0.5\textwidth}%
% \centering \includegraphics[width=0.99\linewidth]{adj_r2.png} \captionof{figure}{Comparison
%of Adjusted R squared} \label{fig: amf_adj_r2} %
%\end{minipage}%
%\begin{minipage}[c]{0.5\textwidth}%
% \centering \includegraphics[width=0.99\linewidth]{InterceptTest.png}
%\captionof{figure}{Comparison of intercept test} \label{fig: amf_int} %
%\end{minipage}
%\end{figure}

%\clearpage{}

\section{Comparison with Alternative Methods}
\label{sec: amf_comparison}

\subsection{Are Fama-French 5 Factors Overfitting?}

We first test whether the Fama-French 5 factors (FF5) overfit the
noise in the data. This can be done by estimating a ``GIBS + FF5''
model. This model is very similar to GIBS, except that it includes
the Fama-French 5 factors to be selected in the last step. That is,
if any of the FF5 factors are not selected by GIBS, we add them back
to our selected basis asset set $\w S_{i}$ and use this set of basis
assets to fit and predict the returns. By comparing the In-Sample
Adjusted $R^{2}$ and the Out-of-Sample $R^{2}$ (see \cite{campbell2008predicting})
of GIBS + FF5 model and the GIBS model, we can determine whether FF5
factors are overfitting. The Out-of-Sample $R^{2}$ (see \cite{campbell2008predicting})
is used to measure the accuracy of prediction of a model. Surprisingly,
the results show that some of the FF5 factors are over-fitting! As
shown in Table \ref{tab: amf_compare_methods}, compared to our GIBS model,
the GIBS + FF5 achieves a better in-sample Adjusted $R^{2}$, with
more significant basis assets, but gives a much worse Out-of-Sample
$R^{2}$. This indicates that the FF5 factors not selected by GIBS
are ``false discoveries'' - they overfit the training data, but
do a poor job in predicting. Therefore, those FF5 factors should not
be used for a company if they are not selected by GIBS. Table \ref{tab: amf_compare_methods}
not only provides evidence of the superior performance of GIBS over
FF5 by comparing the In-Sample Adjusted $R^{2}$ and Out-of-Sample
$R^{2}$ of GIBS and FF5 model, but it also indicates an overfitting
of FF5 by comparing the In-Sample Adjusted $R^{2}$ and Out-of-Sample
$R^{2}$ of GIBS and GIBS + FF5.

\subsection{Comparison with Elastic Net}

\begin{figure}[H]
\center \vspace{0cm}
\includegraphics[width=0.67\textwidth]{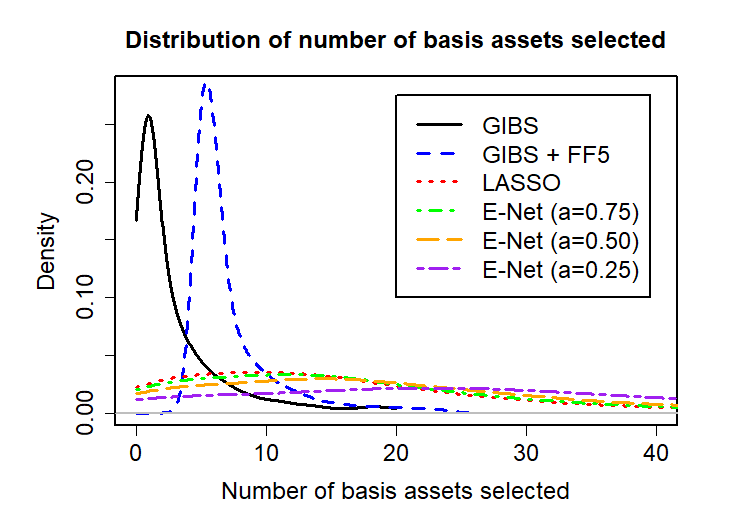}
\caption{Comparison of number of basis assets selected by cross-validation for different methods.}
\label{fig: amf_compare_n_factors}
\end{figure}

Since there are a large number of correlated ETFs it is natural to
employ the RIDGE, LASSO and Elastic Net (E-Net) methods (by Zou and
Hastie (2005) \cite{zou2005regularization}). E-Net is akin to the
ridge regression's treatment of multicollinearity with an additional
tuning (ridge) parameter, $\alpha$, that regularizes the correlations.
We compare GIBS, LASSO, RIDGE and E-Net with different $\alpha$s.
The sparsity inducing parameter $\lambda$ in each model is selected
by the usual 10-fold cross-validation (by Kohavi et al. (1995) \cite{kohavi1995study}).
The comparison results are shown in Table \ref{tab: amf_compare_methods}.
The distribution of the number of basis assets selected by each method
is shown in Figure \ref{fig: amf_compare_n_factors}.

From Table \ref{tab: amf_compare_methods} it is clear that the GIBS model
has better prediction than the FF5 model. The GIBS model increased
the Out-of-Sample $R^{2}$ by 24.07\% compared to FF5. Across all
models, GIBS has the highest Out-of-Sample $R^{2}$, which supports
the fact that the better In-Sample Adjusted $R^{2}$ achieved by the
other models (LASSO, RIDGE, E-Net etc.) is due to overfitting. Furthermore,
from Table \ref{tab: amf_compare_methods} and Figure \ref{fig: amf_compare_n_factors},
we see that GIBS selects the least number of factors.

\begin{table}[H]
%\resizebox{6in}{!}{
\centering
\begin{adjustbox}{width={0.95\textwidth}, totalheight={0.95\textheight},keepaspectratio}
\begin{tabular}{| *{5}{|r} ||}
\hhline{*{5}{|=}|}
Model & Select & Signif. & In-Sample Adj. $R^2$  & Out-of-Sample $R^2$ \\
\hhline{*{5}{|=}|}
FF5 & 5.00 & 1.78 & 0.229 (00.00\%) & 0.030 (00.00\%) \\
\hline
GIBS & 2.98 & 1.92 & 0.319 (39.18\%) & \textbf{0.038 (+24.07\%)} \\
\hline
GIBS + FF5 & 7.20 & 2.50 & 0.350 (52.55\%) & 0.025 (-16.71\%) \\
\hline
LASSO & 15.66 & 5.72 & 0.466 (103.46\%) & 0.018 (-40.97\%) \\
\hline
E-Net ($\alpha$=0.75) & 16.84 & 5.81 & 0.470 (105.22\%) & 0.018 (-40.09\%) \\
\hline
E-Net ($\alpha$=0.50) & 19.28 & 6.05 & 0.479 (109.05\%) & 0.015 (-49.92\%) \\
\hline
E-Net ($\alpha$=0.25) & 26.36 & 6.51 & 0.498 (117.20\%) & 0.009 (-70.33\%) \\
\hline
Ridge & 182.00 & NA & NA & -6$\times10^4$ (-2$\times10^8$\%) \\
\hhline{*{5}{|=}|}
\end{tabular}
\end{adjustbox}
\caption{Comparison table for Alternative Methods. The ``Select'' column gives the average count of the factors selected by the model. The ``Signif.'' column gives the average count of the significant factors selected by the model. The column ``In-sample Adj. $R^2$'' gives the average in-sample Adjusted $R^2$ for each model, the percentage in the bracket is the percentage change compared to the FF5 model. The column ``Out-of-Sample $R^2$'' gives the average Out-of-Sample $R^2$ for each model, the percentage in the bracket is the percentage change compared to the FF5 model.}
\label{tab: amf_compare_methods}
\end{table}

The $\lambda$ of LASSO is traditionally selected by cross-validation
and with this $\lambda$, the model selects an average of 15.66 factors,
as shown in Table \ref{tab: amf_compare_methods}. However, most of the
factors selected by cross-validation are ``false-positive''. Therefore,
instead of the cross-validation, we use the ``1se rule'' with a
hard threshold 20 basis assets at most. The ``1se rule'' use the
largest $\lambda$ such that the cross-validation error is within
one standard error of the minimum error achieved by the cross-validation.
In another word, $\lambda_{1se}$ gives the most regularized model
such that error is within one standard error of the minimum error
achieved by the cross-validation (see \cite{friedman2010regularization,simon2011Regularization,tibshirani2012strong}).
To further avoid over-fitting, we include the threshold that each
company can not be related to more than 20 basis assets. As shown
in the Table \ref{tab: amf_compare_methods}, our method used in GIBS
works well and achieves the best prediction power. The reason for
the superior performance of GIBS compared with the cross-validation
LASSO is that cross-validation often overfits, especially when the
sample size is small, or when the data is not sufficiently independent
and identically distributed. In addition, our results with GIBS are
both stable and interpretable.

Our choice of dimension reduction techniques, using a combination
of prototype clustering and a modified version of LASSO, was motivated
by our desire to select, from a collection of strongly correlated
ETFs, a sparse and interpretable set of basis assets that explains
the cross-sectional variation among asset returns. These two steps
were used as model selection tools to identify basis assets, and we
subsequently estimated the model coefficients using OLS. In future
research, a more integrative method may be designed to combine the
model selection and estimation steps.

The motivation for using prototype clustering is two fold. First,
it can be used to derive the cluster structure of the ETFs so that
the redundant ones are removed. This reduces the correlation and validates
the use of LASSO. Second, this method gives a clear interpretation
of the prototypes, which is important for our interpretation of the
basis assets. The traditional methods of dealing with empirical asset
models are based on variance decomposition of the basis assets (the
$X$ matrix) as in, for example, the principal component analysis
(PCA) approach. More recently, there are modern statistical methods
that introduce sparsity and high-dimensional settings in these traditional
methods (see Zou et al. (2006) \cite{zou2006sparse}). However, as
we argued in the introduction, these methods are not optimal for basis
asset models due to their difficulties in interpreting the basis assets.
Furthermore, it is the correlation that is important in the determination
of the basis assets not the variance itself. Therefore, methods that
focus on finding the rotation with the largest variance (like PCA)
are not optimal in this setting. Instead, we use correlation as our
metric in the prototype clustering step, which gives a clearer interpretation
and a direct analysis of the candidate basis assets, rather than linear
combinations of the basis assets. For this reason we believe that
prototype clustering is preferred to PCA in this setting.

For future work, modern refinements of model-selection and inference
methods may be used. For high-dimensional models obtaining valid p-values
is difficult. This is in part due to the fact that fitting a high-dimensional
model often requires penalization and complex estimation procedures,
which implies that characterizing the distribution of such estimators
is difficult. For statistical testing in the presence of sparsity
a number of new methods are appearing in the literature. One alternative
method is the post-selection procedure by Tibshirani et al. (2016)
\cite{tibshirani2016exact}. Another approach for constructing frequentistp
p-values and confidence intervals for high-dimensional models uses
the idea of de-biasing which was proposed in a series of articles
\cite{vandegeer2014asymptotically, javanmard2014confidence, zhang2014confidence}.
In the de-biasing method, starting from a regularized estimator one
first constructs a de-biased estimator and then makes inference based
on the asymptotic normality of low-dimensional functionals of the
de-biased estimator. In principle, this approach also provides asymptotically
valid p-values for hypotheses testing each of the parameters in the
model. However, in our numerical explorations of these methods we
found that the confidence intervals are too large for the current
application and no meaningful insights could be obtained. The p-values
for these methods are also not very stable. So we use the OLS after
LASSO instead of the post-selection methods with the theoretical guarantee
in the paper by Zhao et al. (2017) \cite{zhao2017defense}. Some details can be found in \cite{zhu2020high}. Some related work can be found in \cite{zhu2021time, zhu2021clustering, zhu2021news, jarrow2021low, li2021frequentnet, bo2021subspace, xie2018data, huang2021staying, huang2020time, hu2021revealing, huang2019identifying, lu2017detrending, zhang2021form, du2020multiple, prashanth2016cumulative, zhao2018examining, jie2018stochastic, lin2018probabilistically, jie2021bidding, jie2018decision}.

%\clearpage
\section{Risk-Factor Determination}

\label{sec: amf_risk_premium}

We focus on the same three year time period 2014 - 2016 and compute the average annal excess returns on the basis assets to determine which have non-zero risk premium (average excess returns), i.e. which are risk-factors in the traditional sense. In this time period there are 182 basis assets selected, including the Fama-French 5 factors and 177 ETFs. The risk premium of the Fama-French 5 factors are shown in Table \ref{tab: amf_ff5_risk_premium}.

\begin{table}[H]
\centering
\begin{tabular}{||r|r|r|r|r|r||}
  \hhline{|=|=|=|=|=|=|}
Fama-French 5 Factors & Market Return & SMB & HML & RMW & CMA \\
  \hline
Annual Excess Return (\%) & 10.0 & -2.1 & 1.7 & 1.1 & -1.2 \\
   \hhline{|=|=|=|=|=|=|}
\end{tabular}
\caption{Risk Premium of Fama-French 5 factors}
\label{tab: amf_ff5_risk_premium}
\end{table}

Out of the 177 selected ETFs, 136 of them have absolute risk premiums larger than the minimum of that of the FF5 factors (which is the RMW, with absolute risk premium 1.1\%). Therefore, at least $ (136 + 5) / (177 + 5) = 77.47 \% $ basis assets are risk factors. Furthermore, 29 out of the 177 selected ETFs have absolute risk premiums larger than 10.0\% (the absolute risk premium of the market return, which is the biggest absolute risk premium of all FF5 factors). The list of the 29 ETFs are in Table \ref{tab: amf_etf_risk_premium}.

\begin{table}[H]
%\resizebox{6in}{!}{
\centering
\begin{adjustbox}{width={0.95\textwidth}, totalheight={0.95\textheight},keepaspectratio}
\begin{tabular}{||l|l|r||}
  \hhline{|=|=|=|}
 & & Risk Pre-  \hspace{0.05cm}  \\
ETF name & Category & mium (\%) \\
\hhline{|=|=|=|}
ProShares Ultra Semiconductors & Leveraged Equities & 55.1 \\
\hline
ProShares Ultra Bloomberg Natural Gas & Leveraged Commodities & -30.4 \\
\hline
ProShares VIX Short-Term Futures ETF & Volatility & -29.2 \\
\hline
ProShares Ultra Real Estate & Leveraged Real Estate & 26.2 \\
\hline
Global X MSCI Nigeria ETF & Foreign Large Cap Equities & -24.0 \\
\hline
Invesco DB Oil Fund & Oil \& Gas & -21.3 \\
\hline
Global X FTSE Greece 20 ETF & Emerging Markets Equities & -20.8 \\
\hline
Invesco S\&P SmallCap Information Technology ETF & Technology Equities & 19.7 \\
\hline
Direxion Daily Energy Bull 3X Shares & Leveraged Equities & -17.5 \\
\hline
Invesco S\&P SmallCap Consumer Staples ETF & Consumer Staples Equities & 15.8 \\
\hline
VanEck Vectors Rare Earth/Strategic Metals ETF & Materials & -15.5 \\
\hline
Global X Uranium ETF & Global Equities & -15.3 \\
\hline
SPDR S\&P Health Care Equipment ETF & Health \& Biotech Equities & 15.0 \\
\hline
SPDR SSGA US Small Cap Low Volatility Index ETF & Volatility Hedged Equity & 15.0 \\
\hline
Vanguard Utilities ETF & Utilities Equities & 15.0 \\
\hline
VanEck Vectors Egypt Index ETF & Emerging Markets Equities & -15.0 \\
\hline
Global X MSCI Colombia ETF & Latin America Equities & -14.6 \\
\hline
SPDR S\&P Insurance ETF & Financials Equities & 14.2 \\
\hline
iShares U.S. Aerospace \& Defense ETF & Industrials Equities & 13.8 \\
\hline
iShares North American Tech-Software ETF & Technology Equities & 13.4 \\
\hline
iShares North American Tech-Multimedia Networking ETF & Communications Equities & 13.3 \\
\hline
FLAG-Forensic Accounting Long-Short ETF & Long-Short & 12.2 \\
\hline
SPDR SSGA US Large Cap Low Volatility Index ETF & Volatility Hedged Equity & 12.0 \\
\hline
Global X MSCI Portugal ETF & Europe Equities & -11.7 \\
\hline
Vanguard Consumer Staples ETF & Consumer Staples Equities & 10.9 \\
\hline
VanEck Vectors Poland ETF & Europe Equities & -10.8 \\
\hline
VanEck Vectors Morningstar Wide Moat ETF & Large Cap Blend Equities & 10.6 \\
\hline
Invesco Russell Top 200 Equal Weight ETF & Large Cap Growth Equities & 10.2 \\
\hline
First Trust NASDAQ CEA Smartphone Index Fund & Technology Equities & 10.1 \\
\hhline{|=|=|=|}
\end{tabular}
\end{adjustbox}
\caption{List of ETFs with large absolute risk premium.}
\label{tab: amf_etf_risk_premium}
\end{table}

%\clearpage
\section{Illustrations}
\label{sec: amf_illustration}

In this section we illustrate our multi-factor estimation process
and compare the results with the Fama-French 5-factor (FF5) model
for three securities: Adobe, Bank of America, and Apple.

\subsection{Adobe}

This section contains the results for Adobe. Using Equation (\ref{eq: ff5}),
we estimate the Fama-French 5 factor (FF5) model as shown in Table
\ref{tab: adobe_ff5}. For our Adaptive Multi-Factor (AMF) model, the final
results are shown in Table \ref{tab: adobe_amf} with the description of the
ETF basis assets selected by GIBS in Table \ref{tab: adobe_etf_sgn}. The adjusted
$R^{2}$ for FF5 is 0.38, while the adjusted $R^{2}$ for AMF is 0.57.
From the Tables it is clear that different significant basis assets
are selected and the ones selected by GIBS gives much better explanation
and prediction power. Only the market return is significant among
the FF5 factors. Additionally, Adobe's returns are related to the
iShares North American Tech-Software ETF, indicating that Adobe is
sensitive to risks in the technology software sector. For Adobe, both
models do not have a significant intercept, indicating that the securities
are properly priced.

\begin{table}[H]
\centering
%\begin{adjustbox}{width={0.95\textwidth}, totalheight={0.95\textheight},keepaspectratio}
\begin{tabular}{|| *{5}{r|} |}
\hhline{*{5}{|=}|}
& $\beta$ & SE & t value & P-value \\
\hhline{*{5}{|=}|}
(Intercept) & 0.002 & 0.002 & 1.227 & 0.222 \\
\hline
Market Return & 1.036 & 0.124 & 8.382 & 0.000 \\
\hline
SMB & -0.168 & 0.191 & -0.883 & 0.379 \\
\hline
HML & -0.480 & 0.247 & -1.942 & 0.054 \\
\hline
RMW & -0.378 & 0.310 & -1.217 & 0.226 \\
\hline
CMA & -0.234 & 0.424 & -0.551 & 0.583 \\
\hhline{*{5}{|=}|}
\end{tabular}
%\end{adjustbox}
\caption{Adobe with the FF5 model. The column $\beta$ provides the coefficients in the OLS regression of Adobe on FF5 factors. The standard error (SE), t value, and P-value related to each coefficient are also provided.}
\label{tab: adobe_ff5}
\end{table}

\begin{table}[H]
%\resizebox{6in}{!}{ %
\centering
\begin{adjustbox}{width={0.95\textwidth}, totalheight={0.95\textheight},keepaspectratio}
\begin{tabular}{|| *{5}{r|} |}
\hhline{*{5}{|=}|}
& $\beta$ & SE & t value & P-value \\
\hhline{*{5}{|=}|}
(Intercept) & 0.002 & 0.002 & 1.110 & 0.269 \\
\hline
Market Return & -0.518 & 0.194 & -2.662 & 0.009 \\
\hline
iShares North American Tech-Software ETF & 1.377 & 0.150 & 9.162 & 0.000 \\
\hhline{*{5}{|=}|}
\end{tabular}
\end{adjustbox}
\caption{Adobe with the AMF. The column $\beta$ provides the coefficients
in the second-step OLS regression of Adobe on basis assets selected
in the AMF. The standard error (SE), t value, and P-value related
to each coefficient are also provided.}
\label{tab: adobe_amf}
\end{table}

\begin{table}[H]
%\resizebox{6in}{!}{ %
\centering
\begin{adjustbox}{width={0.95\textwidth}, totalheight={0.95\textheight},keepaspectratio}
\begin{tabular}{||l|l|l||}
\hhline{|=|=|=|}
ETF Name  & Category  & Big Class \\
\hhline{|=|=|=|}
iShares North American Tech-Software ETF & Technology Equities & Equity \\
\hhline{|=|=|=|}
\end{tabular}
\end{adjustbox}
\caption{Significant ETF basis assets for Adobe. This table shows the category
and big class of each ETF basis asset selected in the AMF.}
\label{tab: adobe_etf_sgn}
\end{table}

\subsection{Bank of America}

This section contains the results for the Bank of America (BOA). The
results are found in Tables \ref{tab: boa_ff5}, \ref{tab: boa_amf}, and \ref{tab: boa_etf_sgn}.
The adjusted $R^{2}$ for FF5 is 0.72 while the adjusted $R^{2}$
for AMF is 0.82. For BOA, it is related to all of the FF5 factors
except SMB. It is related to VanEck Vectors Investment Grade Floating
Rate ETF, FlexShares Ready Access Variable Income Fund, and the WisdomTree
Barclays Negative Duration U.S. Aggregate Bond Fund, indicating that
BOA's security returns (as a bank) are subject to risks related to
the term structure of interest rates and the credit risk embedded
in corporate bonds. It is also related to the SPDR SSgA Multi-Asset
Real Return ETF which is correlated to the health of the economy.
Finally, the $\alpha$'s in both models are not significantly different
from zero, indicating that no arbitrage opportunities exist for this
security.

\begin{table}[H]
\center %
%\begin{adjustbox}{width={0.95\textwidth}, totalheight={0.95\textheight},keepaspectratio}
\begin{tabular}{|| *{5}{r|} |}
\hhline{*{5}{|=}|}
& $\beta$ & SE & t value & P-value \\
\hhline{*{5}{|=}|}
(Intercept) & 0.000 & 0.002 & 0.274 & 0.784 \\
\hline
Market Return & 1.042 & 0.098 & 10.580 & 0.000 \\
\hline
SMB & 0.236 & 0.152 & 1.552 & 0.123 \\
\hline
HML & 1.981 & 0.197 & 10.046 & 0.000 \\
\hline
RMW & -1.145 & 0.247 & -4.627 & 0.000 \\
\hline
CMA & -2.059 & 0.338 & -6.087 & 0.000 \\
\hhline{*{5}{|=}|}
\end{tabular}
\caption{BOA with the FF5 model. The column $\beta$ provides the coefficients in the OLS regression of BOA on FF5 factors. The standard error (SE), t value, and P-value related to each coefficient are also provided.}
\label{tab: boa_ff5}
\end{table}

\begin{table}[H]
%\resizebox{6in}{!}{ %
\centering
\begin{adjustbox}{width={0.95\textwidth}, totalheight={0.95\textheight},keepaspectratio}
\begin{tabular}{|| *{5}{r|} |}
\hhline{*{5}{|=}|}
& $\beta$ & SE & t value & P-value \\
\hhline{*{5}{|=}|}
(Intercept) & 0.001 & 0.001 & 0.405 & 0.686 \\
\hline
Market Return & 2.314 & 0.316 & 7.335 & 0.000 \\
\hline
HML & 1.078 & 0.146 & 7.390 & 0.000 \\
\hline
RMW & -0.752 & 0.221 & -3.398 & 0.001 \\
\hline
VanEck Vectors Investment Grade Floating Rate ETF & 3.347 & 0.871 & 3.842 & 0.000 \\
\hline
ProShares Short 7-10 Year Treasury & -0.359 & 0.318 & -1.129 & 0.261 \\
\hline
SPDR SSgA Multi-Asset Real Return ETF & -0.712 & 0.218 & -3.259 & 0.001 \\
\hline
FlexShares Ready Access Variable Income Fund & -3.421 & 1.710 & -2.000 & 0.047 \\
\hline
IQ Hedge Market Neutral Tracker ETF & -0.773 & 0.513 & -1.507 & 0.134 \\
\hline
SPDR SSGA US Large Cap Low Volatility Index ETF & -0.405 & 0.230 & -1.758 & 0.081 \\
\hline
AdvisorShares Newfleet Multi-Sector Income ETF & -1.460 & 0.852 & -1.714 & 0.089 \\
\hline
WisdomTree Barclays Negative Duration U.S. Aggregate Bond Fund & 0.451 & 0.220 & 2.047 & 0.043 \\
\hline
Vanguard Consumer Staples ETF & -0.290 & 0.181 & -1.604 & 0.111 \\
\hline
Vanguard Utilities ETF & -0.118 & 0.099 & -1.196 & 0.234 \\
\hline
Invesco CurrencyShares Swiss Franc Trust & -0.152 & 0.079 & -1.925 & 0.056 \\
\hline
iShares Short Treasury Bond ETF & -13.968 & 8.592 & -1.626 & 0.106 \\
\hline
Invesco DB Precious Metals Fund & -0.011 & 0.088 & -0.121 & 0.904 \\
\hline
SPDR Barclays Short Term Municipal Bond & -0.011 & 0.927 & -0.012 & 0.990 \\
\hline
iShares Moderate Allocation ETF & -0.087 & 0.600 & -0.146 & 0.885 \\
\hline
ProShares Ultra Yen & -0.008 & 0.067 & -0.123 & 0.902 \\
\hhline{*{5}{|=}|}
\end{tabular}
\end{adjustbox}
\caption{BOA with the AMF. The column $\beta$ provides the coefficients in
the second-step OLS regression of BOA on basis assets selected in
the AMF. The standard error (SE), t value, and P-value related to
each coefficient are also provided.}
\label{tab: boa_amf}
\end{table}

\begin{table}[H]
%\resizebox{6in}{!}{ %
\centering
\begin{adjustbox}{width={0.95\textwidth}, totalheight={0.95\textheight},keepaspectratio}
\begin{tabular}{||l|l|l||}
\hhline{|=|=|=|}
ETF Name  & Category  & Big Class \tabularnewline
\hhline{|=|=|=|}
VanEck Vectors Investment Grade Floating Rate ETF & Corporate Bonds & Bond/Fixed Income \\
\hline
SPDR SSgA Multi-Asset Real Return ETF & Hedge Fund & Alternative ETFs \\
\hline
FlexShares Ready Access Variable Income Fund & Corporate Bonds & Bond/Fixed Income \\
\hline
WisdomTree Barclays Negative Duration U.S. Aggregate Bond Fund & Total Bond Market & Bond/Fixed Income \\
\hhline{|=|=|=|}
\end{tabular}
\end{adjustbox}
\caption{Significant ETF basis assets for BOA. This table shows the category
and big class of each ETF basis asset selected in the AMF.}
\label{tab: boa_etf_sgn}
\end{table}

\subsection{Apple}

This section gives the results for Apple, in Tables \ref{tab: apple_ff5}, \ref{tab: apple_amf},
pand \ref{tab: apple_etf_sgn}. The adjusted $R^{2}$ for FF5 is 0.52, while the
adjusted $R^{2}$ for our AMF model is 0.64. For Apple, the market
return, RMW, CMA are the significant FF5 factors. It is significantly
related to First Trust NASDAQ CEA Smartphone Index Fund, since Apple
produces smartphones. The remaining ETFs capture the health of the
equities and bond markets, documenting that Apple's risk are highly
correlated with the general economy as well. Finally, the $\alpha$'s
in both models are not significantly different from zero.

\begin{table}[H]
\center %
%\begin{adjustbox}{width={0.95\textwidth}, totalheight={0.95\textheight},keepaspectratio}
\begin{tabular}{|| *{5}{r|} |}
\hhline{*{5}{|=}|}
& $\beta$ & SE & t value & P-value \\
\hhline{*{5}{|=}|}
(Intercept) & 0.000 & 0.002 & 0.208 & 0.836 \\
\hline
Market Return & 1.106 & 0.119 & 9.313 & 0.000 \\
\hline
SMB & -0.334 & 0.183 & -1.825 & 0.070 \\
\hline
HML & 0.359 & 0.238 & 1.511 & 0.133 \\
\hline
RMW & 1.242 & 0.298 & 4.163 & 0.000 \\
\hline
CMA & -2.371 & 0.408 & -5.811 & 0.000 \\
\hhline{*{5}{|=}|}
\end{tabular}\caption{Apple with the FF5 model. The column $\beta$ provides the coefficients in the OLS regression of Apple on FF5 factors. The standard error
(SE), t value, and P-value related to each coefficient are also provided.}
\label{tab: apple_ff5}
\end{table}

\begin{table}[H]
%\resizebox{6in}{!}{ %
\centering
\begin{adjustbox}{width={0.95\textwidth}, totalheight={0.95\textheight},keepaspectratio}
\begin{tabular}{|| *{5}{r|} |}
\hhline{*{5}{|=}|}
& $\beta$ & SE & t value & P-value \\
\hhline{*{5}{|=}|}
(Intercept) & -0.002 & 0.002 & -0.972 & 0.333 \\
   \hline
Market Return & 2.062 & 0.380 & 5.430 & 0.000 \\
   \hline
SMB & -0.092 & 0.184 & -0.502 & 0.617 \\
   \hline
RMW & 1.135 & 0.259 & 4.389 & 0.000 \\
   \hline
CMA & -1.780 & 0.295 & -6.043 & 0.000 \\
   \hline
First Trust NASDAQ CEA Smartphone Index Fund & 0.327 & 0.157 & 2.079 & 0.039 \\
   \hline
iShares Floating Rate Bond ETF & 6.831 & 2.412 & 2.832 & 0.005 \\
   \hline
AGFiQ US Market Neutral Momentum Fund & -0.311 & 0.089 & -3.485 & 0.001 \\
   \hline
Invesco Dynamic Media ETF & -0.428 & 0.162 & -2.636 & 0.009 \\
   \hline
Invesco Water Resources ETF & -0.228 & 0.195 & -1.167 & 0.245 \\
   \hline
VanEck Vectors Environmental Services ETF & -0.389 & 0.206 & -1.890 & 0.061 \\
   \hline
ProShares Ultra Semiconductors & 0.014 & 0.060 & 0.233 & 0.816 \\
   \hline
iShares MSCI Israel ETF & -0.615 & 0.128 & -4.818 & 0.000 \\
\hhline{*{5}{|=}|}
\end{tabular}
\end{adjustbox}
\caption{Apple with AMF. The column $\beta$ provides the coefficients in the
second-step OLS regression of Apple on basis assets selected in the
AMF. The standard error (SE), t value, and P-value related to each
coefficient are also provided.}
\label{tab: apple_amf}
\end{table}

\begin{table}[H]
%\resizebox{6in}{!}{ %
\centering
\begin{adjustbox}{width={0.95\textwidth}, totalheight={0.95\textheight},keepaspectratio}
\begin{tabular}{||l|l|l||}
\hhline{|=|=|=|}
ETF Name  & Category  & Big Class \tabularnewline
\hhline{|=|=|=|}
First Trust NASDAQ CEA Smartphone Index Fund & Technology Equities & Equity \\
\hline
iShares Floating Rate Bond ETF & Corporate Bonds & Bond/Fixed Income \\
\hline
AGFiQ US Market Neutral Momentum Fund & Long-Short & Alternative ETFs \\
\hline
Invesco Dynamic Media ETF & All Cap Equities & Equity \\
\hline
iShares MSCI Israel ETF & Large Cap Blend Equities & Equity \\
\hhline{|=|=|=|}
\end{tabular}
\end{adjustbox}
\caption{Significant ETF pbasis assets for Apple. This table shows the category
and big class of each ETF basis asset selected in the AMF.}
\label{tab: apple_etf_sgn}
\end{table}

%\clearpage
\section{Conclusion}

\label{sec: amf_conclusion}

Using a collection of basis assets, the purpose of this paper is to test the new Adaptive Multi-Factor (AMF) model implied by the generalized arbitrage pricing theory (APT) recently developed by Jarrow and Protter (2016) \cite{jarrow2016positive} and Jarrow (2016) \cite{jarrow2016bubbles}. The idea is to obtain the collection of all possible basis assets and to provide a simultaneous test, security by security, of which basis assets are significant. Since the collection of basis assets selected for investigation is large and highly correlated, we propose a new high-dimensional algorithm -- the Groupwise Interpretable Basis Selection (GIBS) algorithm to do the analysis. For comparison with the existing literature, we compare the performance of AMF (using the GIBS algorithm) with the Fama-French 5-factor model and all other alternative methods. Both the Fama-French 5-factor and the AMF model are consistent with the behavior of ``large-time scale'' security
returns. In a goodness-of-fit test comparing the AMF with Fama-French 5-factor model, the AMF model has a substantially larger In-Sample adjusted $R^2$ and Out-of-Sample $R^2$. This documents the AMF model's superior performance in characterizing security returns. Last, as a robustness test, for those securities whose intercepts were non-zero (although insignificant), we tested the AMF model to see if positive alpha trading strategies generate arbitrage opportunities. They do not, thereby confirming that the multi-factor model provides a reasonable characterization of security returns.

\clearpage
\chapter{Time-Invariance Coefficients Tests with the Adaptive Multi-Factor Model}

\section{Introduction}

The purpose of this paper is to test the multi-factor beta model implied by the generalized arbitrage pricing theory (APT) of Jarrow and Protter (2016) \cite{jarrow2016positive}, and recently tested by Zhu et al. (2018) \cite{zhu2020high}, without imposing the exogenous assumption of constant betas. The assumption of time-invariant beta coefficients in multi-factor models is an often employed one in the empirical literature, see Jagannathan et al. (2010) \cite{jagannathan2010cross} and Harvey et al. (2016) \cite{harvey2016and} for reviews. In the paper by Zhu et al. (2018) \cite{zhu2020high}, they address the restrictive nature of this assumption by estimating the model over a short time horizon - three years. For many applications, restricting the time horizon to such a short time period is problematic. Alternative approaches for fitting time-varying betas over longer time horizons have been proposed and estimated, such as the conditional factor models. See paper Adrian et al. (2015) \cite{adrian2015regression}, Cooper \& Maio (2019) \cite{cooper2019new}, and Avramov \& Chordia (2006) \cite{avramov2006asset}. In contrast to these approaches, we show herein that this constant beta assumption can be avoided if we use the Adaptive Multi-Factor (AMF) model with the Groupwise Interpretable Basis Selection (GIBS) algorithm in a dynamic manner using a comparably short rolling window. This insight is a direct corollary of the generalized APT\footnote{It is shown in Jarrow and Protter (2016) \cite{jarrow2016positive} that both Merton's (1973) \cite{merton1973intertemporal} intertemporal capital asset pricing model and Ross's APT (1976) \cite{ross1976arbitrage} is a special case of the generalized APT.}.

Using the Adaptive Multi-Factor (AMF) model with the Groupwise Interpretable Basis Selection (GIBS) algorithm proposed in Zhu et al. (2018) \cite{zhu2020high}, but fitting price differences instead of returns, we estimated a multi-factor model to stock prices over the time period 2007 - 2018. Employing the collection of Exchange Traded Funds (ETFs) as potential factors, we use the high dimensional GIBS algorithm to select the factors for each company. No-arbitrage tests confirm the validity of the generalized APT. As a robustness check, we also show that the estimated model performs better than the traditional Fama-French 5-factor model. After this validation, we perform the time-invariance tests for the $\beta$ coefficients for various time periods. At last, we found that for time periods no more than 5 years, the beta coefficients are time-invariant for the AMF model. Using a dynamic AMF model with a rolling window with a length of no more than 5 years gives a good fit. The arbitrage test confirms the validity of generalized APT's theory.

An outline for this paper is as follows. Section \ref{sec: ti_generalized_apt} presents the Generalized APT theory. Section \ref{sec: ti_estimation} provides the estimation methodology. Section \ref{sec: ti_test_methodology} provides the testing methodology and results. Section \ref{sec: ti_conclude} concludes.

\section{The Generalized APT}
\label{sec: ti_generalized_apt}

Jarrow and Protter (2016) derive a testable multi-factor model over a finite horizon {[}0, T{]} in the context of a continuous time, continuous trading market assuming only frictionless and competitive markets
that satisfy no-arbitrage and no dominance, i. e. the existence of an equivalent martingale measure. As in the traditional asset pricing models, adding a non-zero alpha to this relation (Jensen's alpha) implies a violation of the no-arbitrage condition.

The generalized APT uses linear algebra to prove the existence of an algebraic basis in the security's payoff space at some future time T. Since this is is a continuous time and trading economy, this payoff space is infinite-dimensional. The algebraic basis at time T constitutes the collection of tradeable basis assets and it provides the multi-factor model for a security's price at time T.\footnote{An algebraic basis means that any risky asset's return can be written as a linear combination of a \emph{finite} number of the basis asset returns, and different risky assets may have a \emph{different finite} combination of basis asset explaining their returns. Since the space of random variables generated by the admissible trading strategies is infinite-dimensional, this algebraic basis representation of the relevant risks is parsimonious and sparse.} The coefficients of the time T multi-factor model are constants (non-random). No-arbitrage, the existence of the martingale measure, implies that the arbitrage-free prices of the risky assets at all earlier dates $t\in[0, T)$ will satisfy the same factor model and with the same constant coefficients. Transforming prices into returns (dividing prices at time $t+1$ by time $t$ prices to get the return at $t+1$),
makes the resulting coefficients in the multi-factor model stochastic when viewed at time $0$. However, this is not the case for the multi-factor model specified in a security's price (or price differences). The multi-factor model's beta coefficients in the security's price process are time-invariant.

The generalized APT is important for practice because it provides
an exact identification of the relevant set of basis assets characterizing
a security's \emph{realized} (emphasis added) returns. This enables
a more accurate risk-return decomposition facilitating its use in
trading (identifying mispriced assets) and for risk management. Taking
expectations of this realized return relation with respect to the
martingale measure determines which basis assets are \emph{risk-factors},
i.e. which basis assets have non-zero expected excess returns (risk
premiums) and represent a systematic risk. Since the traditional models
are nested within the generalized APT, an empirical test of the generalized
APT provides an alternative method for testing the traditional models
as well.

Let $B(t)$ denote the time $t$value of a money market account (mma) with initial value of 1 dollar at time 0, i.e.
\begin{equation}
B(t)=1\cdot\prod_{k=0}^{t-1}\Big[1+r_{0}(k)\Big]
\label{eq: ti_mma}
\end{equation}
where $r_{0}(k)$ is the default-free spot rate (the risk-free rate)
from time $k$ to time $k+1$.

Let $A_{i}(t)$ denote the market price of the $i^{th}$ stock at
time $t$ for $i=1,\ldots,N_{c}$. To include cumulative cash flows
and stock splits into the valuation methodology, we need to compute
the adjusted price\footnote{This is sometimes called the gains process from investing in the security
over $[0,T]$.} $Y_{i}$, which is reconstructed by the using the security's returns,
after being adjusted for dividends and stock splits\footnote{These are the returns provided in the available data bases. }:
\begin{equation}
Y_{i}(t)=A_{i}(0)\prod_{k=0}^{t-1}\Big[1+R_{i}(k)\Big]
\label{eq: ti_prc}
\end{equation}
where $A_{i}(0)$ is the initial price and $R_{i}(k)$ its return
over $[k,k+1]$.

Let $V_{j}(t)$ be the adjusted price of the $j^{th}$ basis asset
at time $t$ for $j=1,\ldots,N_{f}$. Here, the sources of the basis asset
in our model are the Fama-French 5 factors and the Exchange-Traded
Funds (ETF). We include within this set of risk-factors the MMA. For
notational simplicity, we let $V_{1}(t)=B(t)$.

Given this notation, the generalized APT implies the following multi-factor
model for the $i^{th}$ security's time $t$ price:
\begin{equation}
Y_{i}(t)=\sum_{j=1}^{N_{f}}\beta_{i,j}V_{j}(t)+ \epsilon_{i}(t)
\label{eq: ti_theory}
\end{equation}
where $\beta_{i,j}$ for all $i,j$ are constants and $\tilde{\epsilon}_{i}(t)$
is an i.i.d. error term with zero mean and constant variance.

The goal of our estimation is two-fold. First, we want to test to
see whether expression (\ref{eq: ti_theory}) provides a good fit to
historical stock price data. Second, we want to investigate whether
the multi-factors coefficients $\beta_{i,j}$'s are time-invariant.
Given that prices are known to be autocorrelated, instead of estimating
expression (\ref{eq: ti_theory}), we fit the first order price differences
of this expression:
\begin{equation}
\Delta Y_{i}(t)=Y_{i}(t+1)-Y_{i}(t)
= \sum_{j=1}^{N_{f}}\beta_{i,j}\Delta V_{j}(t) + \Delta\epsilon_{i}(t)
\label{eq: ti_theory_diff}
\end{equation}
where $\Delta\epsilon_{i}(t)=\epsilon_{i}(t+1)-\epsilon_{i}(t)$.

For a given time period $(t,T)$, letting $n=T-t+1$, $p=N_{f}$ we
can rewrite the expression (\ref{eq: ti_theory}) using time series vectors
asp
\begin{equation}
\bm{Y}_{i}=\bm{V}\bm{\beta}_{i}+\bm{\epsilon}_{i}
\label{eq: ti_theory_vec}
\end{equation}
where $\bm{V}=(\bm{V}_{1},\bm{V}_{2},\ldots,\bm{V}_{p})$ and
\[
\bm{Y}_{i}=\begin{pmatrix}Y_{i}(t)\\
Y_{i}(t+1)\\
\vdots\\
Y_{i}(T)
\end{pmatrix}_{n\times1},\;\bm{V}_{j}=\begin{pmatrix}V_{j,t}\\
V_{j,t+1}\\
\vdots\\
V_{j,T}
\end{pmatrix}_{n\times1},\bm{\beta}_{i}=\begin{pmatrix}\beta_{i,1}\\
\beta_{i,2}\\
\vdots\\
\beta_{i,p}
\end{pmatrix}_{p\times1},\;\bm{\epsilon}_{i}=\begin{pmatrix}\epsilon_{i}(t)\\
\epsilon_{i}(t+1)\\
\vdots\\
\epsilon_{i}(T)
\end{pmatrix}_{n\times1}
\]
Taking the first-order difference of each vector, the equation (\ref{eq: ti_theory_diff})
can be rewritten as
\begin{equation}
\Delta\bm{Y}_{i}=\Delta\bm{V}\bm{\beta}_{i}+\Delta\bm{\epsilon}_{i}
\label{eq: ti_vec_diff}
\end{equation}
where $\Delta\bm{V}=(\Delta\bm{V}_{1},\Delta\bm{V}_{2},\ldots,\Delta\bm{V}_{p})$
and
\[
\Delta\bm{Y}_{i}=\begin{pmatrix}\Delta Y_{i}(t)\\
\Delta Y_{i}(t+1)\\
\vdots\\
\Delta Y_{i}(T-1)
\end{pmatrix}_{(n-1)\times1},\;\Delta\bm{V}_{j}=\begin{pmatrix}\Delta V_{j,t}\\
\Delta V_{j,t+1}\\
\vdots\\
\Delta V_{j,T-1}
\end{pmatrix}_{(n-1)\times1}
\]

\[
\bm{\beta}_{i}=\begin{pmatrix}\beta_{i,1}\\
\beta_{i,2}\\
\vdots\\
\beta_{i,p}
\end{pmatrix}_{p\times1},\;\Delta\bm{\epsilon}_{i}=\begin{pmatrix}\Delta\epsilon_{i}(t)\\
\Delta\epsilon_{i}(t+1)\\
\vdots\\
\Delta\epsilon_{i}(T-1)
\end{pmatrix}_{(n-1)\times1}
\]

Based on the generalized APT theory, we will estimate and test on the time-invariance of the $\beta$'s using the Adaptive Multi-Factor (AMF) model with Groupwise Interpretable Basis Selection (GIBS) algorithm \cite{zhu2020high} in the following sections.

\section{The Estimation Methodology}
\label{sec: ti_estimation}

This section gives the estimation and testing methodology. We first specify the data we use to estimate and test. Then we pick many time periods with various lengths with a good amount of Exchange-Traded Funds (ETF). For each time period, we use the Groupwise Interpretable Basis Selection (GIBS) algorithm to estimate the Adaptive Multi-Factor (AMF) model. Within each time period, we propose several ways of time-variance testing to test if the $\beta$'s are constant during the time period. We will also compare the results for AMF with the benchmark Fama-French 5-Factor (FF5) model.

\subsection{Data and Time Periods}
\label{sec: ti_data}

We use the Exchange-Traded Funds (ETF) to select the risk-factors to be included in the multi-factor model. For comparison to the literature and as a robustness test, we include the Fama-French 5 factors into this set\footnote{For the Fama-French 5 (FF5) factors, we add back the risk-free rate if it is already subtracted in the raw data as in the market return in Fama-French's website.}.

The data used in this paper consists of all the stocks and all the ETFs available in the CRSP (The Center for Research in Security Prices) database over the years 2007 - 2018. We start from 2007 because the number of ETFs with a good amount of market capital (in another word, tradable) available before 2007 is very limited, which make it hard to fit the AMF model.

To avoid the influence of market microstructure effects, we use a weekly observation interval. A security is included in our sample only if it has prices and returns available for more than 2/3 of all the trading weeks. This is consistent with the tradition of the empirical asset pricing literature. Our final sample consists of over 4000 companies listed on the NYSE. To avoid survivorship bias, we include the delisted returns (see Shumway (1997) \cite{shumway1997delisting} for more explanation). For each regression time period, we form the adjusted price using Equation (\ref{eq: ti_prc}).

We repeat our analysis on all sub-periods of 2007 - 2018 (both ends inclusive, the same below) starting on Jan. 1st, ending on Dec. 31th, and with a length at least 3 years. This means we have 55 sub-periods in total, such as 2007 - 2009, 2008 - 2010, ..., 2008 - 2018, 2007 - 2018. We only use the periods with length at least 3 years to ensure that there are sufficient observations to fit the models. As stated above, we use the weekly observations to avoid the market microstructure effects, so 3 years means $ n \geq 156 $ observations in a regression.

The number of ETFs in our sample is quite large, slightly over 2000 at the end of 2018. So the number of basis assets $p_1 > 2000$. The set of potential factors is denoted $V_{j}(1\leq j\leq p_{1})$. Since $n = 156 < 2000 < p_1$ for 3-year time periods, we are in high-dimensional situation and high-dimensional statistical methods, including the GIBS algorithm, need to be used.

\subsection{High-Dimensional Statistics and the GIBS Algorithm}

This section provides a brief review of the high-dimensional statistical methodologies used in this paper, including the GIBS algorithm.

We start from some notation definitions. Let $\norm{\bm{v}}_{d}$ denote the standard $l_{d}$ norm of a vector, then
\begin{equation}
\norm{\bm{v}}_{d}=\left(\sum_{i}|\bm{v}_{i}|^{d}\right)^{1/d}\qquad\text{where}\;\;0<d<\infty.
\label{eq: norm}
\end{equation}
specifically, $\norm{\bm{v}}_{1}=\sum_{i}|\bm{v}_{i}|$. Suppose $\bm{\beta}$
is a vector with dimension $p\times1$. Given a set $S\subseteq\{1,2,...,p\}$,
we let $\beta_{S}$ denote the $p\times1$ vector with i-th element
\begin{equation}
(\bm{\beta}_{S})_{i}=\begin{cases}
\beta_{i}, & \text{if}\;\;i\in S\\
0, & otherwise.
\end{cases}
\label{eq: set_index}
\end{equation}
Here the index set $S$ is called the support of $\bm{\beta}$, in
other words, $supp(\bm{\beta})=\{i:\bm{\beta}_{i}\neq0\}$. Similarly,
if $\bm{X}$ is a matrix instead of a vector, then $\bm{X}_{S}$ are
the columns of $\bm{X}$ indexed by $S$. Denote $\vone_{n}$ as a
$n\times1$ vector with all elements being 1, $\bm{J_{n}}=\vone_{n}\vone_{n}'$,
and $\bm{\bar{J}_{n}}=\frac{1}{n}\bm{J_{n}}$. $\bm{I_{n}}$ denotes
the identity matrix with diagonal 1 and 0 elsewhere. The subscript
$n$ is always omitted when the dimension $n$ is clear from the context.
The notation $\#S$ means the number of elements in the set $S$.

For any matrix $M = \{m_{i, j}\}_{i, j=1}^n$ define the following terms
\begin{align}
\text{k-th skew-diagonal} & = \{ m_{i, j} | i = j + k \}
\label{eq: skew_diag} \\
\text{k-th skew-anti-diagonal} & = \{ m_{i, j} | i = n + k + 1 - j \}
\label{eq: skew_anti_diag}
\end{align}
where $ - n + 1 \leq k \leq n - 1 $.

Because of this high-dimension problem and the high-correlation among the basis assets, traditional methods fail to give an interpretable and systematic way to fit the Adaptive Multi-Factor (AMF) model. Therefore, we employ the Groupwise Interpretable Basis Selection (GIBS) algorithm proposed in the paper Zhu et al. (2018) \cite{zhu2020high} to select the basis assets set for each stock.

We give a brief review of the GIBS algorithm in this section. In the GIBS algorithm, a procedure using the Minimax-Linkage Prototype Clustering is employed to obtain low-correlated ETFs (denoted as $U$ in the paper). The high-dimensional statistical methods (the Minimax-Linkage Prototype Clustering and LASSO) used in the GIBS algorithm can be found in Section \ref{sec: amf_high_dim_method}. The sketch of the GIBS algorithm is shown in Table \ref{tab: ti_gibs_algo}. The details of the GIBS algorithm can be found in the paper Zhu et al. (2018) \cite{zhu2020high}.

For notational simplicity, denote $V_1(t) = B(t)$ as the money market account, and $V_2(t)$ for the market index. It can be checked that most of the ETFs $\Delta\bm{V}_{i}$ are correlated with $\Delta\bm{V}_{2}$, the market portfolio. And we note that this pattern is not true for the other 4 Fama-French factors. Therefore, we first orthogonalize every other basis asset (excluding $\Delta\bm{V}_{1}$ and $\Delta\bm{V}_{2}$) to $\Delta\bm{V}_{2}$. By orthogonalizing with respect to the market return, we avoid choosing redundant risk-factors similar to it and meanwhile, increase the accuracy of fitting. Note that for Ordinary Least-Square (OLS) regression, projection does not affect the estimation since it only affects the coefficients, not the estimation $\Delta\bm{\hat{Y}}$. However, in LASSO, projection does affect the set of selected risk-factors because it changes the magnitude of the coefficients before shrinking.
Thus, we compute
\begin{equation}
\widetilde{\Delta\bm{V}}_{i}
=(\bm{I} - P_{\Delta\bm{V}_{2}})\Delta\bm{V}_{i}
=(\bm{I} - \Delta\bm{V}_{2}(\Delta\bm{V}'_{2}\Delta\bm{V}_{2})^{-1}\Delta\bm{V}_{2}')\Delta\bm{V}_{i}
\qquad\text{where}\quad3\leq i\leq p_{2}
\label{eq: ti_proj1}
\end{equation}
where $P$ denotes the projection operator. Let
\begin{equation}
\widetilde{\Delta\bm{V}}
=(\bm{V}_{1},\bm{V}_{2}, \widetilde{\Delta\bm{V}_{3}},...,\widetilde{\Delta\bm{V}_{p_{2}}})
\label{eq: ti_proj2}
\end{equation}
Note that this is equivalent to the residuals after regressing other risk factors on $\Delta\bm{V}_{2}$.

The transformed ETF basis assets $\widetilde{\Delta\bm{V}}$ still contain highly correlated members. We first divide these basis assets into categories $C_{1}, C_{2}, ..., C_{k}$ based on a financial characterization. Note that $C \equiv \cup_{i=1}^{k} C_{i} = \{1,2,...,p_{1}\}$. The list
of categories with more descriptions can be found in Appendix \ref{ap: etf_classes}. The categories are (1) bond/fixed income, (2) commodity, (3) currency, (4) diversified portfolio, (5) equity, (6) alternative ETFs, (7) inverse, (8) leveraged, (9) real estate, and (10) volatility.

Next, from each category, we need to choose a set of representatives. These representatives should span the categories they are from, but also have a low correlation with each other. This can be done by using the prototype-clustering method with a distance measure defined in Section \ref{sec: amf_high_dim_method}, which yields the ``prototypes'' (representatives) within each cluster (intuitively, the prototype is at the center of each cluster) with low-correlations.

Within each category, we use the prototype clustering methods previously discussed to find the set of representatives. The number of representatives in each category can be chosen according to a correlation threshold. This gives the sets $D_{1}, D_{2}, ..., D_{k}$ with $D_{i}\subset C_{i}$ for $1\leq i\leq k$. Denote $D\equiv\cup_{i=1}^{k} D_{i}$. Although this reduction procedure guarantees low-correlation between the elements in each $D_{i}$, it does not guarantee low-correlation across the elements in the union $D$. So, an additional step is needed, which is prototype clustering on $D$ is used to find a low-correlated representatives set $U$. Note that $U\subseteq D$. Denote $p_{2}\equiv\#U$.

Recall from the notation definition that $\widetilde{\Delta\bm{V}_U}$ means the columns of the matrix $\widetilde{\Delta\bm{V}}$ indexed by the set $U$. Since basis assets in $\widetilde{\Delta\bm{V}_U}$ are not highly correlated, a LASSO regression can be applied. Therefore, we have that
\begin{equation}
\bm{\widetilde{\beta}}_{i}
=\underset{\bm{\beta}_{i}\in\mathbb{R}^{p},(\bm{\beta}_{i})_{j}=0(\forall j\in U^{c})}{\arg\min}
\left\{
\frac{1}{2n}\norm{\Delta\bm{Y}_{i}
- \widetilde{\Delta\bm{V}_U} \bm{\beta}_{i}}_{2}^{2}
+\lambda\norm{\bm{\beta}_{i}}_{1}
\right\}
\end{equation}
where $U^{c}$ denotes the complement of $U$. However, here we use
a different $\lambda$ as compared to the traditional LASSO. Normally
the $\lambda$ of LASSO is selected by cross-validation. However this
will overfit the data as discussed in the paper Zhu et al. (2018)
\cite{zhu2020high}. So here we use a modified version of the $\lambda$
selection rule and set
\begin{equation}
\lambda=\max\{\lambda_{1se},min\{\lambda:\#supp(\bm{\widetilde{\beta}}_{i})\leq20\}\}
\end{equation}
where $\lambda_{1se}$ is the $\lambda$ selected by the ``1se rule''.
The ``1se rule'' gives the most regularized model such that error
is within one standard error of the minimum error achieved by the
cross-validation (see \cite{friedman2010regularization,simon2011Regularization,tibshirani2012strong}).
Therefore we can derive the the set of basis assets selected as
\begin{equation}
S_i \equiv supp(\bm{\widetilde{\beta}}_{i})
\label{eq: ti_factor_select}
\end{equation}
In this way, we construct the set of basis assets $S_i$.

Next, we fit an Ordinary Least-Square (OLS) regression on the selected basis assets, to estimate $\bm{\hat{\beta}}_{i}$,
the OLS estimator from
\begin{equation}
\Delta\bm{Y}_{i}
=\Delta\bm{V}_{S_i}(\bm{\beta}_{i})_{S_i} + \Delta\bm{\epsilon}_{i}.
\label{eq: ti_ols_subset}
\end{equation}
Since this is an OLS regression, we use the original basis assets $\Delta\bm{V}_{S_i}$ rather than the orthogonalized basis assets $\widetilde{\Delta\bm{V}}_{S_i}$. Note that $supp(\bm{\hat\beta}_{i})\subset S_i$. Since we are in the OLS regime, significance tests can be performed on $\bm{\beta}_{i}$. This yields the significant set of coefficients
\begin{equation}
S_{i}^* \equiv\{j:P_{H_{0}}(|\bm{\beta}_{i,j}|\geq|\bm{\hat\beta}_{i,j}|)<0.05\}\quad\text{where}\quad H_{0}:\text{True value}\,\,\bm{\beta}_{i,j}=0.
\end{equation}
Note that the significant basis asset set is a subset of the selected
basis asset set. In another words,
\begin{equation}
S_{i}^{*}\subseteq supp(\bm{\hat{\beta}}_{i})\subseteq S_{i}\subseteq\{1,2,...,p\}.
\end{equation}

A sketch of the GIBS algorithm is shown in Table \ref{tab: ti_gibs_algo}. Recall from the notation definition that for an index set $S\subseteq\{1,2,...,p\}$, $ \Delta\bm{V}_{S_i} $ means the columns of the matrix $ \Delta\bm{V} $ indexed by the set $ S_i $.

\begin{table}[H]
\centering
  \begin{tabular}{|| l ||}
  \hhline{|=|}
  \\[-1.6ex]
  \textbf{The Groupwise Interpretable Basis Selection (GIBS) algorithm }  \\[1ex]
  \hhline{|=|}
  \\[-1.6ex]
  Inputs: Stocks to fit $ \Delta\bm{Y}_{i} $ and basis assets $ \Delta\bm{V} $.
  \\[0.6ex]
  \hline
  \\[-1.6ex]
  1. Derive $ \widetilde{\Delta\bm{V}} $ using $ \Delta\bm{V} $ and the Equation (\ref{eq: ti_proj1}, \ref{eq: ti_proj2}).
  \\[0.6ex]
  2. Divide the transformed basis assets $ \widetilde{\Delta\bm{V}} $ into k groups $C_{1}, C_{2},\cdots C_{k}$ using a
  \\[0.6ex]
  \hspace{0.4cm} financial interpretation.
  \\[0.6ex]
  3. Within each group, use prototype clustering to find prototypes $ D_i \subset C_i $.
  \\[0.6ex]
  4. Let $ D = \cup_{i = 1}^k D_i $, use prototype clustering in D to find prototypes $ U \subset D $.
  \\[0.6ex]
  5. For each stock $ \Delta\bm{Y}_{i} $, use a modified version of LASSO to reduce $\bm{\widetilde{X}}_{U}$
  \\[0.6ex]
  \hspace{0.4cm} to the selected basis assets $\widetilde{\Delta\bm{V}_{S_i}}$.
  \\[0.6ex]
  6. For each stock $ \Delta\bm{Y}_{i} $, fit linear regression on $ \Delta\bm{V}_{S_i} $.
  \\[0.6ex]
  \hline
  \\[-1.6ex]
  Outputs: Selected factors $ S_i $, significant factors $ S_i^*$, and coefficients in step 6.
  \\[0.6ex]
  \hhline{|=|}
  \end{tabular}
  \caption{The sketch of Groupwise Interpretable Basis Selection (GIBS) algorithm}
  \label{tab: ti_gibs_algo}
\end{table}

\section{Testing Methodologies and Results}
\label{sec: ti_test_methodology}

This section gives the testing methodologies and results. We first do an intercept (arbitrage) test, which validates the AMF model we use. Then we use the indicator variable to test the time-invariance of the $\beta$'s in a linear setting. After that, we do a residual test to check whether including more basis assets can give a better fit. At last, we compare the fitting with a Generalized Additive Model (GAM) to test the time-invariance of the $\beta$'s in a non-linear setting. For each test we repeat it on all the time periods discussed in Section \ref{sec: ti_data} and report the results. Some details can be found in \cite{zhu2021time}.

\subsection{The Intercept Test}

We test the validity of the generalized APT by adding an intercept to model (Jensen's alpha) and see if the intercept is non-zero. A non-zero intercept implies that the securities are mispriced (i.e. the rejection of the existence of an equivalent martingale measure). To test the intercept of our model, we add an intercept term $\alpha_{i}$ to equation (\ref{eq: ti_ols_subset}) and test the null hypothesis
\begin{equation}
H_{0}: \alpha_{i} = 0 \quad vs.\quad H_{a}: \alpha_{i} \neq 0.
\end{equation}
Since using price differences removes the intercept, the intercept test has to done using prices
\begin{equation}
\bm{Y}_{i}
= \alpha_{i}\vone + \bm{V}_{S_i}(\bm{\beta}_{i})_{S_i} + \bm{\epsilon}_{i}
\label{eq: ti_int}
\end{equation}
where $\vone$ is an $n\times1$ vector with all elements 1. Where the $S_i$ are the basis assets selected by the GIBS algorithm in the AMF model, defined in Equation (\ref{eq: ti_factor_select}). For the FF5, the $S_i$ are the Fama-French 5-factors and the risk free rate.

Our initial idea to test this hypothesis was to fit the Ordinary Least-Square (OLS) regression regression on the selected risk-factors for each company and then report the p-values for the significance of $\alpha_{i}$. However, we observed that the mma's value $\bm{V}_{1}$ is highly correlated to the constant vector $\vone$ because the risk-free rate is close to (or equal to) 0 for a long time. Therefore, including both $\vone$ and $\bm{V}_{1}$ in the regression leads to the inverse of a nearly-singular matrix, which gives unreliable results. In this case, since the correlation is so large, even projecting out $\vone$ from $\bm{V}_{1}$ does not solve this problem. So we used a two-step procedure instead. First, we estimate the OLS coefficient $(\bm{\hat\beta}_i)_{S_i}$ from the the non-intercept model
\begin{equation}
\bm{Y}_{i}
= \bm{V}_{S_i}(\bm{\beta}_{i})_{S_i} + \bm{\epsilon}_{i}
\end{equation}
and calculate the estimation of residuals
\begin{equation}
\bm{\hat{\epsilon}}_i = \bm{Y}_{i}-\bm{\hat{Y}}_{i}
\quad \text{where} \quad \bm{\hat{Y}}_{i} =\bm{V}_{S_i}(\bm{\hat\beta}_{i})_{S_i}
\end{equation}
Then, we fit an intercept-only regression on the residuals
\begin{equation}
\bm{\hat{\epsilon}}_i = \alpha_{i}\vone+\bm{\delta}_{i}
\end{equation}
and report the p-value for the significance of the intercept. Using this technique, we avoided the collinearity issue in a test of the intercept. The results show that for all time periods, for all stocks, we can not reject the null hypothesis in either AMF or FF5. In other words, there is no significant non-zero intercept for either AMF or FF5 for any time period and for any stock. This evidence provides a validation of the generalized APT and the use of AMF and FF5 model.

\subsection{Time-invariance Test in Linear Setting}
\label{sec: ti_test_linear}

For each time period, this section tests for the time-invariance of the multi-factors beta coefficients in a linear setting. For these tests we use the first-order differences of the prices as described in equations (\ref{eq: ti_vec_diff}) and (\ref{eq: ti_ols_subset}). We use price differences to avoid autocorrelation and any non-stationarities in the price process. Here we only focus on the selected factors. In other words, we only test the time-invariance of $\beta_{i,j}$ where $j\in S_{i}$ and $S_{i}$ is the one defined in equation (\ref{eq: ti_factor_select}). Our null hypothesis for each stock $i$ is that ``$H_{0}$: $\beta_{i,j}$ are time-invariant over the 4 years.''

Denote $\bm{h}=(h_{1},h_{2},...,h_{n})'$ where $h_{i}=0$ for all rows related to first half of the time period and $h_{i}=1$ for all rows related to the second half of the time period. Testing for the significance of interaction of each basis asset with $\bm{h}$ is a way to test whether the coefficients are the same for the first and last two years. To be more specific, consider the regression model:
\begin{equation}
\Delta\bm{Y}_{i}
= \Delta\bm{V}_{S_i} (\bm{\beta}_{i})_{S_i}
+ \left[ \Delta\bm{V}_{S_i} \odot (\bm{h} \vone_n') \right] \bm{\theta}_{i}
+ \Delta\bm{\epsilon}_{i}
\label{eq: ti_test_linear}
\end{equation}
Note that the sign ``$\odot$'' means the element-wise multiplication for two vectors, and $\vone_{n}$ as a $n\times1$ vector with all elements being 1. Here $ \theta_{i,j} = 0 $ indicates that $\beta_{i,j}$ is time-invariant during the time period. Our null hypothesis becomes
\begin{equation}
H_{0}:\theta_{i,j}=0\;\;\text{for}\;\;\forall j\quad vs.\quad H_{1}:\exists j,\;\;\text{such that}\;\;\theta_{i,j}\neq0.
\end{equation}

\begin{figure}[H]
\centering
\includegraphics[width=0.73\textwidth]{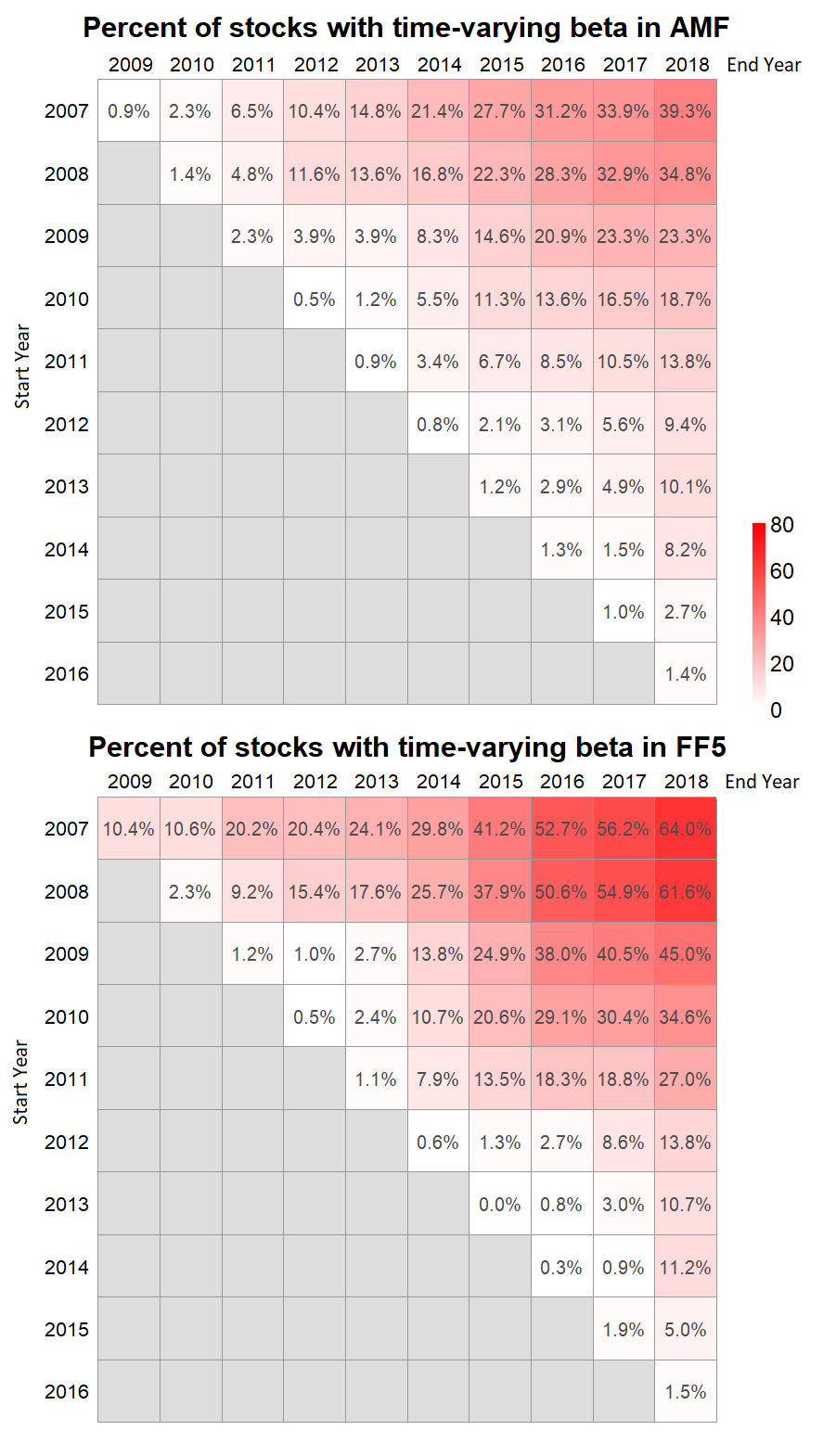} \\
\vspace{-0.2cm}
\caption{Percentage of stocks with time-varying beta using the time-invariance test in a linear setting for each time period. The y-axis is the start year of each time period and the x-axis is the end year. The percentage in each grid is the percentage of stocks with FDR Q-value less than 0.05 in Section \ref{sec: ti_test_linear} ANOVA test comparing the models in Equation (\ref{eq: ti_test_linear}) and (\ref{eq: ti_ols_subset}).}
\label{fig: ti_test_linear_combine}
\end{figure}

An ANOVA test is employed to compare the model in equation (\ref{eq: ti_test_linear}) and (\ref{eq: ti_ols_subset}). A p-value of less than 0.05 rejects the null hypothesis that the $\beta_{i,j}$'s are all time-invariant. We also want to control for the False Discovery Rate (FDR) (see \cite{benjamini1995controlling}). The Benjamini-Hochberg (BH) \cite{benjamini1995controlling} FDR adjusting procedure does not account for the correlation between tests, while the Benjamini-Hochberg-Yekutieli (BHY) \cite{benjamini2001control} FDR adjusting method does. Since we may have a correlation between the basis assets, we use the BHY method to adjust the p-values into the FDR Q-values and then report the percentage of stocks with Q-values less than 0.05 in Figure \ref{fig: ti_test_linear_combine}.

Figure \ref{fig: ti_test_linear_combine} reports the percentage of stocks with time-varying beta using the time-invariance test in a linear setting for each time period. The y-axis is the start year of each time period and the x-axis is the end year of the time period. The percentage in each grid is the percentage of stocks with FDR Q-value less than 0.05 in the ANOVA test comparing the models in Equation (\ref{eq: ti_test_linear}) and (\ref{eq: ti_ols_subset}). The larger the percentage is, the darker the grid will be. The upper heatmap is the result of the AMF model, while the bottom heatmap is the result of the FF5 model.

In the heatmaps in Figure \ref{fig: ti_test_linear_combine} (and same below), all elements on the k-th skew-diagonal (see definition in Equation (\ref{eq: skew_diag})) correspond to time periods of the same length, which is $ k + 3 $ years. For example, the diagonal (0-th skew-diagonal) elements are related to the time periods with 3 years, the 1st skew-diagonal elements are of the time periods 4 years. Comparing the different skew-diagonals, we can see that the AMF model is very stable in all time periods of less than 5 years. For most time periods no more than 5 years, only less than 5\% companies have a least 1 time-varying $\beta$. In other words, for more than 95\% of the companies, the $\beta$'s in the AMF model are time-invariant. However, for the FF5 model, even some 3-year time periods are not stable, such as 2007 - 2009.

In the heatmaps in Figure \ref{fig: ti_test_linear_combine} (and same below), all elements on the k-th skew-anti-diagonal (see definition in Equation (\ref{eq: skew_anti_diag})) correspond to time periods with the same ``mid-year''. For example, the 1st skew-anti-diagonal elements are all related to time periods centered in the mid-week of 2012, with different time lengths. By comparing different skew-anti-diagonals, we can compare if the stability pattern is the same for different mid-years. For the AMF model, we can see that the percentage for a fixed time length does not change much for different mid-years. For example, all time periods of 4 years have time-varying percent less than 5\%, which does not change much for different mid-years. However, the stability of the $\beta$'s for the FF5 model highly depends on the mid-year, not just the length of the time period. FF5 is more volatile with mid-year 2012 - 2013, and 2008 - 2009. In another word, FF5 can not capture the factors as accurate as AMF, therefore the $\beta$'s change a lot during the financial crisis.

\begin{figure}[H]
\centering
\includegraphics[width=0.9\textwidth]{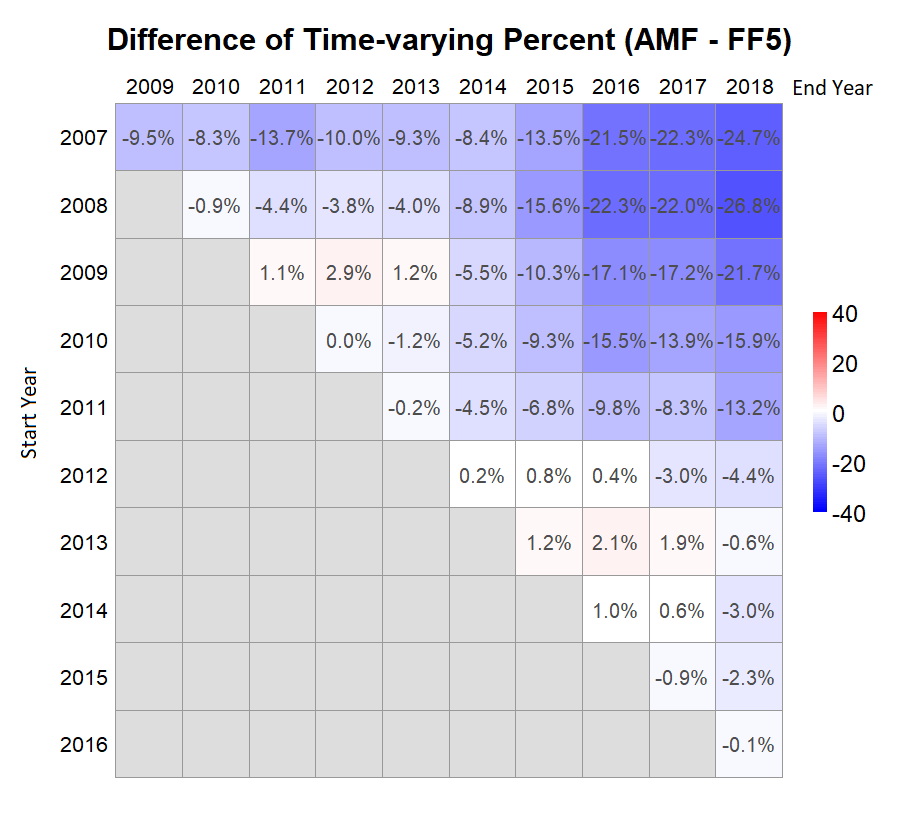} \\
\vspace{-0.2cm}
\caption{Difference of the two heatmaps in Figure \ref{fig: ti_test_linear_combine} to compare AMF and FF5. Each grid is the percent of time-varying stocks in AMF model minus the percent of time-varying stocks in FF5 model shown in Figure \ref{fig: ti_test_linear_combine}.}
\label{fig: ti_test_linear_diff}
\end{figure}

In general, AMF is much more stable than the FF5, which can be seen in Figure \ref{fig: ti_test_linear_diff}. The table in Figure \ref{fig: ti_test_linear_diff} is the difference between the two tables in Figure \ref{fig: ti_test_linear_combine} (AMF - FF5). For most time periods, the grid is blue, meaning that the AMF model is more stable than FF5 by giving less percentage of companies with time-varying $\beta$'s, sometimes the decrease is high and over 20\%. In the only few time periods when FF5 is a little more stable than AMF, both AMF and FF5 are quite stable, giving less than 5\% companies with time-varying $\beta$'s. In all other time periods when FF5 is unstable, AMF performs much better than FF5.

To sum up, AMF outperforms the FF5 in terms of the stability in two folds. First, for all time periods, AMF is much more stable (or at least similarly stable) compared to FF5.  AMF either gives much more stable $\beta$'s than FF5 or gives similar stable $\beta$'s if FF5 is already stable. Second, the stability of the AMF model for each time length is more robust across all mid-years compared to FF5. The stability of the AMF model almost merely depends on the length of the time period and is not affected much about the start and end year, while the stability of the FF5 model depends highly on the mid-year. This indicates that AMF is more insightful and is not too vulnerable to the financial crisis.

\subsection{Residual analysis}
\label{sec: ti_resid_test}

In this section, we test if including more basis assets can help improve the fitting in each time period. Since the number of ETFs available always increases over time, if we focus on the second half of the time period, we will have more ETFs available compared to the whole time period. In other words, for each time period, we want to test if the ETFs newly introduced to the market in the first half of the time period can help to make a better fit for the AMF model.

\begin{figure}[H]
\centering
\includegraphics[width=0.73\textwidth]{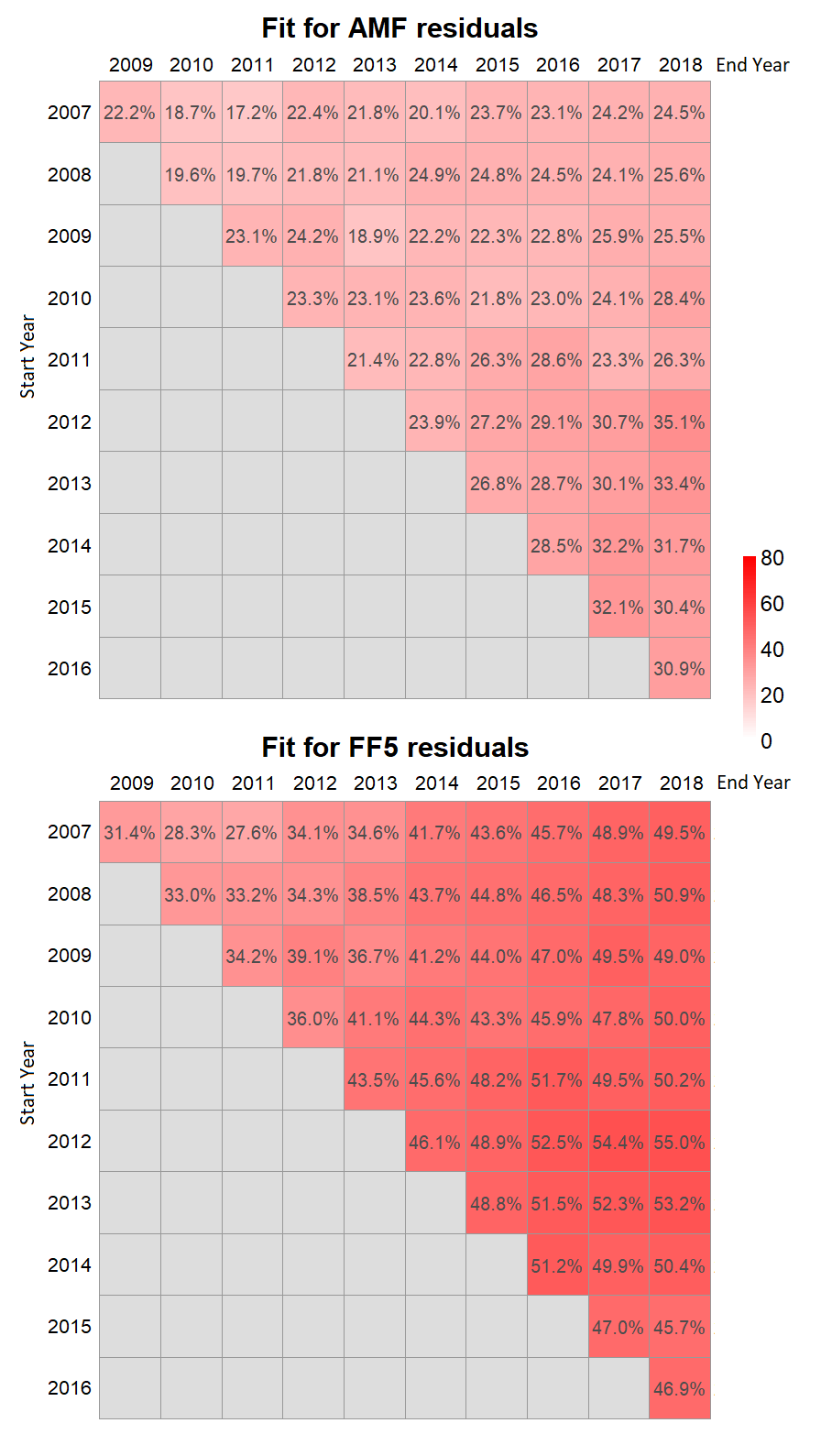} \\
\vspace{-0.2cm}
\caption{Percentage of stocks where the model in Equation (\ref{eq: ti_resid2}) has significantly better fit compared to Equation (\ref{eq: ti_resid2}) for each time period. The y-axis is the start year of each time period and the x-axis is the end year. The percentage in each grid is the percentage of stocks with FDR Q-value less than 0.05.}
\label{fig: ti_test_resid_combine}
\end{figure}

To be more specific, for a time period $[t_a, t_b] $. Let $ t_{mid} = (t_a + t_b) / 2 $. Then we divide $\Delta \bm{Y}_i$ and $\Delta \bm{V}$ into two parts, one in the time period $[t_a, t_{mid})$ and the other one in the time period $[t_{mid}, t_b)$.
\begin{equation}
\Delta \bm{Y}_i = \begin{pmatrix}
\Delta \bm{Y}_{i, a} \\
\Delta \bm{Y}_{i, b}
\end{pmatrix}, \quad
\Delta \bm{V} = \begin{pmatrix}
\Delta \bm{V}_{a} \\
\Delta \bm{V}_{b}
\end{pmatrix}
\end{equation}
We first derive the basis assets set $S_i$ in Equation (\ref{eq: ti_factor_select}) using the GIBS algorithm on the whole time period. Then we fit the Ordinary Least-Square (OLS) regression on the second half of the data
\begin{equation}
\Delta\bm{Y}_{i, b}
=(\Delta\bm{V}_b)_{S_{i}}(\bm{\gamma}_{i})_{S_i} + \Delta\bm{\epsilon}_{i, b}
\label{eq: ti_resid1}
\end{equation}
and derive the estimation of the coefficient as $(\bm{\hat\gamma}_i)_{S_i}$. So the residuals are
\begin{equation}
\Delta \bm{\hat\epsilon}_{i, b}
= \Delta \bm{Y}_{i, b} - (\Delta \bm{V}_b)_{S_i} (\bm{\hat\gamma}_{i})_{S_i}
\end{equation}
We fit an AMF model with the GIBS algorithm again, using the residuals $ \Delta \bm{\hat\epsilon}_{i, b} $ as our new response, using all the basis assets available for the time period $[t_{mid}, t_b)$ except the basis assets already selected in $S_i$ as our basis assets. The new AMF model on the residuals will give use another selected basis asset set $S_{i, b}$. If $ S_{i, b} \neq \emptyset $, we merge the two sets together
\begin{equation}
S_{i, union} = S_i \cup S_{i, b}
\end{equation}
Note that since we remove all the basis assets in $S_i$ in our GIBS fitting on residuals, $ S_i \cap S_{i, b} = \emptyset $. Then we fit another OLS regression on the second half of the data using the $ S_{i, union}$
\begin{equation}
\Delta\bm{Y}_{i, b}
=(\Delta\bm{V}_b)_{S_{i, union}}(\bm{\gamma}_{i})_{S_i, union} + \Delta\bm{\epsilon}_{i, b}
\label{eq: ti_resid2}
\end{equation}
Then we use an Analysis of Variance (ANOVA) test to compare the model in Equation (\ref{eq: ti_resid1}) and Equation (\ref{eq: ti_resid2}). For each time period, this test is done on all stocks and we can have a list of p-values. Similarly to the previous sections, we adjust the False Discovery Rate (FDR) by the Benjamini-Hochberg-Yekutieli (BHY) \cite{benjamini2001control} method and count the number of companies with FDR Q-values less than 0.05. And report the percentage in Figure \ref{fig: ti_test_resid_combine}.

Note that if for $i$-th stock the set $ S_{i, b} = \emptyset $, there will not be p-value for this stock since the two models in Equation (\ref{eq: ti_resid1}) and (\ref{eq: ti_resid2}) are the same. So this stock will not be counted as the company with FDR Q-values less than 0.05. However, when presenting the percentage, we use the total count of companies available in that time period as the denominator. In another word, if $S_{i, b} = \emptyset$, then $i$ will not be counted in the numerator, but will still be counted in the denominator. This makes our result conservative.

For the FF5 model, we use $S_i$ as the FF5 5-factors and the risk-free rate instead of the set selected by the GIBS algorithm. All the other steps are the same. The results for FF5 residuals are in comparison to our main result.

From Figure \ref{fig: ti_test_resid_combine} we see that for all time periods, we can give a significantly better fit using the new basis assets in the second half of the time period for around 20\% of the stocks in the AMF model. Comparing this result with Figure \ref{fig: ti_test_linear_combine}, we know that there are stocks that have constant $\beta$'s but can be fitted better using the new basis assets available in the second half of the time period. This gives an interesting insight into the market: For some stocks, its $\beta$'s related to the old risk-factors are still time-invariant. However, including new factors does give a significantly better fit. In other words, with more basis assets introduced to the market, we can have better information about the $\epsilon$ noises that were not able to be explained using the old risk-factors, while the effects of the old risk-factors are still the same.

\begin{figure}[H]
\centering
\includegraphics[width=0.9\textwidth]{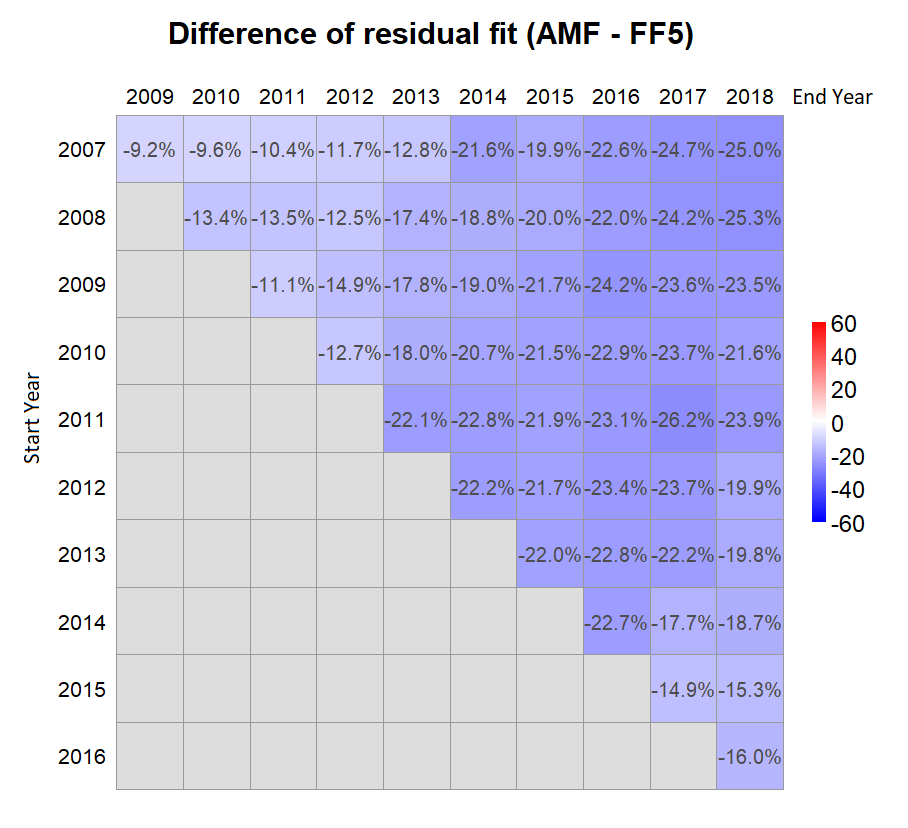} \\
\vspace{-0.2cm}
\caption{Difference of the two heatmaps in Figure \ref{fig: ti_test_resid_combine}. Each grid is the percentage in AMF model minus the percentage in FF5 model shown in Figure \ref{fig: ti_test_resid_combine}.}
\label{fig: ti_test_resid_diff}
\end{figure}

Finally, comparing the results for AMF residuals and FF5 residuals, it is clear that AMF can capture more significant basis assets and leave less information in the residuals compared to the FF5. From Figure \ref{fig: ti_test_resid_diff} we know that the percentage of stocks that can be better fitted in the second half time period is always much less for AMF compared to FF5. This indicates that AMF is more powerful than FF5 in capturing useful basis assets.p

\subsection{Time-invariance Test in Non-Linear GAM Setting}
\label{sec: ti_test_gam}

Instead of testing the time-invariance of the $\beta$'s based on a linear model, we can test based on the Generalized Additive Model (GAM). This is a more strict test, since this tests not only on the time-invariance of the $\beta$'s, but the linearity of pthem as well. We first fit a Time-Varying Coefficient model as a special case of the Generalized Additive Model (GAM). The equation can be written as:
\begin{equation}
\Delta\bm{Y}_{i}
= \Delta\bm{V}_{S_i} [\bm{\beta}_{i}(t)]_{S_i} + \Delta\bm{\epsilon}_{i}
\label{eq: ti_test_gam}
\end{equation}
where $t$ is the time. Note that the only difference between Equation (\ref{eq: ti_test_gam}) and Equation (\ref{eq: ti_ols_subset}) is whether we allow $\beta$'s to be functions of time $t$. The GAM model estimate each $\beta$ as a combination of splines or kernels with regard to $t$. This can be done by the \textit{gam()} function in the R package \textit{mgcv}.

\begin{figure}[H]
\centering
\includegraphics[width=0.73\textwidth]{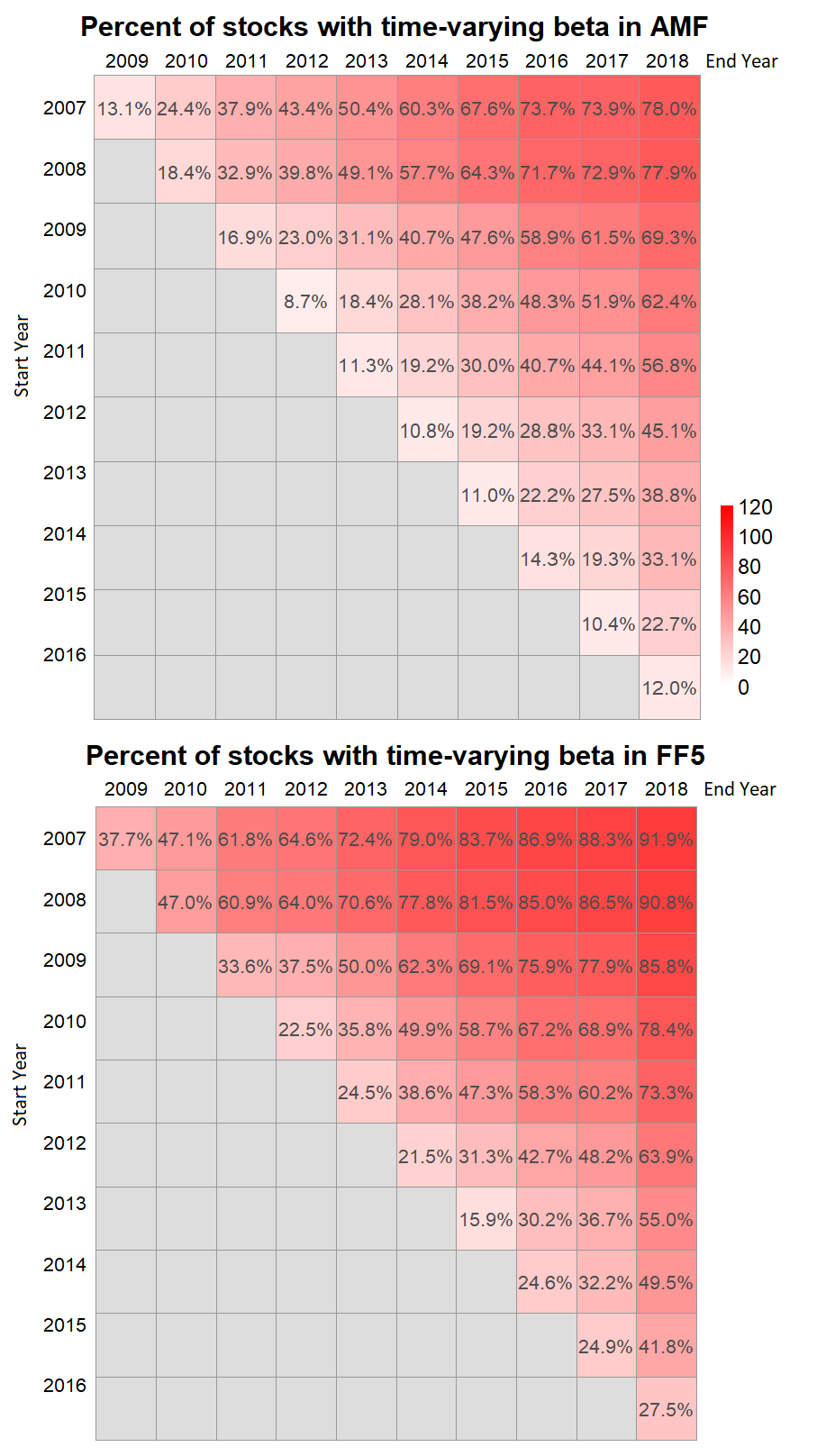} \\
\vspace{-0.2cm}
\caption{Percentage of stocks with time-varying beta using the time-invariance test in a non-linear GAM setting for each time period. The y-axis is the start year of each time period and the x-axis is the end year. The percentage in each grid is the percentage of stocks with FDR Q-value less than 0.05 in Section \ref{sec: ti_test_linear} ANOVA test comparing the models in Equation (\ref{eq: ti_test_gam}) and (\ref{eq: ti_ols_subset}).}
\label{fig: ti_test_gam_combine}
\end{figure}

After that, do a variance analysis (ANOVA) test between the GAM model in Equation (\ref{eq: ti_test_gam}) and the linear model in Equation (\ref{eq: ti_ols_subset}) where each $\bm{\beta}_i$ are constants over time. In this way, we will get a p-value for each stock. Similar to the section before, we adjust the p-values by the Benjamini-Hochberg-Yekutieli (BHY) \cite{benjamini2001control} method to account for the False Discovery Rate (FDR). We report the percentage of stocks with Q-values less than 0.05 in Figure \ref{fig: ti_test_gam_combine}.

Figure \ref{fig: ti_test_gam_combine} reports the percentage of stocks with time-varying beta using the time-invariance test in a nonlinear GAM setting for each time period. The y-axis is the start year of each time period and the x-axis is the end year of the time period. The percentage in each grid is the percentage of stocks with FDR Q-value less than 0.05 in the ANOVA test comparing the models in Equation (\ref{eq: ti_test_gam}) and (\ref{eq: ti_ols_subset}). The larger the percentage is, the darker the grid will be. The upper heatmap is the result of the AMF model, while the bottom heatmap is the result of the FF5 model.

\begin{figure}[H]
\centering
\includegraphics[width=0.9\textwidth]{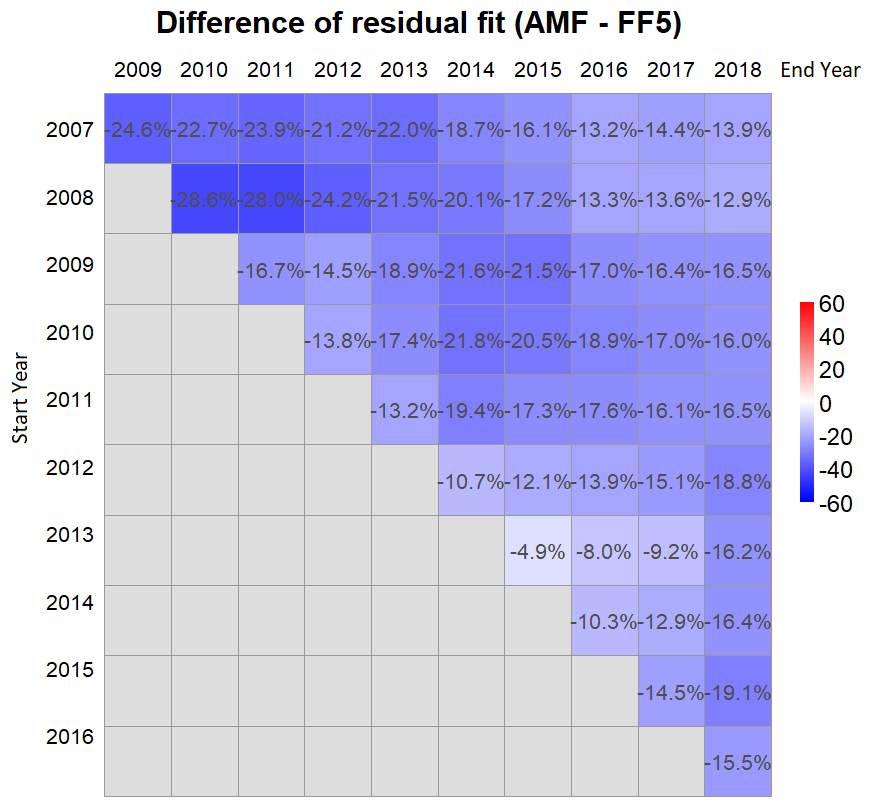} \\
\vspace{-0.2cm}
\caption{Difference of the two heatmaps in Figure \ref{fig: ti_test_gam_combine} to compare AMF and FF5. Each grid is the percent of time-varying stocks in AMF model minus the percent of time-varying stocks in FF5 model shown inp Figure \ref{fig: ti_test_gam_combine}.}
\label{fig: ti_test_gam_diff}
\end{figure}

Similarly, by comparing the different skew-diagonals, we can see that both models are more stable in short time periods. AMF outperforms the FF5 in all time periods, which is also shown in Figure \ref{fig: ti_test_gam_diff}. The Figure \ref{fig: ti_test_gam_diff} is the difference between the percentage in two heatmaps in Figure \ref{fig: ti_test_gam_combine} (AMF - FF5). All the grids are blue, indicating that AMF is more stable than FF5 in all periods in the GAM setting.

However, comparing the results from GAM and the results in the previous Section \ref{sec: ti_test_linear}, we do find that the GAM test is more strict than the linear test. For any period and both AMF and FF5 models, more companies are shown to have time-varying $\beta$'s in the GAM test, although AMF still outperforms the FF5.

Note that this may not necessarily mean that the $\beta$'s are time-varying in all periods in both tests, since it is really easy for GAM to overfit, especially in short periods. To be more specific, the number of observations for 3 year time periods is $n = 156$. For the FF5 model, there are 5 basis assets. For the AMF model, there are sometimes more basis assets selected. For each basis assets, the GAM model selects some splines or kernels, say 10 splines, which can easily boom the real dimension of parameters to $ p = 5 \times 10 = 50 $, which is already too large for regression with $ n = 156 $ observations. In other words, the number of parameters is still too close to the number of observations. Also, these new variables can be highly correlated, making the fitting more unstable. Therefore, some of these GAM fittings may have severe overfitting.

Therefore, the testing based on the GAM model in this section is explorative. Considering this high-dimensional issue, penalization and constraints need to be introduced to traditional GAM for our application. This extension is left for future research.

\section{Conclusion}
\label{sec: ti_conclude}

The purpose of this paper is to test the multi-factor beta model implied by the generalized arbitrage pricing theory (APT) and the Adaptive Multi-Factor (AMF) model with the Groupwise Interpretable Basis Selection (GIBS) algorithm, without imposing the exogenous assumption of constant betas. The intercept (arbitrage) tests show that there are no significant non-zero intercepts in either AMF or FF5 model, which validates the 2 models.

We do the time-invariance tests for the $\beta$'s for both the AMF model and the FF5 in various periods. We show that the constant-beta assumption holds in the AMF model in all periods with a length of less than 6 years and is quite robust regardless of the start year. However, even for short periods, FF5 sometimes gives very unstable estimation, especially in the financial crisis. This indicates that AMF is more insightful and can capture the risk-factors to explain the market shift during the financial crisis.

For periods with length longer than 6 years, both AMF and FF5 fail to provide time-invariance $\beta$'s. However, the $\beta$'s estimate by the AMF is more time-invariant than the FF5 for nearly all periods. This shows the superior performance of the AMF model.

Considering the two results above, using the dynamic AMF model with a decent rolling window (such as 5 years) is more powerful and stable than the FF5 model.

\clearpage
\chapter{Low-volatility Anomaly and the Adaptive Multi-Factor Model}

\section{Introduction}

This paper relates to two branches of the asset pricing literature.
The first is multi-factor model estimation based on Ross' (1976) \cite{ross1976arbitrage}
Arbitrage Pricing Theory (APT) and Merton's (1973) \cite{merton1973intertemporal}
Inter-temporal Capital Asset Pricing Model (ICAPM). The second is
the growing literature related to the low-risk anomaly. The purpose
of this paper to understand the low-risk anomaly. The low-risk anomaly
is the empirical observation that stocks with lower risk yield higher
returns than stocks with higher risk. Here risk is quantified as either
a security's return volatility or a security's beta as derived from
a Capital Asset Pricing Model (CAPM). This paper focuses only on the
low-volatility anomaly.\footnote{In analyzing the low-beta anomaly, an issue is the non synchronized
trading of small stocks, where a small stock may trade less frequently
than the index it is regressed on (market return) to obtain the stock's
beta (see Mcinish and Wood (1986) \cite{mcinish1986adjusting}).}

To study this anomaly, we use the Adaptive Multi-Factor (AMF) model
developed in Zhu et al. (2018) \cite{zhu2020high} which includes
both the APT and ICAPM are special cases. The AMF is derived under
a weaker set of assumptions. Its three main benefits are: 1) it is
consistent with a large number of risk factors being needed to explain
all security returns, 2) yet, the set of risk factors is small for
any single security and different for different securities, and 3)
the risk factors are traded. These benefits imply that the underlying
model estimated is more robust than those used in the existing literature.
%\bigskip

The low-risk anomaly contradicts accepted APT or CAPM theories that
higher risk portfolios earn higher returns. The low-risk anomaly is
not a recent empirical finding but an observation documented by a
large body of literature dating back to the 1970s. Despite its longevity,
the academic community differs over the causes of the anomaly. The
two main explanations are: 1) it is due to leverage constraints that
retail, pension and mutual fund investors face which limits their
ability to generate higher returns by owning lower risk stocks, and
2) it is due to behavioral biases ranging from the lottery demand
for high beta stocks, beating index benchmarks with a limits to arbitrage,
and the sell side analysts over-bias on high volatility stocks' earnings.
%\bigskip

The first explanation can be traced back to Black et al. (1972) \cite{black1972capitala}
who showed empirically using stock returns from 1926 to 1966 that
expected excess returns on high-beta assets are lower than and that
expected excess returns on low-beta stocks are larger than those suggested
by the CAPM. In a follow-up paper, accounting for borrowing constraints,
Black (1972) \cite{black1972capital} proved that the slope of the
line between expected returns and $\beta$ must be smaller than when
there are no borrowing restrictions. More recently, Frazzini and Pedersen
(2014) \cite{frazzini2014betting}document the low-beta anomaly in
20 international equity markets and across assets classes including
Treasury bonds, corporate bonds, and futures. They argue that investors
facing leverage and margin constraints increase the prices of high-beta
assets, which generates lower alphas. They show that the Betting Against
Beta (BAB) factor, which is long leveraged low-beta assets and short
high beta assets, yields positive risk adjusted returns. Ang et al.
(2006) \cite{ang2006cross} find that stocks with high-idiosyncratic
volatility after controlling for size, book-to-market, momentum, liquidity
effects, and market-wide volatility risk (VIX) earn lower absolute
and risk-adjusted returns than stocks with lower-idiosyncratic volatility.
Ang et al. (2009) \cite{ang2009high} also show that there is a strong
comovement in the anomaly across 27 developed markets which implies
that easily diversifiable factors cannot explain the out performance
of a low idiosyncratic volatility portfolio. %\bigskip

The behavioral explanations (see Baker et al. (2011) \cite{baker2011benchmarks})
are that irrational investor preferences for lottery-like stocks (more
attention is triggered if you talk about Tesla (TSLA) versus Procter
\& Gamble at a party) and an overconfidence bias pushes the prices
of high risk stocks above their fundamentals. Second, because institutional
investors have a mandate to outperform some market weighted index,
they also over emphasize investments in high-risk stocks. By increasing
the beta exposure of their portfolio in this way, they are more likely
to beat the benchmark in a rising market. And, due to limits to arbitrage,
"smart money" is not able to arbitrage away this low-risk anomaly.
Providing additional support for this explanation, Bali et al. (2017)
\cite{bali2017lottery} show that a proxy for lottery demand stocks
(the average of the five highest daily returns in a given month) explains
this low-beta anomaly. Hong and Sraer (2016) \cite{hong2016speculative}
provide an equilibrium model consistent with this behavior. Last,
it is also argued that the low-risk anomaly is tied to analyst earnings
reports because high-risk stocks are characterized with more inflated
sell side analyst's earnings growth forecasts which produce an investor
overreaction and yields lower returns as shown in Hsu et al. (2013)
\cite{hsu2013when}. %\bigskip

In this paper we study the low-volatility anomaly from a new perspective
based on the Adaptive Multi-Factor (AMF) model proposed in the paper
by Zhu et al. (2018) \cite{zhu2020high} using the recently developed
Generalized Arbitrage Pricing Theory (see Jarrow and Protter (2016)
\cite{jarrow2016positive}). In Zhu et al. (2018) \cite{zhu2020high},
basis assets (formed from the collection of Exchange Traded Funds
(ETF)) are used to capture risk factors in \textit{realized} returns
across securities. Since the collection of basis assets is large and
highly correlated, high-dimension methods (including the LASSO and
prototype clustering) are used. This paper employs the same methodology
to investigate the low-volatility anomaly. We find that high-volatility
and low-volatility portfolios load on different basis assets, which
indicates that volatility is not an independent risk. The out-performance
of the low-volatility portfolio is due to the (equilibrium) performance
of these loaded risk factors. For completeness, we compare the AMF
model with the traditional Fama-French 5-factor (FF5) model, documenting
the superior performance of the AMF model. A brief review of the high
dimensional statistical methods can be found in Section \ref{sec: amf_high_dim_method}.

An outline for this paper is as follows. Section \ref{sec: vol_est_method} discusses the estimation
methodology, and section \ref{sec: vol_estimation_results} presents the results. Section \ref{sec: vol_conclude} concludes.

\section{The Estimation Methodology}

\label{sec: vol_est_method}

This section gives the estimation methodology. We first pick the universe
of stocks and ETFs, so that we only focus on the assets that are easy
to trade. Then, we form the high and low volatility portfolios and
we choose a time period that exhibits the low-volatility anomaly.
We will employ the Adaptive Multi-Factor (AMF) asset pricing model
with the Groupwise Interpretable Basis Selection (GIBS) algorithm
to these two portfolios and analyze the results in a subsequent section.

\subsection{The Stock and ETF Universe}
\label{sec: vol_universe}

\begin{figure}[H]
\centering
\includegraphics[width=0.7\textwidth]{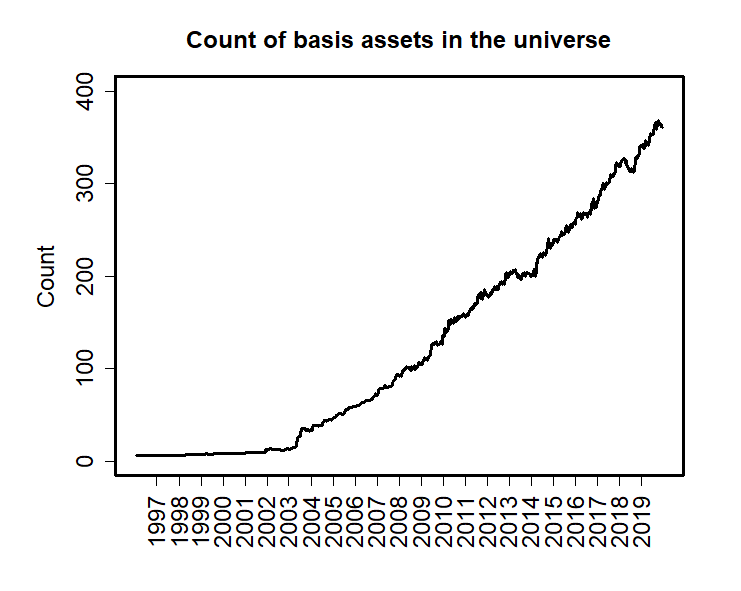}
\caption{Count of the ETFs in the universe.}
\label{fig: vol_etf_count}
\end{figure}

\begin{figure}[H]
\centering
\includegraphics[width=0.7\textwidth]{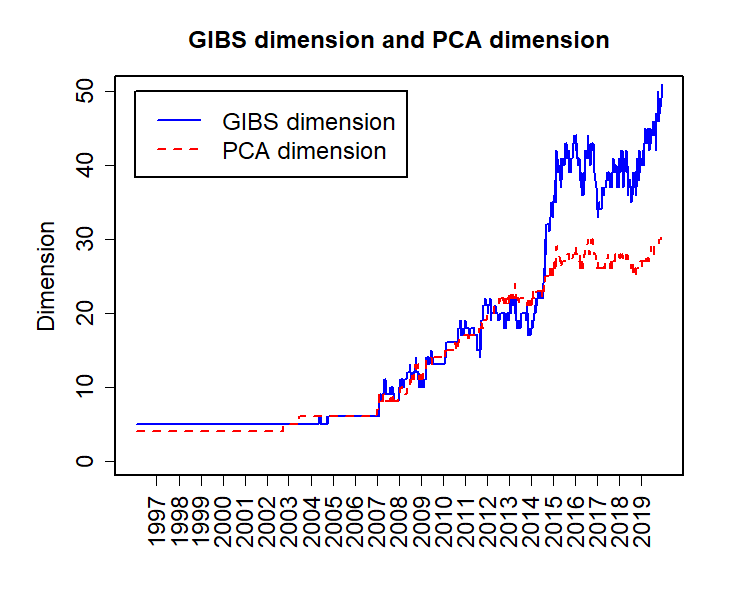}
\caption{GIBS dimension and PCA dimension of ETFs in the universe.}
\label{fig: vol_etf_dim}
\end{figure}

The data initially consists of all stocks and ETFs available in the Center for Research in Security Prices (CRSP) database. To ensure all our securities are actively traded, we focus on stocks and ETFs with a market capitalization ranking in top 2500. This filter excludes small stocks that are more
likely to exhibit the low-volatility anomaly. In addition, we select stocks and ETFs satisfying the following criteria.
\begin{enumerate}
\item According to the description of the CRSP database, ETFs should be with a Share Code (SHRCD) 73. We follow this common practice.
\item We excluded American Depositary Receipts (ADR) from our stock universe. Removing the ADR is a common practice in all empirical finance papers and the main reason is that they are not part of the indices.  This is achieved by using the Share Code (SHRCD) 10 or 11 from the CRSP dataset.
\item We only choose ETFs and stocks which are listed on the NYSE, AMEX and NASDAQ exchanges. This is obtained with the Exchange Code (EXCHCD) 1, 2 or 3 from the CRSP dataset.
\item For a stock to be included at time t, its return has to be observable at least 80\% of trading times in the previous year in order to calculate its volatility. For an ETF to be considered at time t, its return has to be observable during the 3-year regression window before time t.
\end{enumerate}
The number of ETFs in our universe increased rapidly after 2003, see
Figure \ref{fig: vol_etf_count}. To apply the AMF model and GIBS algorithm,
we need enough ETFs to form our collection of basis assets. Hence,
we begin our analysis in 2003. To understand the dimension of the
set of basis assets, we calculate both the GIBS dimension and the
Principal Component Analysis (PCA) dimension of the ETFs in our universe.
The PCA dimension at time $t$ is defined to be the number of principal
components needed to explain 90\% of the variance during the previous
3-years. The GIBS dimension is defined to be the number of ``representatives''
selected using the GIBS algorithm from the basis assets, i.e. the
cardinality of the set $U$ in Table \ref{tab: vol_gibs_algo}. The GIBS
and PCA dimensions of the ETFs across time are shown in Figure \ref{fig: vol_etf_dim}.

Comparing Figures \ref{fig: vol_etf_count} and \ref{fig: vol_etf_dim}, we
see that the number of ETFs and dimensions increase over this time
period. The GIBS and PCA dimension do not increase as fast as the
number of ETFs But, the GIBS dimension increases faster than the PCA
dimension, suggesting that GIBS is able to pick more basis assets
than does the PCA. The reason is that PCA mixes basis assets together
in linear combinations, while the GIBS algorithm does not.

\subsection{Portfolios and the Low-volatility Anomaly}

\label{sec: vol_anomaly}

To study the low-volatility anomaly, we need to form the high- and low-volatility portfolios. To do this, we first calculate the volatility of the stocks in our universe at time $t$ as the standard deviation of their excess returns over the previous year. The excess return is the raw return minus the risk free rate. Using the excess return is a common practice in the empirical finance literature since it removes the risk free rate and focuses on the risk premiums. The high-volatility portfolio is constructed as an equally weighted portfolio using the stocks with the highest 25\% volatilities. Similarly, we take the stocks having the lowest 25\% volatilities to form the low-volatility portfolio. To avoid a survivorship bias, we include the delisted returns (see Shumway (1997) \cite{shumway1997delisting} for more explanation).

\begin{figure}[H]
\centering
\includegraphics[width=0.8\textwidth]{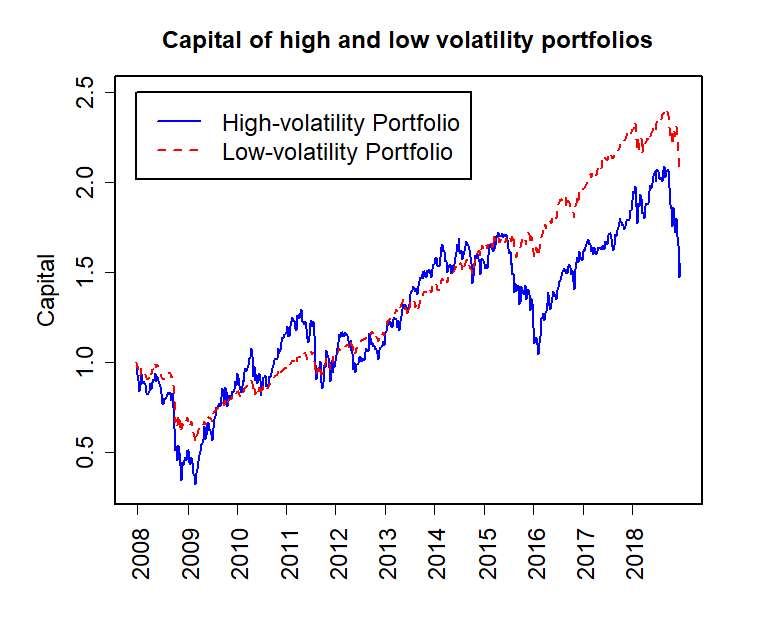}
\caption{Cumulative value of the excess returns from the high- and low-volatility
portfolios.}
\label{fig: vol_caps_ex}
\end{figure}

We then compare the excess returns of the high- and low-volatility
portfolios to verify the existence of the low-volatility anomaly over
our sample period. Given the graph, we selected 2008 as the start
of the anomaly. Between 2008 - 2018, the low-volatility portfolio
had an excess return of 121.4\%, which is higher than the 62.5\% excess
return of the high-volatility portfolio. This documents the existence
of the low-volatility anomaly over our sample period. The superior
performance of the low-volatility portfolio is manifested in Figure
\ref{fig: vol_caps_ex} which graphs the cumulative value of the two portfolios
starting from \$1 at the beginning of 2008.

\subsection{The AMF and GIBS Estimation}
\label{sec: vol_amf_analysis}

This section estimates the Adaptive Multi-Factor (AMF) model based
on the Groupwise Interpretable Basis Selection (GIBS) algorithm proposed
by Zhu et al. (2018) \cite{zhu2020high}. Although a brief review
of the AMF model and the GIBS algorithm are included in this section,
the details can be found in \cite{zhu2020high}. A sketch of the GIBS
algorithm is in Table \ref{tab: vol_gibs_algo} at the end of this section.

To avoid the effects of market micro-structure frictions, we use a
weekly horizon. Because the number of ETFs increase over time and
the market structure changes during the financial crisis, to approximate
stationarity, we pick a 3-year regression window to do the analysis.
In addition, because the number of ETFs exceed the number of observations
within each regression window, we are in the high-dimensional regime.
Therefore, the high-dimensional GIBS algorithm needs to be used to
estimate the AMF model. We use a dynamic version of the GIBS algorithm
applied to rolling windows. For each week $t$ in 2008 - 2018, we
use the time period from 3 years earlier as our current regression
window. We use all the ETFs in our universe as described in Section
\ref{sec: vol_universe} and the FF5 factors as our basis assets. Then,
we apply the GIBS algorithm to select the GIBS determined basis assets
and use them to explain the excess returns of the high- and low-volatility
portfolios using the AMF model. The following is a more detailed introduction
to the AMF model and the GIBS algorithm within each time window.

In the asset pricing theory, given a frictionless, competitive, and
arbitrage free market, a dynamic generalization of Ross's (1976) \cite{ross1976arbitrage}
APT and Merton's (1973) \cite{merton1973intertemporal} ICAPM is contained
in Jarrow and Protter (2016) \cite{jarrow2016positive}. This extension
implies that the Adaptive Multi-Factor (AMF) model holds for any security's
return over $[t,t+1]$:
\begin{equation}
R_{i}(t)-r_{0}(t)=\sum_{j=1}^{p}\beta_{i,j}\big[r_{j}(t)-r_{0}(t)\big]
=\bm{\beta}_{i}'\cdot[\bm{r}(t)-r_{0}(t)\vone]
\label{eq: vol_linear model}
\end{equation}
where $R_{i}(t)$ denotes the return of the i-th security for $1\leq i\leq N$
(where $N$ is the number of securities), $r_{j}(t)$ denotes the
return on the j-th basis asset for $1\leq j\leq p$, $r_{0}(t)$ is
the risk free rate, $\bm{r}(t)=(r_{1}(t),r_{2}(t),...,r_{p}(t))'$
denotes the vector of security returns, $\vone$ is a column vector
with every element equal to one, and $\bm{\beta}_{i}=(\beta_{i,1},\beta_{i,2},...,\beta_{i,p})'$.

In this paper we are only concerned with the low- and high-volatility
portfolios. Let $R_{1}$ denote the raw return of the low-volatility
portfolio and $R_{2}$ the raw return of the high-volatility portfolio.
To empirically test our model, both an intercept $\alpha_{i}$ and
a noise term $\epsilon_{i}(t)$ are added to expression (\ref{eq: vol_linear model}),
i.e.
\begin{equation}
R_{i}(t)-r_{0}(t)=\alpha_{i}+\sum_{j=1}^{p}\beta_{i,j}(t)\big[r_{j}(t)-r_{0}(t)\big]+\epsilon_{i}(t)
=\alpha+\bm{\beta}_{i}'[\bm{r}(t)-r_{0}(t)\vone]+\epsilon_{i}(t)
\label{eq: vol_testable}
\end{equation}
where $\epsilon_{i}(t)\overset{iid}{\sim}N(0,\sigma_{i}^{2})$ and
$1\leq i\leq N$. The intercept enables the testing for mispriced
securities, and the error term allows for noise in the return observations.

As mentioned earlier, using weekly returns over a 3-year time period
necessitates the use of high-dimensional statistics. To understand
why, consider the following. For a given time period $(t,T)$, letting
$n=T-t+1$, we can rewrite expression (\ref{eq: vol_testable}) as
\begin{equation}
\bm{R}_{i}-\bm{r}_{0}
=\alpha_{i}\vone_{n}+(\bm{r}-\bm{r}_{0}\vone'_{p})\bm{\beta}_{i}+\bm{\epsilon}_{i}
\label{eq: vol_vec}
\end{equation}
where $1\leq i\leq N$, $\bm{\epsilon}_{i}\sim N(0,\sigma_{i}^{2}\bm{I}_{n})$
and {\small{}
\[
\bm{R}_{i}=\begin{pmatrix}R_{i}(t)\\
R_{i}(t+1)\\
\vdots\\
R_{i}(T)
\end{pmatrix}_{n\times1},\,\bm{r}_{0}=\begin{pmatrix}r_{0}(t)\\
r_{0}(t+1)\\
\vdots\\
r_{0}(T)
\end{pmatrix}_{n\times1},\,\bm{\epsilon}_{i}=\begin{pmatrix}\epsilon_{i}(t)\\
\epsilon_{i}(t+1)\\
\vdots\\
\epsilon_{i}(T)
\end{pmatrix}_{n\times1}
\]
}{\small\par}

{\small{}
\[
\bm{\beta}_{i}=\begin{pmatrix}\beta_{i,1}\\
\beta_{i,2}\\
\vdots\\
\beta_{i,p}
\end{pmatrix}_{p\times1},\,\bm{r}_{i}=\begin{pmatrix}r_{i}(t)\\
r_{i}(t+1)\\
\vdots\\
r_{i}(T)
\end{pmatrix}_{n\times1},\,\bm{r}(t)=\begin{pmatrix}r_{1}(t)\\
r_{2}(t)\\
\vdots\\
r_{p}(t)
\end{pmatrix}_{p\times1}\,
\]
}{\small\par}

{\small{}
\begin{equation}
\bm{r}_{n\times p}=(\bm{r}_{1},\bm{r}_{2},...,\bm{r}_{p})_{n\times p}=\begin{pmatrix}\bm{r}(t)'\\
\bm{r}(t+1)'\\
\vdots\\
\bm{r}(T)'
\end{pmatrix}_{n\times p},\,\bm{R}=(\bm{R}_{1},\bm{R}_{2},...,\bm{R}_{N})\label{eq: vol_vec_end}
\end{equation}
}Recall that the coefficients $\beta_{ij}$ are assumed to be constants.
This assumption is only reasonable when the time period $(t,T)$ is
small (say 3 years), so the number of observations is $n\approx150$.
However, the number of of basis assets $p$, is around 300 in recent
years, implying that the independent variables exceed the observations.

Because of this high-dimension problem and the high-correlation among
the basis assets, traditional methods fail to give an interpretable
and systematic way to fit the Adaptive Multi-Factor (AMF) model. Therefore,
we employ the Groupwise Interpretable Basis Selection (GIBS) algorithm
to select the basis assets set $S\subseteq\{1,2,...,p\}$ (the derivation
of $S$ is provided later). Then, the model becomes
\begin{equation}
\bm{R}_{i}-\bm{r}_{0}
=\alpha_{i}\vone_{n}+(\bm{r}_{S}-(\bm{r}_{0})_{S}\vone'_{p})(\bm{\beta}_{i})_{S}+\bm{\epsilon}_{i}.
\label{eq: vol_theory_ols_subset}
\end{equation}
The notation $\bm{r}_{S}$ denotes the columns in the matrix $\bm{r}_{n\times p}$
indexed by the index set $S\subseteq\{1,2,...,p\}$, and $(\bm{r}_{0})_{S}$
denotes the elements in the vector $(\bm{r}_{0})_{n\times1}$ indexed
by the index set $S\subseteq\{1,2,...,p\}$. We will use this notation
for any matrices, vectors and indices sets throughout this paper.
An example of expression (\ref{eq: vol_theory_ols_subset}) is the Fama-French
(2015) \cite{fama2015five} 5-factor (FF5) model where all of the
basis assets are risk-factors, earning non-zero expected excess returns.
However, FF5 assumes that the number of risk-factors is small and
common to all the securities, whereas the AMF and GIBS does not.

Next, we give a brief review of the GIBS algorithm. For notation simplicity,
denote
\begin{equation}
\bm{Y}_{i}=\bm{R}_{i}-\bm{r}_{0}\,,\quad\bm{X}_{i}=\bm{r}_{i}-\bm{r}_{0}\,,\quad\bm{Y}
=\bm{R}-\bm{r}_{0}\,,\quad\bm{X}=\bm{r}-\bm{r}_{0}
\label{eq: vol_simplify_notation}
\end{equation}
where the definitions of $\bm{R}_{i},\,\bm{R},\,\bm{r}_{i},\,\bm{r}$
are in equations (\ref{eq: vol_vec} - \ref{eq: vol_vec_end}). Let $\bm{r}_{1}$
denote the market return. It is easy to check that most of the ETF
basis assets $\bm{X}_{i}$ are correlated with $\bm{X}_{1}$ (the
market return minus the risk free rate). We note that this pattern
is not true for the other four Fama-French factors. Therefore, we
first orthogonalize every other basis asset to $\bm{X}_{1}$. By orthogonalizing
with respect to the market return, we avoid choosing redundant basis
assets similar to it and also increase the accuracy of fitting. Note
that for the Ordinary Least-Square (OLS) regression, projection does not affect the estimation since it only affects the coefficients, not the estimated $\bm{\hat{y}}$. However,
in high-dimension methods such as LASSO, projection does affect the
set of selected basis assets because it changes the magnitude of shrinking.
Thus, we compute
\begin{equation}
\bm{\widetilde{X}}_{i}=(\bm{I}-P_{\bm{X}_{1}})\bm{X}_{i}
=(\bm{I}-\bm{X}_{1}(\bm{X}'_{1}\bm{X}_{1})^{-1}\bm{X}_{1}')\bm{X}_{i}
\qquad\text{where}\quad2\leq i\leq p_{1}
\label{eq: vol_proj1}
\end{equation}
where $P_{\bm{X}_{1}}$ denotes the projection operator, and $p_{1}$
is the number of columns in $\bm{X}_{n\times p_{1}}$. Denote the
vector
\begin{equation}
\bm{\widetilde{X}}=(\bm{X}_{1},\bm{\widetilde{X}}_{2},\bm{\widetilde{X}}_{3},...,\bm{\widetilde{X}}_{p_{1}}).
\label{eq: vol_proj2}
\end{equation}
Note that this is equivalent to the residuals after regressing other
basis assets on the market return minus the risk free rate.

The transformed ETF basis assets $\bm{\widetilde{X}}$ still contain
highly correlated members. We first divide these basis assets into
categories $A_{1},A_{2},...,A_{k}$ based on a financial characterization.
Note that $A\equiv\cup_{i=1}^{k}A_{i}=\{1,2,...,p_{1}\}$. The list
of categories with more descriptions can be found in Appendix \ref{ap: etf_classes}.
The categories are (1) bond/fixed income, (2) commodity, (3) currency,
(4) diversified portfolio, (5) equity, (6) alternative ETFs, (7) inverse,
(8) leveraged, (9) real estate, and (10) volatility.

Next, from each category we need to choose a set of representatives.
These representatives should span the categories they are from, but
also have low correlation with each other. This can be done by using
the prototype-clustering method with a distance measure defined by
equation (\ref{eq: corr_dist}), which yield the ``prototypes'' (representatives)
within each cluster (intuitively, the prototype is at the center of
each cluster) with low-correlations.

Within each category, we use the prototype clustering methods previously
discussed to find the set of representatives. The number of representatives
in each category can be chosen according to a correlation threshold.
This gives the sets $B_{1},B_{2},...,B_{k}$ with $B_{i}\subset A_{i}$
for $1\leq i\leq k$. Denote $B\equiv\cup_{i=1}^{k}B_{i}$. Although
this reduction procedure guarantees low-correlation between the elements
in each $B_{i}$, it does not guarantee low-correlation across the
elements in the union $B$. So, an additional step is needed, in which
is prototype clustering on $B$ is used to find a low-correlated representatives
set $U$. Note that $U\subseteq B$. Denote $p_{2}\equiv\#U$.

Recall from the notation definition in equation \ref{eq: vol_ols_subset}
that $\bm{\w{X}}_{U}$ means the columns of the matrix $\bm{\w{X}}$
indexed by the set $U$. Since basis assets in $\bm{\widetilde{X}}_{U}$
are not highly correlated, a LASSO regression can be applied. By equation
(\ref{eq: lasso_beta}), we have that
\begin{equation}
\bm{\widetilde{\beta}}_{i}
=\underset{\bm{\beta}_{i}\in\mathbb{R}^{p},(\bm{\beta}_{i})_{j}=0 (\forall j\in U^{c})}{\arg\min}
\left\{ \frac{1}{2n}\norm{\bm{Y}_{i}-\bm{\widetilde{X}}_U (\bm{\beta}_{i})_U}_{2}^{2}
+ \lambda\norm{\bm{\beta}_{i}}_{1}
\right\}
\end{equation}
where $U^{c}$ denotes the complement of $U$. However, here we use
a different $\lambda$ as compared to the traditional LASSO. Normally
the $\lambda$ of LASSO is selected by cross-validation. However this
will overfit the data as discussed in the paper Zhu et al. (2018)
\cite{zhu2020high}. So here we use a modified version of the $\lambda$
selection rule and set
\begin{equation}
\lambda=\max\{\lambda_{1se},min\{\lambda:\#supp(\bm{\widetilde{\beta}}_{i})\leq20\}\}
\end{equation}
where $\lambda_{1se}$ is the $\lambda$ selected by the ``1se rule''.
The ``1se rule'' gives the most regularized model such that error
is within one standard error of the minimum error achieved by the
cross-validation (see \cite{friedman2010regularization,simon2011Regularization,tibshirani2012strong}). Therefore we can derive the the set of basis assets selected as
\begin{equation}
S_i \equiv supp(\bm{\widetilde{\beta}}_{i})
\label{eq: vol_factor_select}
\end{equation}

Next, we fit an Ordinary Least-Square (OLS) regression on the selected basis assets, to estimate $\bm{\hat{\beta}}_{i}$, the OLS estimator from
\begin{equation}
\bm{Y}_{i}=\alpha_{i}\vone_{n}+\bm{X}_{S_{i}}(\bm{\beta}_{i})_{S_{i}}+\bm{\epsilon}_{i}.
\label{eq: vol_ols_subset}
\end{equation}
Note that $supp(\bm{\hat{\beta}}_{i})\subseteq{S}_{i}$. The adjusted
$R^{2}$ is obtained from this estimation. Since we are in the OLS
regime, significance tests can be performed on $\bm{\hat{\beta}}_{i}$.
This yields the significant set of coefficients
\begin{equation}
S_{i}^{*}\equiv\{j:P_{H_{0}}(|\bm{\beta}_{i,j}|\geq|\bm{\hat{\beta}}_{i,j}|)<0.05\}\quad\text{where}\quad H_{0}:\text{True value}\,\,\bm{\beta}_{i,j}=0.
\end{equation}
Note that the significant basis asset set is a subset of the selected
basis asset set. In another words,
\begin{equation}
S_{i}^{*}\subseteq supp(\bm{\hat{\beta}}_{i})\subseteq S_{i}\subseteq\{1,2,...,p\}.
\end{equation}
Then we look at the significant basis assets for the high- and the
low-volatility portfolios separately by creating heatmaps. Each heatmap
presents the percentage of selected factors in all of the ETF sectors.

A sketch of the GIBS algorithm is shown in Table \ref{tab: vol_gibs_algo}.
Recall from the notation definition in equation \ref{eq: vol_ols_subset}
that for an index set $S\subseteq\{1,2,...,p\}$, $\bm{\w{X}}_{S}$
means the columns of the matrix $\bm{\w{X}}$ indexed by the set $S$.

\begin{table}[H]
\centering
  \begin{tabular}{|| l ||}
  \hhline{|=|}
  \\[-1.6ex]
  \textbf{The Groupwise Interpretable Basis Selection (GIBS) algorithm }  \\[1ex]
  \hhline{|=|}
  \\[-1.6ex]
  Inputs: Stocks to fit $\bm{Y}$ and basis assets $\bm{X}$.
  \\[0.6ex]
  \hline
  \\[-1.6ex]
  1. Derive $\bm{\widetilde{X}}$ using $\bm{X}$ and the Equation (\ref{eq: vol_proj1}, \ref{eq: vol_proj2}).
  \\[0.6ex]
  2. Divide the transformed basis assets $\bm{\widetilde{X}}$ into k groups $A_{1},A_{2},\cdots A_{k}$ using a
  \\[0.6ex]
  \hspace{0.4cm} financial interpretation.
  \\[0.6ex]
  3. Within each group, use prototype clustering to find prototypes $ B_i \subset A_i $.
  \\[0.6ex]
  4. Let $ B = \cup_{i = 1}^k B_i $, use prototype clustering in B to find prototypes $ U \subset B $.
  \\[0.6ex]
  5. For each stock $\bm{Y}_{i}$, use a modified version of LASSO to reduce $\bm{\widetilde{X}}_{U}$
  \\[0.6ex]
  \hspace{0.4cm} to the selected basis assets $\bm{\widetilde{X}}_{S_{i}}$
  \\[0.6ex]
  6. For each stock $ \bm{Y}_i $, fit linear regression on $ \bm{X}_{S_i} $.
  \\[0.6ex]
  \hline
  \\[-1.6ex]
  Outputs: Selected factors $ S_i $, significant factors $ S_i^*$, and coefficients in step 6.
  \\[0.6ex]
  \hhline{|=|}
  \end{tabular}
  \caption{The sketch of Groupwise Interpretable Basis Selection (GIBS) algorithm}
  \label{tab: vol_gibs_algo}
\end{table}

We repeat this estimation process for all of the 3 year rolling regression
windows ending with weeks in 2008 - 2018 (the time period we found
in Section \ref{sec: vol_anomaly} with the low-volatility anomaly). Finally,
we compare the significant basis assets selected for the high- and
low-volatility portfolios. The results are given in the following
Section \ref{sec: vol_estimation_results}. Some details can be found in \cite{jarrow2021low}.

\section{Estimation Results}
\label{sec: vol_estimation_results}

This section provides the results
from our regressions employing both the FF5 and the AMF model.

\subsection{Residual Analysis: Can FF5 explain the low-volatility anomaly?}

We first look at the time series plot of the cumulative capital from
investing in the high- and low-volatility portfolios. Figure \ref{fig: vol_caps_ex}
in Section \ref{sec: vol_anomaly} shows the cumulative capital from the
excess returns for both portfolios from 2008 to 2018. As evidenced
in these graphs, the two portfolios have different volatilities and
the low-volatility portfolio outperforms the high-volatility portfolio.

\begin{figure}[H]
\centering
\includegraphics[width=0.81\textwidth]{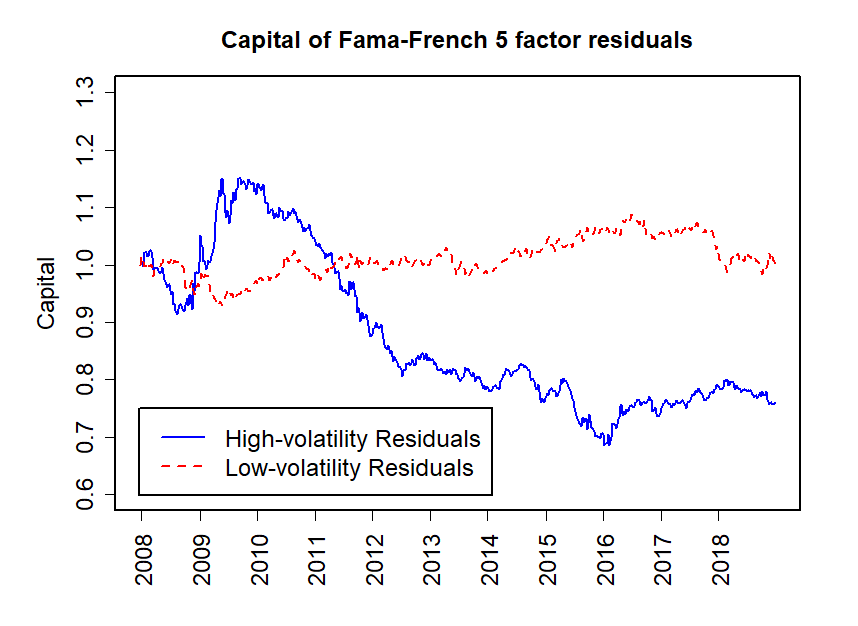}
\caption{Cumulative capital plot of the FF5 residuals of the high-and low-volatility portfolios.}
\label{fig: vol_caps_ff5_resid}
\includegraphics[width=0.81\textwidth]{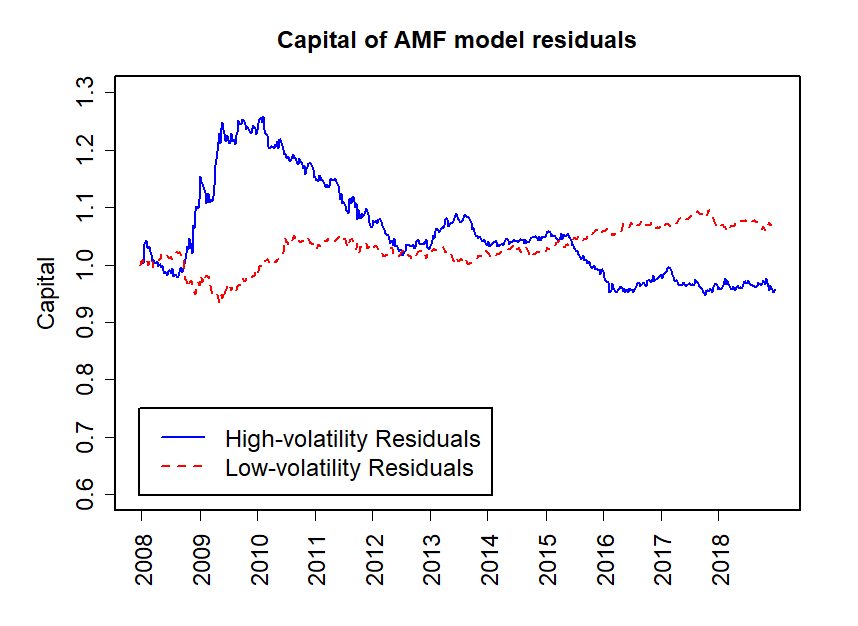}
\caption{Cumulative capital plot of the AMF residuals of the high- and low-volatility portfolios.}
\label{fig: vol_caps_gibs_resid}
\end{figure}

Next, we calculate the cumulative capital from investing in the residual
returns of the high- and the low-volatility portfolios. Figure \ref{fig: vol_caps_ff5_resid}
plots the FF5 model and Figure \ref{fig: vol_caps_gibs_resid} plots the
AMF model. Comparing Figures \ref{fig: vol_caps_ex}, \ref{fig: vol_caps_ff5_resid},
and \ref{fig: vol_caps_gibs_resid}, it is clear that the low-volatility
anomaly is more pronounced in the excess returns as compared to the
residuals. It is still obvious in the FF5 residuals, however, it almost
disappears in the AMF residuals.

Formally, we can test for the differences between the cumulative capitals
from these two portfolios. Since they have different volatilities,
we use Welch's Two-sample t-test corrected for unequal variances.
The hypotheses are
\begin{equation}
H_{0}:\mu_{l}\leq\mu_{h}\qquad H_{A}:\mu_{l}>\mu_{h}\label{eq: vol_resid_test}
\end{equation}
where $\mu_{l}$ indicates the capital of the low-volatility portfolio,
and the $\mu_{h}$ indicates the capital of the high-volatility portfolio.
We do 3 tests where the cumulative capital is calculated with the
excess returns, the residual returns from FF5, and the residual returns
from AMF. If we reject the null-hypothesis $H_{0}$, then there is
strong evidence that the low-volatility anomaly exists.

The p-values of the tests are reported in the parentheses in Table
\ref{tab: vol_resid}. In the Table \ref{tab: vol_resid}, the first row is
the excess return, the second row is the FF5 residual return, and
the third row is the AMF residual return. The first column gives the
return to the low-volatility portfolios from 2008 - 2018. The second
column is that of the high-volatility portfolios. The third column
reports the difference between the two portfolios and gives the p-values
of the tests in equation (\ref{eq: vol_resid_test}).

\begin{table}[H]
\centering
\begin{tabular}{||l|r|r|l||}
\hhline{|*{4}{=|}}
& Low Portfolio & High Portfolio & Low - High (P value) \\
\hhline{|*{4}{=|}}
Excess return & 1.21 & 0.62 & 0.59 ($1.17 \times 10^{-5}$) \\
\hline
FF5 residual return & -0.13 & -0.31 & 0.18 ($9.25 \times 10^{-95}$) \\
\hline
AMF residual return & -0.05 & -0.18 & 0.12 (1.00) \\
\hhline{|*{4}{=|}}
\end{tabular}
\caption{Residual analysis comparing the FF5 and AMF models.}
\label{tab: vol_resid}
\end{table}

The p-value of the FF5 residual is still close to 0, rejecting the
null hypothesis, thereby indicating that the low-volatility anomaly
still exists after adjusting for risk using the FF5 model. In another
words, the FF5 model cannot explain the low-volatility anomaly. However,
the p-value of the AMF residual is close to 1, which implies that
the low-volatility anomaly is not significant after adjusting for
risk with the AMF model. Thus, the AMF model explains the low-volatility
anomaly. Indeed, as we will see in Section \ref{sec: vol_sgn_assets}
below, the AMF model shows that the two portfolios load on different
basis assets (and implied risk factors). It is the (equilibrium) performance
of these factors underlying the low-volatility portfolio that generates
the low-volatility anomaly. Because the FF5 model makes the strong
assumption that every security loads on the same 5 factors, it is
not able to capture the low-volatiity risk premium differences. This
highlights the superior performance of the AMF model.

\subsection{Factor Comparisons}

\label{sec: vol_sgn_assets}

\begin{figure}
\centering
\includegraphics[height=10.5cm, width=0.8\textwidth]{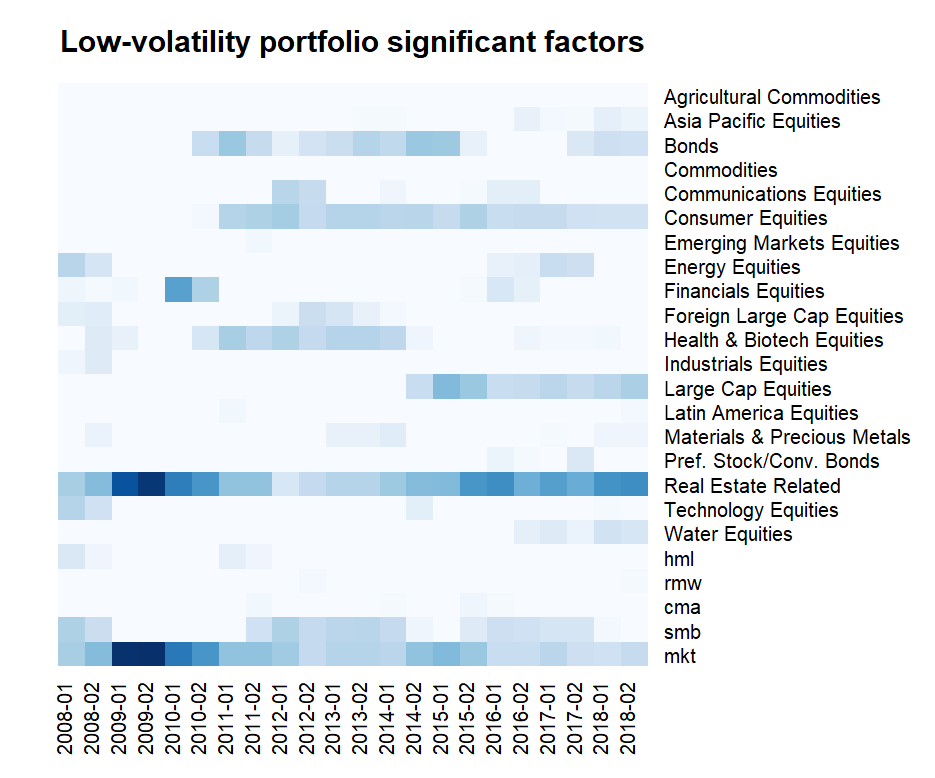} \\
\vspace{-0.2cm}
\caption{Heatmap for the low-volatility portfolio selected factors.}
\label{fig: vol_heatmap_low}
\vspace{0.1cm}
\includegraphics[height=10.5cm, width=0.8\textwidth]{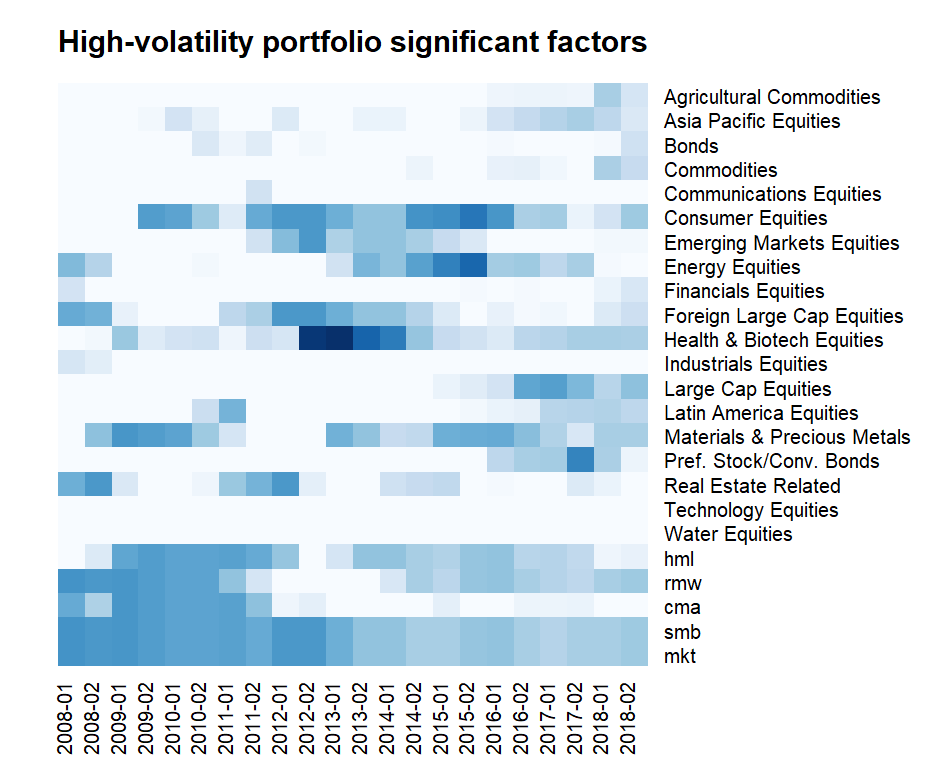} \\
\vspace{-0.2cm}
\caption{Heatmap for the high-volatility portfolio selected factors}
\label{fig: vol_heatmap_high}
\end{figure}

In this section we compare the significant basis assets or ``factors''
selected by the GIBS algorithm for the two high- and low-volatility
portfolios over 2008 - 2018. Figures \ref{fig: vol_heatmap_low} and \ref{fig: vol_heatmap_high}
show the percentage of the significant factors for each ETF class
selected by the GIBS algorithm for the low- and high-volatility portfolios
each half-year across 2008 - 2018 (see Appendix \ref{ap: etf_classes}
for more details on the ETF classifications).

As evidenced in the figures, the two portfolios load on very different
factors. The low-volatility portfolio is mainly related to the ETFs
in Bonds, Consumer Equities, and Real Estate. This is intuitive because
bonds and real estate are of lower risk. Among the FF5 factors, the
low-volatility portfolio only relates to the market return and the
SMB factor. For the high-volatility portfolio, it mainly loads on
ETFs in Materials \& Precious metals, Consumer Equities, Health \&
Biotech Equities, and all the FF5 factors, except CMA. The common
factors for both portfolios are the market return, the SMB factor,
and the Consumer Equity ETFs. Consequently, the out-performance of
low-volatility portfolio is due to the different factor loadings,
and reflects the differential performance of Bonds and Real Estate
relative to materials, precious metals, and the healthcare industry.

Next, we provide a statistical test of whether the two volatility
portfolios load on the same factors. Recalling the notation in equation
\ref{eq: vol_simplify_notation}, denote $\bm{Y}_{1}$ as the excess return
to the low-volatility portfolio, and denote $\bm{Y}_{2}$ as the excess
return to the high-volatility portfolio over the estimation period.
From equation (\ref{eq: vol_factor_select}), denote $S_{1}$ and $S_{2}$
as the basis assets selected by GIBS for the low- and high-volatility
portfolios, respectively. Then we let
\begin{equation}
\bm{Z}=\begin{pmatrix}\bm{Y}_{1}\\
\bm{Y}_{2}
\end{pmatrix}_{2n\times1},\,\bm{W}=\begin{pmatrix}\bm{X}\\
\bm{X}
\end{pmatrix}_{2n\times1},\,S=S_{1}\cup S_{2}
\end{equation}
\begin{equation}
\bm{W}_{S}=\begin{pmatrix}\bm{X}_{S}\\
\bm{X}_{S}
\end{pmatrix}_{2n\times1},\,\bm{h}=\begin{pmatrix}\bm{0}_{n\times1}\\
\vone_{n\times1}
\end{pmatrix}_{2n\times1}
\end{equation}
The vector $\bm{h}$ is an indicator vector which takes the value
1 for the high-volatility portfolio returns and 0 elsewhere.

The testing for the difference between the basis assets selected by
the two volatility portfolios can be transformed to a testing for
the significance of the interaction between $\bm{h}$ and the selected
factors $\bm{W}_{S}$. We fit two models
\begin{equation}
\text{Model 1: }\bm{Z}=\bm{W}_{S}\bm{\beta}_{S}+\bm{\epsilon}
\end{equation}
\begin{equation}
\text{Model 2: }
\bm{Z}=\bm{W}_{S}\bm{\beta}_{S}+\left[\bm{W}_{S}\odot(\bm{h}\vone'_{2n})\right]\bm{\gamma}_{S}+\bm{\epsilon}
\end{equation}
where $\odot$ means the element-wise product for two matrices with
the same dimension. Under the null hypothesis that the two portfolios
have the same coefficients on the same factors, the goodness of fit
of model 1 should be same as that of model 2. So we do an ANOVA(Model1,
Model2) test to compare the two models. The results are in Table \ref{tab: vol_diff_test}.

\begin{table}[H]
\centering
\begin{tabular}{||*{7}{r|}|}
\hhline{|*{7}{=|}}
& Res.Df & RSS & Df difference & Sum of Sq & F statistic & Pr($>$F) \\
\hhline{|*{7}{=|}}
Model 1 & 1139 & 0.24 &  &  &  &  \\
\hline
Model 2 & 1130 & 0.06 & 9 & 0.18 & 355.25 & 0.000 \\
\hhline{|*{7}{=|}}
\end{tabular}
\caption{The ANOVA test of the difference of the factors for the two portfolios.}
\label{tab: vol_diff_test}
\end{table}

The p-value of the test is approximately 0 to 3 decimal places, much
smaller than 0.05, which means that the difference between model 1
and model 2 is significant. This is a strong evidence that the two
volatility portfolios have different factor loadings, which validates
the implications from the heatmap.

\begin{figure}[H]
\centering
\includegraphics[height=10.5cm, width=0.8\textwidth]{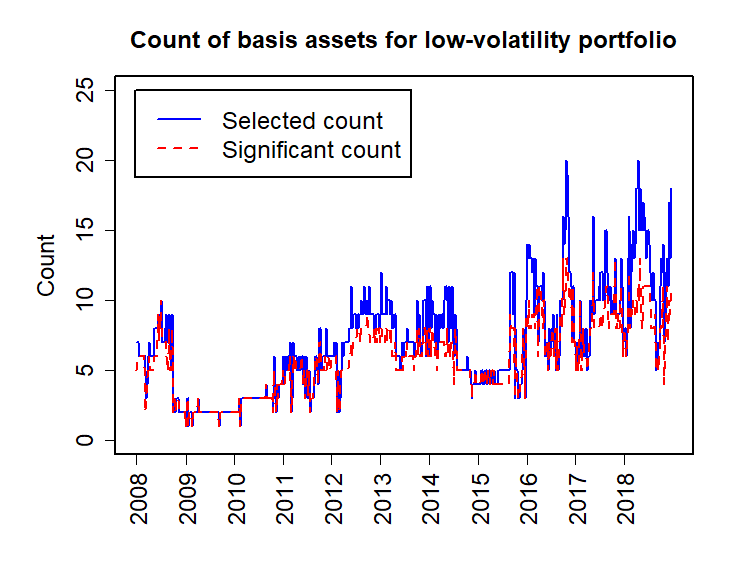} \\
\vspace{-0.2cm}
\caption{The number of selected and significant basis assets for the low-volatility portfolio.}
\label{fig: vol_count_low}
\includegraphics[height=10.5cm, width=0.8\textwidth]{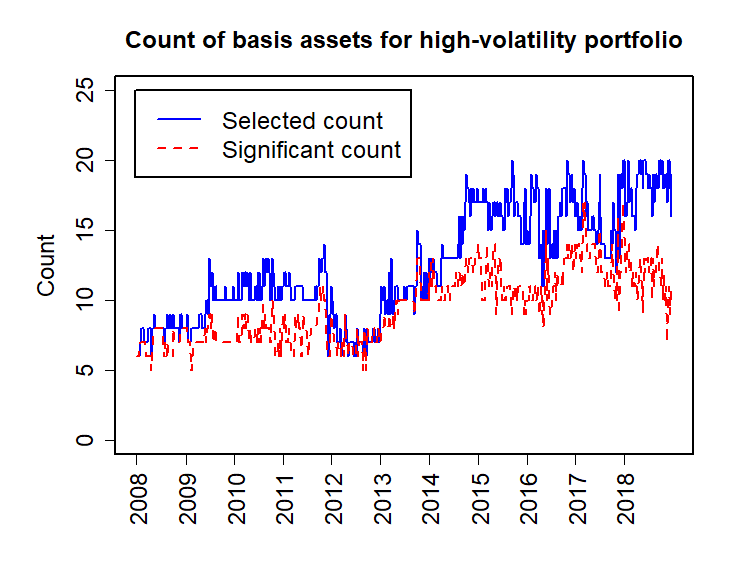} \\
\vspace{-0.2cm}
\caption{The number of the selected and significant basis assets for the high-volatility portfolio.}
\label{fig: vol_count_high}
\end{figure}

Another observation from the heatmaps is that the high-volatility
portfolio is related to more basis assets than is the low-volatility
portfolio. The number of basis assets selected by the GIBS algorithm
and the number of significant basis assets among them is tabulated
in Table \ref{tab: vol_count} and graphed in Figures \ref{fig: vol_count_low}
and \ref{fig: vol_count_high}. We see that more basis assets are selected
and significant for the high-volatility portfolio. This is intuitive
because the larger volatility is generated by the uncorrelated risks
of more basis assets in more volatile industries. More significant
basis assets come from the ETF factors than from the FF5 factors.
On average only 1.54 of the FF5 factors are significant for the low-volatility
portfolio and 3.83 of the FF5 factors are significant for the high-volatility
portfolio, manifesting a limitation of the FF5 model. Furthermore,
most of the ETFs selected by the GIBS algorithm turn out to be significant,
indicating that GIBS is more able to find relevant basis assets.

\begin{table}[ht]
\centering
\begin{tabular}{|| *{7}{r|} |}
\hhline{|*{7}{=|}}
Portfolio & Select & FF5 Select & ETF Select & Signif. & FF5 Signif. & ETF Signif. \\
\hhline{|*{7}{=|}}
Low & 6.96 & 1.72 & 5.23 & 5.82 & 1.54 & 4.29 \\
\hline
High & 12.50 & 4.12 & 8.38 & 9.53 & 3.83 & 5.70 \\
\hline
Difference & 5.54 & 2.39 & 3.15 & 3.71 & 2.29 & 1.41 \\
\hhline{|*{7}{=|}}
\end{tabular}
\caption{The number of the selected or significant basis assets / FF5 factors
/ ETFs for the two volatility portfolios. The ``Select'' column
gives the mean number of basis assets selected by the GIBS algorithm.
The ``Signif.'' column gives the mean number of significant basis
assets among the selected ones. The number of the select / significant
basis assets is the sum of the number of FF5 factors selected / significant
and the ETFs selected / significant. The row ``Low'' is the results
for the low-volatility portfolio, while the row ``High'' is for
high-volatility portfolio. The row ``Difference'' gives the differences
between two portfolios numbers using High - Low.}
\label{tab: vol_count}
\end{table}

In summary, our estimation results show that the two volatility portfolios
load on very different factors, which implies that volatility is not
an independent risk measure, but that it is related to the identified
basis assets (risk factors) in the relevant industries. It is the
(equilibrium) performance of the loaded factors that results in the
low-volatility anomaly and not disequilibrium (abnormal) returns.

\subsection{Intercept Test}

This section provides the tests for a non-zero intercept for both
the FF5 and AMF models for both volatility portfolios. This intercept
test is a test for abonormal performance, i.e. performance inconsistent
with the FF5 and AMF risk models. The null hypothesis is that the
$\alpha$'s in equation \ref{eq: vol_testable} are 0.

\begin{figure}[H]
\centering
\includegraphics[height=10.5cm, width=0.8\textwidth]{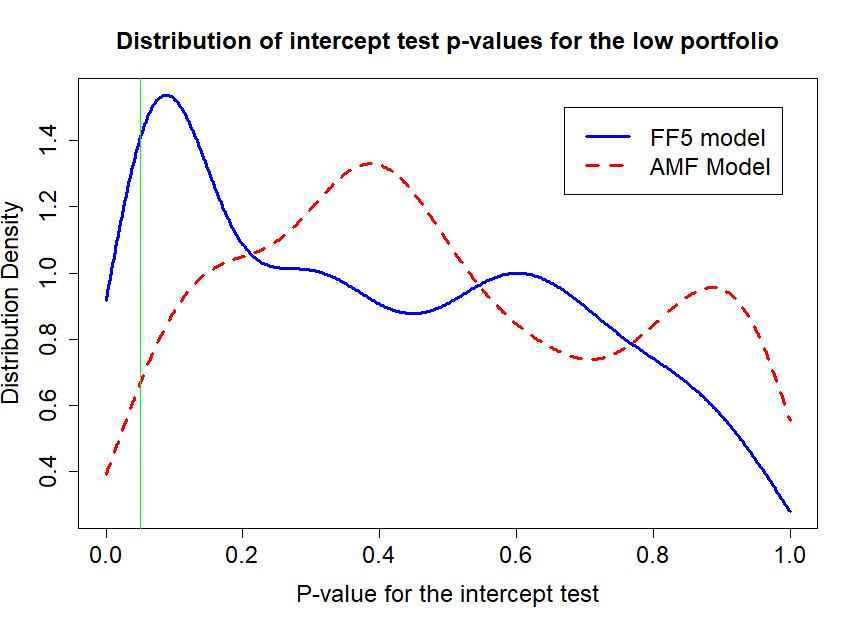} \\
\vspace{-0.2cm}
\caption{Distribution of intercept test p-values for the low-volatility portfolio.}
\label{fig: vol_int_low}
\vspace{0.1cm}
\includegraphics[height=10.5cm, width=0.8\textwidth]{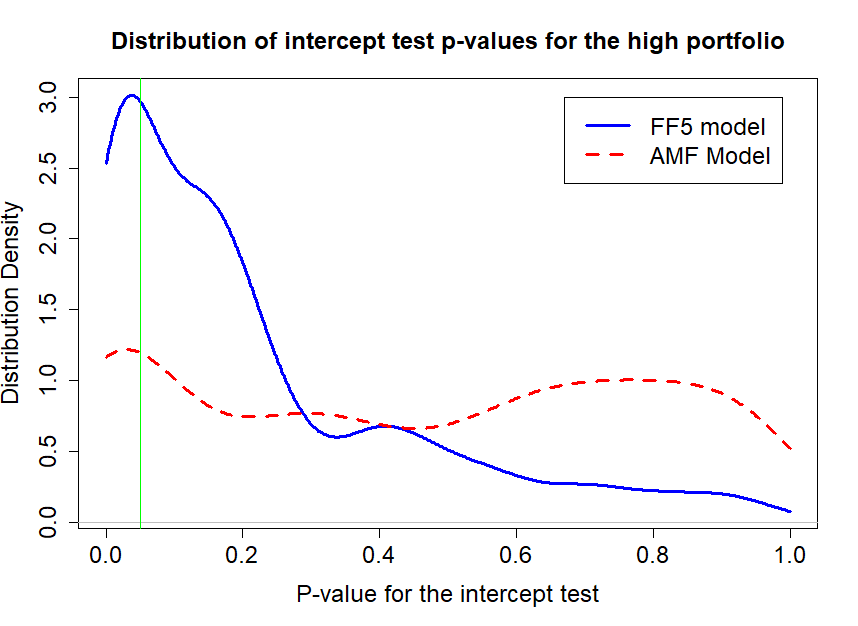} \\
\vspace{-0.2cm}
\caption{Distribution of intercept test p-values for the high-volatility portfolio.}
\label{fig: vol_int_high}
\end{figure}

Figure \ref{fig: vol_int_low} compares the distribution of the intercept
test p-values for the FF5 and AMF models for the low-volatility portfolio,
while Figure \ref{fig: vol_int_high} is for the high-volatility portfolio.
As we can see from the distribution plots, for both portfolios, the
AMF model gives much larger p-values than does the FF5 model. Indeed,
there are more weeks where a significant non-zero intercept is exhibited
with the FF5 model as compared to the AMF model. This suggests that
the AMF model is more consistent with market returns than is the FF5
model. However, this difference could be due to the fact that we repeat
this test 520 times (520 weeks in 2008 - 2018), and a percentage of
the violations will occur and be observed even if the null hypothesis
is true. Hence, it is important to control for a False Discovery Rate
(FDR).

We adjust for the false discovery rate using the Benjamini-Hochberg-Yekutieli
(BHY) procedures \cite{benjamini2001control} since it accounts for
any correlation between the intercept tests. Note that the Benjamini-Hochberg
(BH) procedure \cite{benjamini1995controlling} does not account for
correlations, and in our case, weekly returns may be correlated. After
adjusting for the false discover rate, for all weeks we fail to reject
the non-zero hypothesis for both the FF5 and AMF models. The results
are in Table \ref{tab: vol_intercept_test}. This evidence is consistent
with the low- and high-volatility portfolios exhibiting no abnormal
performance over our observation period.

\begin{table}[H]
\center %
\begin{tabular}{|| *{5}{r|} |}
\hhline{|*{5}{=}}
\multirow{2}{*}{Portfolio}  & \multicolumn{4}{c||}{Percentage of Significant Weeks}\tabularnewline
\multirow{2}{*}{}  & FF5 p$<$0.05 & AMF p$<$0.05 & FF5 FDR q$<$0.05 & AMF FDR q$<$0.05 \\
\hhline{|*{5}{=|}}
High & 30.1\% & 20.6\% & 0.00\% & 0.00\% \\
\hline
Low & 6.3\% & 2.6\% & 0.00\% & 0.00\% \\
\hhline{|*{5}{=|}}
\end{tabular}
\caption{Intercept Test with control of the False Discovery Rate. The first
row is for the high-volatility portfolio, while the second row is
for the low-volatility portfolio. The columns are related to p-values
and False Discovery Rate (FDR) q-values for the FF5 model and the
AMF model. For each column, we listed the percentage of weeks with
significant non-zero intercept out of all the weeks in the 2008 -
2018 time period.}
\label{tab: vol_intercept_test}
\end{table}

\subsection{In- and Out-of-Sample Goodness-of-fit Tests}

This section compares both the In- and the Out-of-Sample Goodness-of-Fit
tests for both the FF5 and the AMF model. For the In-Sample Goodness-of-Fit
test, for each volatility portfolio, we record the adjusted $R^{2}$'s
(see \cite{theil1961economic}) for both models for each rolling regression.
Then, we calculate the mean of the adjusted $R^{2}$'s. The results
are in the Table \ref{tab: vol_in_sample_gf}.

As shown in this table, even though the FF5 model does a good job
in fitting the data, the AMF model is able to increase the adjusted
$R^{2}$'s. We fit an ANOVA test comparing the FF5 model to the model
using all the basis assets selected by GIBS and FF5. For all the rolling-window
regressions over 2008 - 2018, the p-value of this ANOVA test is close
to 0, less than 0.05. In another words, for all the weeks over 2008
- 2018, the AMF model has a significantly better fit than does the
FF5 model, for both the low- and high-volatility portfolios. Since
we do the tests 520 times (520 weeks in 2008 - 2018), we again need
to adjust for the False Discovery Rate (FDR). However, even after
adjusting the false discovery rate using the most strict BHY method
accounting for the correlation between weeks, the FDR q-value is still
smaller than 0.05 for all of the weeks. This is a strong evidence
that the AMF out-performs the FF5.

\begin{table}[H]
\centering
\begin{adjustbox}{width={0.95\textwidth}, totalheight={0.95\textheight},keepaspectratio}
\begin{tabular}{|| *{5}{r|} |}
\hhline{|*{5}{=|}}
Portfolio & FF5 Adj. $R^2$ & AMF Adj. $R^2$ (change) & $p<0.05$ Ratio & FDR $q<0.05$ Ratio \\
\hhline{|*{5}{=|}}
Low & 0.905 & 0.961 (+6.18\%) & 100\% & 100\% \\
\hline
High & 0.931 & 0.973 (+4.56\%) & 100\% & 100\% \\
\hhline{|*{5}{=|}}
\end{tabular}
\end{adjustbox}
\caption{In-Sample Goodness-of-Fit Tests.}
\label{tab: vol_in_sample_gf}
\end{table}

For the Out-of-Sample Goodness-of-Fit test, we compare the 1-week
ahead Out-of-Sample $R^{2}$ for the FF5 and AMF models for both volatility
portfolios. The results are in Table \ref{tab: vol_out_of_sample_gf}.
It seen in the table, the AMF model out-performs the FF5 model in
the prediction period, which indicates that the better performance
of the AMF model is not due to over-fitting.

\begin{table}[H]
\centering
\begin{tabular}{|| *{3}{r|} |}
\hhline{|*{3}{=|}}
Portfolio & FF5 Out-of-Sample $R^2$ & AMF Out-of-Sample $R^2$ (change) \\
\hhline{|*{3}{=|}}
Low & 0.951 & 0.973 (+2.25\%) \\
\hline
High & 0.973 & 0.982 (+1.01\%) \\
\hhline{|*{3}{=|}}
\end{tabular}
\caption{Out-of-Sample Goodness-of-Fit Tests.}
\label{tab: vol_out_of_sample_gf}
\end{table}

\subsection{Risk Factor Determination}
\label{sec: vol_risk_factor}

The AMF model selects the basis assets that best span a security's
return. Some of these basis assets may not earn risk premium, and
therefore, would not contribute to the risk premium earned by the
volatility portfolios. This section studies which of the ETF basis
assets correspond to risk-factors, i.e. factors that have non-zero
expected excess returns.

At the end of our estimation period 2008 - 2018, there are 335 ETFs
in the universe. Among them, the GIBS algorithm selects 35 ETF representatives
in total. The list of these 35 ETF representatives is contained in
Appendix \ref{ap: etf_representatives} Table \ref{tab: vol_etf_representatives}. Therefore, there are $35+5=40$ basis assets considering both the FF5 factors and ETFs together. In
another words, $p_{1}=335$ and $p_{2}=40$ in Section \ref{sec: vol_amf_analysis}.
For each of these basis assets we compute the sample mean excess return
over our sample period. This is an estimate of the basis assets' risk
premium.

The risk premium estimates for the Fama-French 5 factors are shown
in Table \ref{tab: vol_ff5_risk_premium}. All of these are non-zero.

\begin{table}[H]
\centering
\begin{tabular}{|| *{6}{r|} |}
\hhline{|*{6}{=|}}
Fama-French 5 Factors & Market Return & SMB & HML & RMW & CMAP \\
\hhline{|*{6}{=|}}
Annual Excess Return (\%) & 13.2 & -0.7 & -1.7 & 0.9 & -1.7 \\
\hhline{|*{6}{=|}}
\end{tabular}
\caption{Risk Premium of Fama-French 5 factors}
\label{tab: vol_ff5_risk_premium}
\end{table}

The estimated risk premium for the ETFs are in Table \ref{tab: vol_etf_risk_premium}.
Out of the 35 ETF representatives selected by the GIBS algorithm,
23 of them have absolute risk premium larger than the minimum of that
of the FF5 factors (which is the RMW with absolute risk premium 0.9\%).
This means that at least $(23+5)/(35+5)=70\%$ basis assets are risk
factors in the traditional sense. The list of the 23 ETFs that earns
nonn-zero risk premium are listed in Table \ref{tab: vol_etf_risk_premium}.

\begin{table}[H]
%\resizebox{6in}{!}{
\centering
\begin{adjustbox}{width={0.95\textwidth}, totalheight={0.95\textheight},keepaspectratio}
\begin{tabular}{||l|l|r||}
\hhline{|=|=|=|}
& & Risk Pre-  \hspace{0.05cm}  \\
ETF name & Category & mium (\%) \\
\hhline{|=|=|=|}
iShares MSCI Brazil ETF & Latin America Equities & 44.0 \\
\hline
VanEck Vectors Russia ETF & Emperging Markets Equities & 20.6 \\
\hline
iShares Mortgage Real Estate ETF & Real Estate & 14.2 \\
\hline
Invesco Water Resources ETF & Water Equities & 13.9 \\
\hline
Vanguard Financials ETF & Financials Equities & 13.2 \\
\hline
Vanguard FTSE Emerging Markets ETF & Emerging Markets Equities & 12.3 \\
\hline
VanEck Vectors Agribusiness ETF & Large Cap Blend Equities & 12.3 \\
\hline
Industrial Select Sector SPDR Fund & Industrials Equities & 12.1 \\
\hline
iShares Select Dividend ETF & Large Cap Value Equities & 11.5 \\
\hline
Materials Select Sector SPDR ETF & Materials & 10.9 \\
\hline
Vanguard Healthcare ETF & Health \& Biotech Equities & 10.2 \\
\hline
iShares U.S. Home Construction ETF & Building \& Construction & 9.5 \\
\hline
iShares MSCI Canada ETF & Foreign Large Cap Equities & 9.4 \\
\hline
SPDR Barclays Capital Convertible Bond ETF & Preferred Stock/Convertible Bonds & 9.2 \\
\hline
First Trust Amex Biotechnology Index & Health \& Biotech Equities & 8.1 \\
\hline
Vanguard FTSE All-World ex-US ETF & Foreign Large Cap Equities & 7.5 \\
\hline
SPDR Barclays High Yield Bond ETF & High Yield Bonds & 6.1 \\
\hline
Invesco DB Agriculture Fund & Agricultural Commodities & -5.9 \\
\hline
iShares MSCI Japan ETF & Japan Equities & 5.8 \\
\hline
iShares MSCI Malaysia ETF & Asia Pacific Equities & 5.7 \\
\hline
Invesco DB Commodity Index Tracking Fund & Commodities & 4.9 \\
\hline
iShares Gold Trust & Precious Metals & 4.2 \\
\hline
Vanguard Consumer Staples ETF & Consumer Staples Equities & 3.8 \\
\hhline{|=|=|=|}
\end{tabular}
\end{adjustbox}
\caption{List of ETFs with large absolute risk premium.}
\label{tab: vol_etf_risk_premium}
\end{table}

\section{Conclusion}
\label{sec: vol_conclude}

In this paper we construct high- and low-volatility portfolios within
the investable universe to explain the low-volatility anomaly using
a new model, the Adaptive Multi-Factor (AMF) model estimated by the
Groupwise Interpretable Basis Selection (GIBS) algorithm proposed
in the paper by Zhu et al. (2018) \cite{zhu2020high}. For comparison
with the literature, we compare the AMF model with the traditional
Fama-French 5-factor (FF5) model. Our estimation shows the superior
performance of the AMF model over FF5. Indeed, we show that the FF5
cannot explain the low-volatility anomaly while the AMF can. The AMF
results show that the two volatility portfolios load on very different
factors, which indicates that the volatility is not an independent
measure of risk. It is the performance of the underlying risk factors
that results in the low-volatility anomaly. Alternatively stated,
the out-performance of the low-volatility portfolio reflects the equilibrium
compensation for the risk of its underlying risk factors.

%\bibliography{bib_liao}

\bibliographystyle{plain}
\bibliography{bib_liao}

\begin{appendices}

\clearpage
\chapter{Summary of Generalized APT}
\label{ap: summary_apt}
This appendix provides a summary of the key results
from the generalized APT contained in Jarrow and Protter (2016) \cite{jarrow2016positive}
that are relevant to the empirical estimation and different from the
traditional approach to the estimation of factor models. Prior to
that, however, we first discuss the traditional approach to the estimation
of factor models.

The traditional approach to the estimation of factor models, based
on Merton (1973) \cite{merton1973intertemporal} and Ross (1976) \cite{ross1976arbitrage},
starts with an expected return relation between a risky asset and
a finite collection of risk-factors. Merton's relation is derived
from the first order conditions of an investor's optimization problem,
given in expectations. Ross's relation is derived from a limiting
arbitrage pricing condition satisfied by the asset's expected returns.
Given these are expected return relations, the included risk factor's
excess expected returns are all non-zero by construction. These non-zero
expected excess returns are interpreted as risk premiums earned for
systematic risk exhibited by the risk factors. Jensen's alpha is then
viewed as a wedge between the risky assets expected return and those
of the risk factors implying a mispricing in the market.

To get an empirical relation in realized returns for estimation, the
traditional approach rewrites these risk-factor expected returns as
a realized return less an error. Substitution yields an expression
relating a risky asset's realized return to the realized return's
of the risk factors plus a cumulative error. Since the risk factors
do not reflect idiosyncratic risk, the cumulative error in this relation
may not satisfy the standard assumptions needed for a regression model.
In particular, the error term may be autocorrelated and/or correlated
with other idiosyncratic risk terms not included in the estimated
equation. Because the estimated relation is derived from a relation
involving expectations using risk-factors, the $R^2$ of the realized return regression can be small.

Jarrow and Protter (2016) derive a testable multi-factor model in
a different and more general setting. With trading in a finite number
for risky assets in the context of a continuous time, continuous trading
market assuming only frictionless and competitive markets that satisfy
no-arbitrage and no dominance, i. e. the existence of an equivalent
martingale measure, they are able to derive a no-arbitrage condition
satsifed by any trading strategy's realized returns. Adding additional
structure to the economy and then taking expectations yields both
Merton's (1973) \cite{merton1973intertemporal} and Ross's (1976) \cite{ross1976arbitrage}
models as special cases.

The generalize APT uses linear algebra to prove the existence of an
algebraic basis in the risky asset's payoff space at some future time
T. Since this is is a continuous time and trading economy, this payoff
space is infinite dimensional. The algebraic basis at time T constitutes
the collection of basis assets. It is important to note that this
set of basis assets are tradable. An algebraic basis means that any
risky asset's return can be written as a linear combination of a \emph{finite}
number of the basis asset returns, and different risky assets may
have a \emph{different finite }combination of basis assets explaining
their returns. Since the space of random variables generated by the
admissible trading strategies is infinite dimensional, this algebraic
basis representation of the relevant risks is parsimonious and sparse.
Indeed, only a finite number of basis assets in an infinite dimensional
space explains any trading strategy's return process.

No arbitrage, the martingale relation, implies that the same set of
basis assets explains any particular risky asset's realized returns
at all earlier times $t\in[0,T]$ as well. This is the no arbitrage
relation satisfied by the realized return processes. Adding a non-zero
alpha to this relation (Jensen's alpha) implies a violation of the
no-arbitrage condition. This no-arbitrage condition is valid and testable
in realized return space.

In contrast to the traditional approach (discussed above), deriving
the relation in realized return space implies that the error structure
is more likely to satisfy the standard assumptions of regression models.
This follows because including basis assets with zero excess expected
returns (which are excluded in the traditional approach based on risk-factors)
will reduce correlations between the error terms across time and cross-sectionally
between the error terms and the other basis assets (independent variables).
The result is the estimated equation should have a larger $R^{2}$.
After fitting the multi-factor model in realized returns, estimating
the relevant basis assets expectations determines which are risk-factors,
i.e. which earn a risk premium for systematic risk.

\clearpage
\chapter{Company Classes by SIC Code}
\label{sec: amf_comp_classes}

The following list of company classes based on the first two digits of Standard Industrial Classification (SIC) code is from the website of United States Department of Labor
(\url{https://www.osha.gov/pls/imis/sic_manual.html}):

{\small
\begin{longtable}{|| m{0.18\textwidth} | m{0.72\textwidth} ||}

\endhead
\multicolumn{2}{c}{{\footnotesize{} Continued on the next page }}  \\
\endfoot
\endlastfoot

\multicolumn{2}{c}{} \\
\hhline{|=|=|}
Division A  &  Agriculture, Forestry, And Fishing\\
\hhline{|=|=|}
Major Group 01 &  Agricultural Production Crops\\
\hline
Major Group 02 &  Agriculture production livestock and animal specialties\\
\hline
Major Group 07 &  Agricultural Services\\
\hline
Major Group 08 &  Forestry\\
\hline
Major Group 09 &  Fishing, hunting, and trapping\\
\hhline{|=|=|}

\multicolumn{2}{c}{} \\
\hhline{|=|=|}
Division B &  Mining\\
\hhline{|=|=|}
Major Group 10 &  Metal Mining\\
\hline
Major Group 12 &  Coal Mining\\
\hline
Major Group 13 &  Oil And Gas Extraction\\
\hline
Major Group 14 &  Mining And Quarrying Of Nonmetallic Minerals, Except Fuels\\
\hhline{|=|=|}

\multicolumn{2}{c}{} \\
\hhline{|=|=|}
Division C &  Construction\\
\hhline{|=|=|}
\hline
Major Group 15 &  Building Construction General Contractors And Operative Builders \\
\hline
Major Group 16 &  Heavy Construction Other Than Building Construction Contractors \\
\hline
Major Group 17 &  Construction Special Trade Contractors\\
\hhline{|=|=|}

\multicolumn{2}{c}{} \\
\hhline{|=|=|}
Division D &  Manufacturing\\
\hhline{|=|=|}
Major Group 20 &  Food And Kindred Products\\
\hline
Major Group 21 &  Tobacco Products\\
\hline
Major Group 22 &  Textile Mill Products\\
\hline
Major Group 23 &  Apparel And Other Finished Products Made From Fabrics And Similar Materials\\
\hline
Major Group 24 &  Lumber And Wood Products, Except Furniture\\
\hline
Major Group 25 &  Furniture And Fixtures\\
\hline
Major Group 26 &  Paper And Allied Products\\
\hline
Major Group 27 &  Printing, Publishing, And Allied Industries\\
\hline
Major Group 28 &  Chemicals And Allied Products\\
\hline
Major Group 29 &  Petroleum Refining And Related Industries\\
\hline
Major Group 30 &  Rubber And Miscellaneous Plastics Products\\
\hline
Major Group 31 &  Leather And Leather Products\\
\hline
Major Group 32 &  Stone, Clay, Glass, And Concrete Products\\
\hline
Major Group 33 &  Primary Metal Industries\\
\hline
Major Group 34 &  Fabricated Metal Products, Except Machinery And \hspace{2cm} Transportation Equipment\\
\hline
Major Group 35 &  Industrial And Commercial Machinery And Computer \hspace{3cm} Equipment\\
\hline
Major Group 36 &  Electronic And Other Electrical Equipment And Components,
Except Computer Equipment\\
\hline
Major Group 37 &  Transportation Equipment\\
\hline
Major Group 38 &  Measuring, Analyzing, And Controlling Instruments;
Photographic, Medical And Optical Goods; Watches And Clocks\\
\hline
Major Group 39 &  Miscellaneous Manufacturing Industries\\
\hhline{|=|=|}

\multicolumn{2}{c}{} \\
\hhline{|=|=|}
Division E  &  Transportation, Communications, Electric, Gas, And Sanitary
Services\\
\hhline{|=|=|}
Major Group 40 &  Railroad Transportation\\ \hline
Major Group 41 &  Local And Suburban Transit And Interurban Highway
Passenger Transportation\\ \hline
Major Group 42 &  Motor Freight Transportation And Warehousing\\ \hline
Major Group 43 &  United States Postal Service\\ \hline
Major Group 44 &  Water Transportation\\ \hline
Major Group 45 &  Transportation By Air\\ \hline
Major Group 46 &  Pipelines, Except Natural Gas\\ \hline
Major Group 47 &  Transportation Services\\ \hline
Major Group 48 &  Communications\\ \hline
Major Group 49 &  Electric, Gas, And Sanitary Services\\
\hhline{|=|=|}

\multicolumn{2}{c}{} \\
\hhline{|=|=|}
Division F &  Wholesale Trade\\
\hhline{|=|=|}
Major Group 50 &  Wholesale Trade-durable Goods\\ \hline
Major Group 51 &  Wholesale Trade-non-durable Goods\\
\hhline{|=|=|}

\multicolumn{2}{c}{} \\
\hhline{|=|=|}
Division G &  Retail Trade\\
\hhline{|=|=|}
Major Group 52 &  Building Materials, Hardware, Garden Supply, And
Mobile Home Dealers\\ \hline
Major Group 53 &  General Merchandise Stores\\ \hline
Major Group 54 &  Food Stores\\ \hline
Major Group 55 &  Automotive Dealers And Gasoline Service Stations\\ \hline
Major Group 56 &  Apparel And Accessory Stores\\ \hline
Major Group 57 &  Home Furniture, Furnishings, And Equipment Stores\\ \hline
Major Group 58 &  Eating And Drinking Places\\ \hline
Major Group 59 &  Miscellaneous Retail\\
\hhline{|=|=|}

\multicolumn{2}{c}{} \\
\hhline{|=|=|}
Division H &  Finance, Insurance, And Real Estate\\
\hhline{|=|=|}
Major Group 60 &  Depository Institutions\\ \hline
Major Group 61 &  Non-depository Credit Institutions\\ \hline
Major Group 62 &  Security And Commodity Brokers, Dealers, Exchanges,
And Services\\ \hline
Major Group 63 &  Insurance Carriers\\ \hline
Major Group 64 &  Insurance Agents, Brokers, And Service\\ \hline
Major Group 65 &  Real Estate\\ \hline
Major Group 67 &  Holding And Other Investment Offices\\
\hhline{|=|=|}

\multicolumn{2}{c}{} \\
\hhline{|=|=|}
Division I &  Services\\
\hhline{|=|=|}
Major Group 70 &  Hotels, Rooming Houses, Camps, And Other Lodging
Places\\ \hline
Major Group 72 &  Personal Services\\ \hline
Major Group 73 &  Business Services\\ \hline
Major Group 75 &  Automotive Repair, Services, And Parking\\ \hline
Major Group 76 &  Miscellaneous Repair Services\\ \hline
Major Group 78 &  Motion Pictures\\ \hline
Major Group 79 &  Amusement And Recreation Services\\ \hline
Major Group 80 &  Health Services\\ \hline
Major Group 81 &  Legal Services\\ \hline
Major Group 82 &  Educational Services\\ \hline
Major Group 83 &  Social Services\\ \hline
Major Group 84 &  Museums, Art Galleries, And Botanical And Zoological
Gardens\\ \hline
Major Group 86 &  Membership Organizations\\ \hline
Major Group 87 &  Engineering, Accounting, Research, Management, And
Related Services\\ \hline
Major Group 88 &  Private Households\\ \hline
Major Group 89 &  Miscellaneous Services\\
\hhline{|=|=|}

\multicolumn{2}{c}{} \\
\hhline{|=|=|}
Division J &  Public Administration\\
\hhline{|=|=|}
Major Group 91 &  Executive, Legislative, And General Government, Except
Finance\\ \hline
Major Group 92 &  Justice, Public Order, And Safety\\ \hline
Major Group 93 &  Public Finance, Taxation, And Monetary Policy\\ \hline
Major Group 94 &  Administration Of Human Resource Programs\\ \hline
Major Group 95 &  Administration Of Environmental Quality And Housing \hspace{1cm} pPrograms \\ \hline
Major Group 96 &  Administration Of Economic Programs\\ \hline
Major Group 97 &  National Security And International Affairs\\ \hline
Major Group 99 &  Nonclassifiable Establishments \\
\hhline{|=|=|}
\caption{Company classes by the Standard Industrial Classification (SIC) code}
\end{longtable}

}

\clearpage
\chapter{ETF Classes and Subclasses}
\label{ap: etf_classes}

ETFs can be divided into 10 classes, 73 subclasses (categories) in total, based on their financial explanations. The classify criteria are found from the ETFdb database: www.etfdb.com. The classes and subclasses are listed below:

\begin{enumerate}
\item \textbf{Bond/Fixed Income}: California Munis, Corporate Bonds, Emerging
Markets Bonds, Government Bonds, High Yield Bonds, Inflation-Protected
Bonds, International Government Bonds, Money Market, Mortgage Backed
Securities, National Munis, New York Munis, Preferred Stock/Convertible
Bonds, Total Bond Market.

\item \textbf{Commodity}: Agricultural Commodities, Commodities, Metals,
Oil \& Gas, Precious Metals.

\item \textbf{Currency}: Currency.

\item \textbf{Diversified Portfolio}: Diversified Portfolio, Target Retirement
Date.

\item \textbf{Equity}: All Cap Equities, Alternative Energy Equities, Asia
Pacific Equities, Building \& Construction, China Equities, Commodity
Producers Equities, Communications Equities, Consumer Discretionary
Equities, Consumer Staples Equities, Emerging Markets Equities, Energy
Equities, Europe Equities, Financial Equities, Foreign Large Cap Equities,
Foreign Small \& Mid Cap Equities, Global Equities, Health \& Biotech
Equities, Industrials Equities, Japan Equities, Large Cap Blend Equities,
Large Cap Growth Equities, Large Cap Value Equities, Latin America
Equities, MLPs (Master Limited Partnerships), Materials, Mid Cap Blend
Equities, Mid Cap Growth Equities, Mid Cap Value Equities, Small Cap
Blend Equities, Small Cap Growth Equities, Small Cap Value Equities,
Technology Equities, Transportation Equities, Utilities Equities,
Volatility Hedged Equity, Water Equities.

\item \textbf{Alternative ETFs}: Hedge Fund, Long-Short.

\item \textbf{Inverse}: Inverse Bonds, Inverse Commodities, Inverse Equities,
Inverse Volatility.

\item \textbf{Leveraged}: Leveraged Bonds, Leveraged Commodities, \\
 Leveraged Currency, Leveraged Equities, Leveraged Multi-Asset, Leveraged
Real Estate, Leveraged Volati-lity.

\item \textbf{Real Estate}: Global Real Estate, Real Estate.

\item \textbf{Volatility}: Volatility.
\end{enumerate}

In Section \ref{sec: vol_sgn_assets}, we merged several categories to
give a better visualization of the significant factors for each portfolio.
The merged categories are
\begin{itemize}
\item Bonds: Corporate Bonds, Government Bonds, High Yield Bonds, Total
Bond Market, Leveraged Bonds.
\item Consumer Equities: Consumer Discretionary Equities, Consumer Staples
Equities.
\item Real Estate Related: Real Estate, Leveraged Real Estate, Global Real
Estate, Utilities Equities,"Building \& Construction.
\item Energy Equities: Energy Equities, Alternative Energy Equities.
\item Materials \& Precious Metals: Materials, Precious Metals
\item Large Cap Equities: Large Cap Blend Equities, Large Cap Growth Equities,
Large Cap Value Equities.
\end{itemize}

\clearpage
\chapter{Low-Correlated ETF Name Lists}
\label{ap: etf_representatives}

The low-correlated ETF name list in Section \ref{sec: gibs_algo} is in Table \ref{tab: amf_etf_representatives}.

{
\small
%\fontsize{1}{1.2}\selectfont

\begin{longtable}{|| m{0.61\textwidth} | m{0.31\textwidth} ||}
\hhline{|=|=|}
ETF Names  & Category  \\
\hhline{|=|=|}
\endhead
\hline
\multicolumn{2}{c}{{\footnotesize{}Continued on next page}}  \\
\endfoot
\endlastfoot
iShares California Muni Bond ETF & California Munis \\
\hline
iShares Emerging Markets Corporate Bond ETF & Corporate Bonds \\
\hline
FlexShares Ready Access Variable Income Fund & Corporate Bonds \\
\hline
Invesco International Corporate Bond ETF & Corporate Bonds \\
\hline
WisdomTree Emerging Markets Corporate Bond Fund & Corporate Bonds \\
\hline
iShares Floating Rate Bond ETF & Corporate Bonds \\
\hline
ProShares Investment Grade-Interest Rate Hedged & Corporate Bonds \\
\hline
iShares iBonds Mar 2020 Corporate ETF & Corporate Bonds \\
\hline
VanEck Vectors Investment Grade Floating Rate ETF & Corporate Bonds \\
\hline
SPDR Barclays Capital Investment Grade Floating Rate ETF & Corporate Bonds \\
\hline
iShares iBonds Mar 2020 Corporate ex-Financials ETF & Corporate Bonds \\
\hline
Vanguard Emerging Markets Government Bond ETF & Emerging Markets Bonds \\
\hline
ProShares Short Term USD Emerging Markets Bond ETF & Emerging Markets Bonds \\
\hline
SPDR Barclays 1-3 Month T-Bill ETF & Government Bonds \\
\hline
iShares Short Treasury Bond ETF & Government Bonds \\
\hline
SPDR Portfolio Short Term Treasury ETF & Government Bonds \\
\hline
SPDR BofA Merrill Lynch Crossover Corporate Bond ETF & High Yield Bonds \\
\hline
VanEck Vectors International High Yield Bond ETF & High Yield Bonds \\
\hline
SPDR Blackstone/ GSO Senior Loan ETF & High Yield Bonds \\
\hline
Highland iBoxx Senior Loan ETF & High Yield Bonds \\
\hline
Invesco Global Short Term High Yield Bond ETF & High Yield Bonds \\
\hline
WisdomTree Interest Rate Hedged High Yield Bond Fund & High Yield Bonds \\
\hline
First Trust Senior Loan Exchange-Traded Fund & High Yield Bonds \\
\hline
PIMCO 0-5 Year High Yield Corporate Bond Index Fund & High Yield Bonds \\
\hline
WisdomTree Negative Duration High Yield Bond Fund & High Yield Bonds \\
\hline
ProShares Inflation Expectations ETF & Inflation-Protected Bonds \\
\hline
WisdomTree Asia Local Debt Fund & International Government Bonds \\
\hline
iShares Ultra Short-Term Bond ETF & Money Market \\
\hline
VanEck Vectors AMT-Free Short Municipal Index ETF & National Munis \\
\hline
Invesco VRDO Tax-Free Weekly ETF & National Munis \\
\hline
Pimco Short Term Municipal Bond Fund & National Munis \\
\hline
VanEck Vectors AMT-Free Intermediate Municipal Index ETF & National Munis \\
\hline
VanEck Vectors Pre-Refunded Municipal Index ETF & National Munis \\
\hline
iShares S\&P Short Term AMT-Free Bond ETF & National Munis \\
\hline
SPDR Barclays Short Term Municipal Bond & National Munis \\
\hline
SPDR Barclays Capital Convertible Bond ETF & Preferred Stock or \hspace{2cm} Convertible Bonds \\
\hline
iShares International Preferred Stock ETF & Preferred Stock or \hspace{2cm} Convertible Bonds \\
\hline
iShares U.S. Preferred Stock ETF & Preferred Stock or \hspace{2cm} Convertible Bonds \\
\hline
iShares Short Maturity Bond ETF & Total Bond Market \\
\hline
Invesco Chinese Yuan Dim Sum Bond ETF & Total Bond Market \\
\hline
Franklin Short Duration U.S. Government ETF & Total Bond Market \\
\hline
Invesco CEF Income Composite ETF & Total Bond Market \\
\hline
AdvisorShares Newfleet Multi-Sector Income ETF & Total Bond Market \\
\hline
SPDR SSgA Ultra Short Term Bond ETF & Total Bond Market \\
\hline
WisdomTree Barclays Interest Rate Hedged U.S. Aggregate Bond Fund & Total Bond Market \\
\hline
WisdomTree Barclays Negative Duration U.S. Aggregate Bond Fund & Total Bond Market \\
\hline
PIMCO Enhanced Short Maturity Strategy Fund & Total Bond Market \\
\hline
Invesco DB Agriculture Fund & Agricultural Commodities \\
\hline
Invesco DB Base Metals Fund & Metals \\
\hline
Invesco DB Oil Fund & Oil \& Gas \\
\hline
Aberdeen Standard Physical Palladium Shares ETF & Precious Metals \\
\hline
Invesco DB Precious Metals Fund & Precious Metals \\
\hline
Invesco CurrencyShares Swiss Franc Trust & Currency \\
\hline
Invesco CurrencyShares Canadian Dollar Trust & Currency \\
\hline
WisdomTree Brazilian Real Fund & Currency \\
\hline
Invesco DB G10 Currency Harvest Fund & Currency \\
\hline
First Trust Dorsey Wright People's Portfolio ETF & Diversified Portfolio \\
\hline
Arrow Dow Jones Global Yield ETF & Diversified Portfolio \\
\hline
First Trust Multi-Asset Diversified Income Index Fund & Diversified Portfolio \\
\hline
iShares Moderate Allocation ETF & Diversified Portfolio \\
\hline
Renaissance IPO ETF & All Cap Equities \\
\hline
Invesco Dynamic Leisure and Entertainment ETF & All Cap Equities \\
\hline
VanEck Vectors Israel ETF & All Cap Equities \\
\hline
Invesco Dynamic Media ETF & All Cap Equities \\
\hline
Invesco Cleantech ETF & Alternative Energy Equities \\
\hline
First Trust ISE Global Wind Energy Index Fund & Alternative Energy Equities \\
\hline
VanEck Vectors Global Alternative Energy ETF & Alternative Energy Equities \\
\hline
First Trust NASDAQ Clean Edge Smart Grid Infrastructure Index Fund & Alternative Energy Equities \\
\hline
Invesco WilderHill Progressive Energy ETF & Alternative Energy Equities \\
\hline
WisdomTree India Earnings Fund & Asia Pacific Equities \\
\hline
Vanguard FTSE Pacific ETF & Asia Pacific Equities \\
\hline
First Trust India NIFTY 50 Equal Weight ETF & Asia Pacific Equities \\
\hline
WisdomTree Australia Dividend Fund & Asia Pacific Equities \\
\hline
iShares MSCI Thailand ETF & Asia Pacific Equities \\
\hline
VanEck Vectors Vietnam ETF & Asia Pacific Equities \\
\hline
iShares MSCI Philippines ETF & Asia Pacific Equities \\
\hline
First Trust ISE Chindia Index Fund & Asia Pacific Equities \\
\hline
iShares MSCI China Small-Cap ETF & Asia Pacific Equities \\
\hline
iShares MSCI New Zealand ETF & Asia Pacific Equities \\
\hline
First Trust ISE Global Engineering and Construction ETF & Building \& Construction \\
\hline
SPDR S\&P Homebuilders ETF & Building \& Construction \\
\hline
Invesco Dynamic Building \& Construction ETF & Building \& Construction \\
\hline
VanEck Vectors ChinaAMC CSI 300 ETF & China Equities \\
\hline
KraneShares CSI China Five Year Plan ETF & China Equities \\
\hline
Invesco Global Agriculture ETF & Commodity Producers \hspace{0.7cm} Equities \\
\hline
iShares North American Tech-Multimedia Network \hspace{1cm} pETF & Communications Equities \\
\hline
iShares U.S. Telecommunications ETF & Communications Equities \\
\hline
First Trust NASDAQ Global Auto Index Fund & Consumer Discretionary \hspace{0.5cm} Equities \\
\hline
VanEck Vectors Gaming ETF & Consumer Discretionary \hspace{0.5cm} Equities \\
\hline
SPDR S\&P Retail ETF & Consumer Discretionary \hspace{0.5cm} Equities \\
\hline
Invesco S\&P SmallCap Consumer Staples ETF & Consumer Staples Equities \\
\hline
IQ Global Agribusiness Small Cap ETF & Consumer Staples Equities \\
\hline
Vanguard Consumer Staples ETF & Consumer Staples Equities \\
\hline
iShares MSCI Frontier 100 ETF & Emerging Markets Equities \\
\hline
VanEck Vectors Russia Small-Cap ETF & Emerging Markets Equities \\
\hline
Global X FTSE Greece 20 ETF & Emerging Markets Equities \\
\hline
WisdomTree Middle East Dividend Fund & Emerging Markets Equities \\
\hline
VanEck Vectors Egypt Index ETF & Emerging Markets Equities \\
\hline
iShares MSCI Turkey ETF & Emerging Markets Equities \\
\hline
VanEck Vectors Coal ETF & Energy Equities \\
\hline
iShares MSCI Ireland ETF & Europe Equities \\
\hline
VanEck Vectors Poland ETF & Europe Equities \\
\hline
iShares MSCI United Kingdom Small-Cap ETF & Europe Equities \\
\hline
WisdomTree Europe Hedged Equity Fund & Europe Equities \\
\hline
Xtrackers MSCI United Kingdom Hedged Equity Fund & Europe Equities \\
\hline
Global X MSCI Portugal ETF & Europe Equities \\
\hline
First Trust Germany AlphaDEX Fund & Europe Equities \\
\hline
Invesco Global Listed Private Equity ETF & Financials Equities \\
\hline
ProShares Global Listed Private Equity ETF & Financials Equities \\
\hline
Invesco KBW High Dividend Yield Financial ETF & Financials Equities \\
\hline
Invesco DWA Financial Momentum ETF & Financials Equities \\
\hline
SPDR S\&P Insurance ETF & Financials Equities \\
\hline
iShares MSCI EAFE ETF & Foreign Large Cap Equities \\
\hline
VanEck Vectors Africa Index ETF & Foreign Large Cap Equities \\
\hline
First Trust S\&P International Dividend Aristocrats ETF & Foreign Large Cap Equities \\
\hline
Invesco S\&P International Developed Momentum ETF & Foreign Large Cap Equities \\
\hline
Global X MSCI Nigeria ETF & Foreign Large Cap Equities \\
\hline
Global X MSCI Argentina ETF & Global Equities \\
\hline
iShares MSCI Peru ETF & Global Equities \\
\hline
ROBO Global Robotics and Automation Index ETF & Global Equities \\
\hline
Global X Uranium ETF & Global Equities \\
\hline
IQ Hedge Macro Tracker ETF & Global Equities \\
\hline
AdvisorShares Dorsey Wright ADR ETF & Global Equities \\
\hline
SPDR S\&P Health Care Services ETF & Health \& Biotech Equities \\
\hline
iShares U.S. Pharmaceuticals ETF & Health \& Biotech Equities \\
\hline
SPDR S\&P Health Care Equipment ETF & Health \& Biotech Equities \\
\hline
VanEck Vectors Environmental Services ETF & Industrials Equities \\
\hline
iShares U.S. Aerospace \& Defense ETF & Industrials Equities \\
\hline
SPDR MSCI ACWI IMI ETF & Large Cap Blend Equities \\
\hline
Invesco S\&P 500 BuyWrite ETF & Large Cap Blend Equities \\
\hline
VanEck Vectors Morningstar Wide Moat ETF & Large Cap Blend Equities \\
\hline
iShares MSCI Israel ETF & Large Cap Blend Equities \\
\hline
Global X NASDAQ 100 Covered Call ETF & Large Cap Growth Equities \\
\hline
AlphaClone Alternative Alpha ETF & Large Cap Growth Equities \\
\hline
Invesco Russell Top 200 Equal Weight ETF & Large Cap Growth Equities \\
\hline
Invesco NASDAQ Internet ETF & Large Cap Growth Equities \\
\hline
Global X MSCI Colombia ETF & Latin America Equities \\
\hline
iShares Global Timber \& Forestry ETF & Materials \\
\hline
VanEck Vectors Rare Earth/Strategic Metals ETF & Materials \\
\hline
Global X Lithium ETF & Mid Cap Blend Equities \\
\hline
Invesco Global Water ETF & Mid Cap Growth Equities \\
\hline
Invesco DWA NASDAQ Momentum ETF & Small Cap Growth Equities \\
\hline
iShares North American Tech-Software ETF & Technology Equities \\
\hline
Invesco S\&P SmallCap Information Technology ETF & Technology Equities \\
\hline
SPDR S\&P Semiconductor ETF & Technology Equities \\
\hline
First Trust NASDAQ CEA Smartphone Index Fund & Technology Equities \\
\hline
iShares Transportation Average ETF & Transportation Equities \\
\hline
Vanguard Utilities ETF & Utilities Equities \\
\hline
SPDR SSGA US Small Cap Low Volatility Index ETF & Volatility Hedged Equity \\
\hline
SPDR SSGA US Large Cap Low Volatility Index ETF & Volatility Hedged Equity \\
\hline
Invesco S\&P 500 Downside Hedged ETF & Volatility Hedged Equity \\
\hline
Invesco Water Resources ETF & Water Equities \\
\hline
First Trust ISE Water Index Fund & Water Equities \\
\hline
WisdomTree Managed Futures Strategy Fund & Hedge Fund \\
\hline
IQ Merger Arbitrage ETF & Hedge Fund \\
\hline
Proshares Merger ETF & Hedge Fund \\
\hline
SPDR SSgA Multi-Asset Real Return ETF & Hedge Fund \\
\hline
First Trust Morningstar Managed Futures Strategy Fund & Hedge Fund \\
\hline
IQ Real Return ETF & Hedge Fund \\
\hline
ProShares RAFI Long/Short & Long-Short \\
\hline
IQ Hedge Market Neutral Tracker ETF & Long-Short \\
\hline
FLAG-Forensic Accounting Long-Short ETF & Long-Short \\
\hline
AGFiQ US Market Neutral Anti-Beta Fund & Long-Short \\
\hline
AGFiQ US Market Neutral Size Fund & Long-Short \\
\hline
AGFiQ US Market Neutral Momentum Fund & Long-Short \\
\hline
ProShares Short 7-10 Year Treasury & Inverse Bonds \\
\hline
Short MSCI Emerging Markets ProShares & Inverse Equities \\
\hline
AdvisorShares Ranger Equity Bear ETF & Inverse Equities \\
\hline
ProShares Ultra High Yield & Leveraged Bonds \\
\hline
ProShares Ultra Bloomberg Natural Gas & Leveraged Commodities \\
\hline
ProShares Ultra Yen & Leveraged Currency \\
\hline
Direxion Daily Energy Bull 3X Shares & Leveraged Equities \\
\hline
ProShares Ultra Basic Materials & Leveraged Equities \\
\hline
ProShares Ultra Semiconductors & Leveraged Equities \\
\hline
ProShares Ultra Real Estate & Leveraged Real Estate \\
\hline
SPDR DJ Wilshire International Real Estate ETF & Global Real Estate \\
\hline
ProShares VIX Short-Term Futures ETF & Volatility \\
\hhline{|=|=|}
\caption{Low-correlated ETF name list in Section \ref{sec: gibs_algo}.}
\label{tab: amf_etf_representatives}
\end{longtable}
}

The low-correlated ETF name list in Section \ref{sec: vol_risk_factor} is in Table \ref{tab: vol_etf_representatives}.

{
\small
\begin{longtable}{|| m{0.61\textwidth} | m{0.31\textwidth} ||}
\hhline{|=|=|}
ETF Names  & Category  \\
\hhline{|=|=|}
\endhead
\hline
\multicolumn{2}{c}{{\footnotesize{}Continued on next page}}  \\
\endfoot
\endlastfoot
iShares Gold Trust & Precious Metals \\
\hline
iShares MSCI Malaysia ETF & Asia Pacific Equities \\
\hline
Vanguard FTSE All-World ex-US ETF & Foreign Large Cap Equities \\
\hline
iShares MSCI Canada ETF & Foreign Large Cap Equities \\
\hline
VanEck Vectors Agribusiness ETF & Large Cap Blend Equities \\
\hline
Vanguard FTSE Emerging Markets ETF & Emerging Markets Equities \\
\hline
VanEck Vectors Russia ETF & Emerging Markets Equities \\
\hline
PIMCO Enhanced Short Maturity Strategy Fund & Total Bond Market \\
\hline
iShares 3-7 Year Treasury Bond ETF & Government Bonds \\
\hline
SPDR Barclays 1-3 Month T-Bill ETF & Government Bonds \\
\hline
iShares Short Treasury Bond ETF & Government Bonds \\
\hline
iShares U.S. Home Construction ETF & Building \& Construction \\
\hline
Alerian MLP ETF & MLPs \\
\hline
SPDR Barclays High Yield Bond ETF & High Yield Bonds \\
\hline
Vanguard Healthcare ETF & Health \& Biotech Equities \\
\hline
SPDR Barclays Short Term Municipal Bond & National Munis \\
\hline
Materials Select Sector SPDR ETF & Materials \\
\hline
iShares MSCI Japan ETF & Japan Equities \\
\hline
WisdomTree Japan Hedged Equity Fund & Japan Equities \\
\hline
iShares Mortgage Real Estate ETF & Real Estate \\
\hline
Invesco DB Commodity Index Tracking Fund & Commodities \\
\hline
SPDR S\&P Retail ETF & Consumer Discretionary \hspace{2cm} Equities \\
\hline
Vanguard Financials ETF & Financials Equities \\
\hline
iShares MSCI Brazil ETF & Latin America Equities \\
\hline
iShares MSCI Mexico ETF & Latin America Equities \\
\hline
iShares Select Dividend ETF & Large Cap Value Equities \\
\hline
Invesco Water Resources ETF & Water Equities \\
\hline
SPDR DJ Wilshire Global Real Estate ETF & Global Real Estate \\
\hline
iShares North American Tech-Software ETF & Technology Equities \\
\hline
Consumer Staples Select Sector SPDR Fund & Consumer Staples Equities \\
\hline
SPDR Barclays Capital Convertible Bond ETF & Preferred Stock / \hspace{3cm} Convertible Bonds \\
\hline
Invesco Preferred ETF & Preferred Stock / \hspace{3cm} Convertible Bonds \\
\hline
Invesco DB Agriculture Fund & Agricultural Commodities \\
\hline
Industrial Select Sector SPDR Fund & Industrials Equities \\
\hline
SPDR FTSE International Government Inflation-Protected Bond ETF & Inflation-Protected Bonds\\
\hhline{|=|=|}
\caption{Low-correlated ETF name list in Section \ref{sec: vol_risk_factor}.}
\label{tab: vol_etf_representatives}
\end{longtable}
}

\end{appendices}

\end{document}